\documentclass[preprint,superscriptaddress,showpacs,preprintnumbers,amsmath,amssymb]{revtex4}

\usepackage{graphicx,color}% Include figure files
\usepackage{epsfig}
\usepackage{dcolumn}% Align table columns on decimal point
\usepackage{bm}% bold math
\usepackage{CJK}
\allowdisplaybreaks
\allowdisplaybreaks[4]
\usepackage[dvipdfm,
            pdfstartview=FitH,
            CJKbookmarks=true,
            bookmarksnumbered=true,
            bookmarksopen=true,
            colorlinks,
            pdfborder=001,
            linkcolor=blue,
            anchorcolor=blue,
            citecolor=blue
            ]{hyperref}

\begin{document}
\begin{CJK*}{GBK}{song}

\title{Collective Hamiltonian for wobbling modes}

\author{Q. B. Chen}
\affiliation{State Key Laboratory of Nuclear Physics and Technology,
             School of Physics,
             Peking University, Beijing 100871, China}%

\author{S. Q. Zhang}\email{sqzhang@pku.edu.cn}
\affiliation{State Key Laboratory of Nuclear Physics and Technology,
             School of Physics,
             Peking University, Beijing 100871, China}%

\author{P. W. Zhao}
\affiliation{State Key Laboratory of Nuclear Physics and Technology,
             School of Physics,
             Peking University, Beijing 100871, China}%
\affiliation{Yukawa Institute for Theoretical Physics,
             Kyoto University, Kyoto 606-8502, Japan}%

\author{J. Meng}\email{mengj@pku.edu.cn}
\affiliation{State Key Laboratory of Nuclear Physics and Technology,
             School of Physics,
             Peking University, Beijing 100871, China}%
\affiliation{School of Physics and Nuclear Energy Engineering, Beihang University,
             Beijing 100191, China}%
\affiliation{Department of Physics, University of Stellenbosch,
             Stellenbosch, South Africa}%

\date{\today}

\begin{abstract}

The simple, longitudinal, and transverse wobblers are systematically
studied within the framework of collective Hamiltonian, where the
collective potential and mass parameter included are obtained based
on the tilted axis cranking approach. Solving the collective Hamiltonian
by diagonalization, the energies and the wave functions of the wobbling
states are obtained. The obtained results are compared with those by
harmonic approximation formula and particle rotor model. The wobbling
energies calculated by the collective Hamiltonian are closer to the exact
solutions by particle rotor model than harmonic approximation formula. It is
confirmed that the wobbling frequency increases with the rotational frequency
in simple and longitudinal wobbling motions while decreases in transverse
wobbling motion. These variation trends are related to the stiffness of
the collective potential in the collective Hamiltonian.

\end{abstract}

\date{\today}

\maketitle

%%%%%%%%%%%%%%%%%%%%%%%%%%%%%%%%%%%%%%%%%%%%%%%%%%%%%%%%%%
%                      Introduction
%%%%%%%%%%%%%%%%%%%%%%%%%%%%%%%%%%%%%%%%%%%%%%%%%%%%%%%%%%

\section{Introduction}\label{sec1}

Atomic nuclei possess a wide variety of shapes in both their ground and
excited states. The shapes may range from spherical to deformed, from
quadrupole to octupole, and even more exotic shapes, such as superdeformed
and tetrahedral. For deformed nuclei, they in general possess axially
symmetric shape. The loss of axially symmetry would lead to triaxial shape.
The triaxiality has been invoked to describe many interesting phenomena
including $\gamma$-band~\cite{Bohr1975}, signature inversion~\cite{Bengtsson1984NPA},
anomalous signature splitting~\cite{Hamamoto1988PLB}, chiral symmetry
breaking~\cite{Frauendorf1997NPA, Frauendorf2001RMP, Meng2010JPG}, and
the wobbling motion~\cite{Bohr1975}. The wobbling motion and
chirality are regarded as fingerprints of stable triaxial nuclei.

The wobbling motion within nuclear rotation was originally introduced by Bohr
and Mottelson~\cite{Bohr1975} in the context of the triaxial rotor model (TRM).
For a rotating triaxial even-even nuclei, the rotation motions about any of
axes are all possible and the corresponding TRM Hamiltonian reads
\begin{align}
  \hat{H}_{\rm rot}=\frac{\hat{I}_1^2}{2\mathcal{J}_1}+\frac{\hat{I}_2^2}{2\mathcal{J}_2}
                   +\frac{\hat{I}_3^2}{2\mathcal{J}_3},
\end{align}
with three distinct moments of inertia $\mathcal{J}_k$ (usually defines
$\mathcal{J}_1$ as maximal) associating with each of the principle axes.
It is pointed out that although the triaxial nucleus energetically favors
the rotation about the axis with the largest moment of inertia (i.e., $1$-axis),
contributions from rotations about the other two axes ($2$ and $3$ axes) would
quantum mechanically disturb this rotation and force the angular momentum
vector off the $1$-axis. As a consequence, besides the uniform rotation about
$1$-axis, there is wobbling motion~\cite{Bohr1975}. The energies of wobbling
states, characterized by the wobbling phonon number $n$ together with total angular
momentum $I$, are
\begin{align}
  E(n,I)=\frac{I(I+1)}{2\mathcal{J}_1}+(n+\frac{1}{2})\hbar\Omega_{\rm wob}.
\end{align}
The quantum number $n$ describes the wobbling motion of the axes
with respect to the direction of $I$. For small amplitudes, this motion
has the character of a harmonic vibration with wobbling frequency given by
\begin{align}\label{eq6}
  \hbar\Omega_{\rm wob}= 2I\sqrt{\Big(\frac{\hbar^2}{2\mathcal{J}_2}
  -\frac{\hbar^2}{2\mathcal{J}_1}\Big)
  \Big(\frac{\hbar^2}{2\mathcal{J}_3}-\frac{\hbar^2}{2\mathcal{J}_1}\Big)},
\end{align}
which is related to the moments of three axes and found to be proportional
to the spin. Similar as Ref.~\cite{Frauendorf2014PRC}, such type of wobbling
motion for a triaxial rotor is also denoted as ``\emph{simple wobbler}''
at the present investigation.

The wobbling motion appears not only in the even-even nuclei but also
in the odd-$A$ nuclei. For rotating odd-$A$ triaxial nuclei, there are two
types of wobbling motions suggested by Frauendorf and D\"{o}nau~\cite{Frauendorf2014PRC}
very recently according to the relation between the orientation of quasiparticle
angular momentum vector with respect to the rotor axis with the largest moment of
inertia. If the quasiparticle angular momentum vector is aligned with the axis
with the largest moment of inertia, it is called ``\emph{longitudinal wobbler}''.
If the quasiparticle angular momentum vector is perpendicular to the axis with the
largest moment of inertia, it is called ``\emph{transverse wobbler}''. Assuming
frozen alignment of the quasiparticle with one of the rotor axes and harmonic
oscillations (HFA), a rather simple analytic expression for wobbling frequency of
these two types of wobbling motions is derived~\cite{Frauendorf2014PRC}.
According to this analytic expression, the increasing trend of wobbling frequency
for longitudinal wobbling motion and decreasing trend for transverse wobbling can
be expected.

On the experimental side, although the wobbling phenomenon has been predicted
for a long time~\cite{Bohr1975}, it was not observed until the beginning
of this century when the first experimental evidence was reported in
$^{163}$Lu~\cite{Odegard2001PRL}. Subsequently, it has been extensively studied in
the triaxial strongly deformed (TSD) region around $N = 94$, where the wobbling bands
have been identified in $^{161,163,165,167}\rm Lu$~\cite{Jensen2002PRL, Jensen2002NPA,
Schonwasser2003PLB, Bringel2005EPJA, Amro2003PLB, Hagemann2004EPJA} and
$^{167}\rm Ta$~\cite{Hartley2009PRC}. All wobbling bands in this mass region
are based on $\pi i_{13/2}$ configuration. Very recently, a new candidate wobbling band
is proposed in $^{135}\rm Pr$~\cite{Frauendorf2014PRC}, which is built on $\pi h_{11/2}$
configuration, differing from the configuration of previous known examples. For
even-even nuclei, however, the wobbling spectra are scarce since stable triaxial
ground states are rare. The best example identified so far is $^{112}\rm Ru$~\cite{S.J.Zhu2009IJMPE}.

On the theoretical side, the wobbling motion was firstly investigated by
TRM~\cite{Bohr1975}. Following the discovery of the first wobbling structure in odd-$A$
$^{163}\rm Lu$~\cite{Odegard2001PRL}, the quantal particle rotor model (PRM) was used to
describe the wobbling mode, see Refs.~\cite{Hamamoto2002PRC, Hamamoto2003PRC, Tanabe2006PRC,
Tanabe2008PRC, Frauendorf2014PRC}. Based on the framework of mean field theory, there are
many efforts to extend the cranking model to study the wobbling motion. Due to the mean-filed
approximation, cranking model yields only the yrast sequence for a given configuration,
Therefore, in order to describe the wobbling excitations, one has to go beyond the
mean-filed approximation. At present, this has been done by incorporating the quantum
correlations by means of random phase approximation (RPA)~\cite{Shimizu1995NPA, Matsuzaki2002PRC,
Matsuzaki2003EPJA, Matsuzaki2004PRCa, Matsuzaki2004PRC, Shimizu2005PRC, Shimizu2008PRC,
Shoji2009PTP} or by the generator coordinate method after angular momentum projection
(GCM+AMP) based on the cranking intrinsic states~\cite{Oi2000PLB}.

Another promising method is to construct a collective Hamiltonian on
the top of cranking mean field solutions. By taking into account the quantum
fluctuation along the collective degree of freedom, the collective Hamiltonian
goes beyond the mean-field approximation and restores the broken symmetry~\cite{Q.B.Chen2013PRC}.
This has been implemented based on the framework of tilted axis cranking (TAC)
single-$j$ shell model to investigate the chiral vibration and rotation
motions~\cite{Mukhopadhyay2007PRL, B.Qi2009PLB}. The chiral symmetry broken in the intrinsic
reference frame is restored and chiral doublet bands are obtained in the
laboratory reference frame. For wobbling motion, the wobbling states are
formed due to the quantum fluctuation of the total angular momentum
deviating from the principle axes of the rotor. It is thus interesting
to extend the collective Hamiltonian to describe the phenomenon of
wobbling motion.

In this work, the collective Hamiltonian will be extended to study the simple,
longitudinal, and transverse wobbling motions, in particular, to examine the trend
of the wobbling frequency with respect to the rotational frequency. In the collective
Hamiltonian, the collective potentials are calculated from TAC model and the mass
parameter is obtained with the assumption of harmonic approximation (HA) for simple
wobbling motion or HFA approximation for longitudinal and transverse wobbling
motions. The energy levels and wave function of wobbling states are obtained
by diagonalizing the collective Hamiltonian. The corresponding energy spectra will
be in comparison with the results obtained by HA (HFA) analytic expression as well
as TRM (PRM) for simple (longitudinal and transverse) wobbling to evaluate the accuracy
of the collective Hamiltonian.

The paper is organized as follows. In Sec.~\ref{sec2}, a brief introduction to
the collective Hamiltonian is given. The corresponding numerical details adopted
in the calculations are presented in Sec.~\ref{sec3}. In Sec.~\ref{sec4}, the
obtained potential energy and the mass parameter are respectively
shown for the three types of wobbling motions and the corresponding energy levels
and wave functions obtained by collective Hamiltonian are discussed in details.
A brief summary is given in Sec.~\ref{sec5}.

\section{Theoretical framework}\label{sec2}

\begin{figure}[!th]
  \begin{center}
    \includegraphics[width=7 cm]{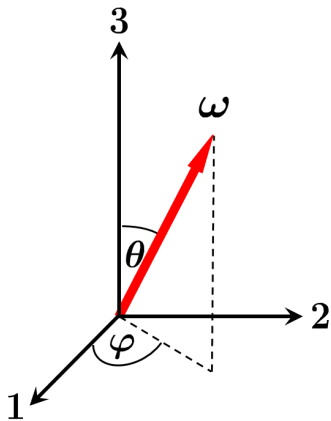}
    \caption{(Color online) Orientation of the rotational frequency $\bm{\omega}$
    with respect to the principal axes.}\label{fig1}
  \end{center}
\end{figure}

The collective Hamiltonian, in terms of a few numbers of collective
coordinates and momenta, is an effective method for describing various collective
processes which involve small velocities. The well-known Bohr Hamiltonian describe
the collective rotational and vibrational degrees of freedom with the five collective
intrinsic variables $\beta$, $\gamma$, and Euler angles $\bm{\Omega}$~\cite{Bohr1975}.
In Ref.~\cite{Q.B.Chen2013PRC}, to describe the chiral motions in triaxial rotational
nuclei, a collective Hamiltonian based on the TAC solutions was constructed.
Therein, the orientation of nucleus in rotating mean-field, described
by polar angle $\theta$ and azimuth angle $\varphi$ in the spherical coordinate
as illustrated in Fig.~\ref{fig1}, is considered as collective variable. As the motion
along $\varphi$ direction is much easier than $\theta$ direction, the collective
Hamiltonian has been restricted to one dimensional motion along $\varphi$
direction~\cite{Q.B.Chen2013PRC}.

For the wobblers caused by the quantum fluctuation of the total angular momentum
orientation, the azimuth angle $\varphi$ can also be taken as collective coordinate
to the wobbling motions and the wobbling excitation is restricted to one dimensional
motion along $\varphi$ direction. It is necessary to mention that in a semi-classical
model for wobbling motion, the azimuth angle $\varphi$ has been interpreted as the
wobbling angle of the total angular momentum vector~\cite{Oi2006PLB}.

The detailed theoretical framework of collective Hamiltonian based on the TAC
solutions has been formulated in Ref.~\cite{Q.B.Chen2013PRC}. The formalism can
be analogized to describe the wobbling motion. Here for completeness, a brief
introduction to the formalism is presented.

\subsection{Collective Hamiltonian}

Taking $\varphi$ as the collective variable, the classical form of collective
Hamiltonian is written as the sum of kinetic and potential terms
\begin{align}
  H_{\rm coll}=T_{\rm kin}(\varphi)+V(\varphi)
              =\displaystyle \frac{1}{2}B(\varphi)\dot{\varphi}^2+V(\varphi),
\label{eq13}
\end{align}
where $V(\varphi)$ is collective potential and $B(\varphi)$ mass parameter.
The quantized form of the collective Hamiltonian is obtained according to
the Pauli prescription~\cite{Pauli1933}
\begin{align}
  \hat{H}_{\rm coll}=-\displaystyle \frac{\hbar^2}{2\sqrt{B(\varphi)}}
  \frac{\partial}{\partial\varphi}\frac{1}{\sqrt{B(\varphi)}}
  \frac{\partial}{\partial \varphi}+V(\varphi).
\end{align}
By solving this Hamiltonian on the basis states with appropriate boundary
condition on $\varphi$, e.g., box boundary condition~\cite{Q.B.Chen2013PRC},
the wobbling levels and corresponding wave functions can be obtained.

\subsection{Collective potential $V(\varphi)$}

Both the collective potential $V(\varphi)$ and the mass parameter $B(\varphi)$
in the collective Hamiltonian~(\ref{eq13}) can be determined based on TAC model.

%\subsubsection{Longitudinal and transverse wobblers}

Let us first discuss $V(\varphi)$ for the cases of longitudinal and transverse wobblers.
For schematic discussions, we consider a system of a high-$j$ particle coupled to
a triaxial rotor. The cases for more than one particle coupled to triaxial rotor
can be easily extended as well. The cranking Hamiltonian reads
\begin{align}
  \hat{h}^\prime &=\hat{h}_{\rm def}-\bm{\omega}\cdot\hat{\bm{j}},\notag\\
  \bm{\omega}    &=(\omega\sin\theta\cos\varphi, \omega\sin\theta\sin\varphi,
                    \omega\cos\theta),
\end{align}
where $\hat{\bm{j}}$ is the single particle angular momentum and deformed single
particle Hamiltonian $\hat{h}_{\rm def}$ is taken as the single-$j$ shell
Hamiltonian
\begin{align}
  \hat{h}_{\rm def}=\frac{1}{2}C\Big\{(\hat{j}_3^2-\frac{j(j+1)}{3})\cos\gamma
                   +\frac{1}{2\sqrt{3}}(\hat{j}_+^2+\hat{j}_-^2)\sin\gamma\Big\}.
\end{align}
Diagonalizing the cranking Hamiltonian, one obtains the total Routhian
\begin{align}\label{eq9}
  E^\prime(\theta,\varphi)=\langle h^\prime\rangle-
  \frac{1}{2}\sum_{k=1}^3\mathcal{J}_k \omega_k^2, \quad \mathcal{J}_k: \textrm{moments of inertia},
\end{align}
Minimizing the total Routhian with respect to $\theta$ for given $\varphi$,
the collective potential $V(\varphi)$ is finally obtained.

%\subsubsection{Simple wobbler}

For simple wobbler, i.e., a simple triaxial rotor without coupling any
particles, the total Routhian (\ref{eq9}) is degenerated to
\begin{align}\label{eq10}
  E^\prime(\theta,\varphi)=-\frac{1}{2}\sum_{k=1}^3\mathcal{J}_k\omega_k^2,
\end{align}
and similarly the collective potential $V(\varphi)$ is obtained by minimizing the total Routhian
with respect to $\theta$ for given $\varphi$.

\subsection{Mass parameter $B$}

%\subsubsection{Simple wobbler}

Before discussing how to calculate the mass parameter, it is worth noting once more
that as pointed out by Bohr and Mottelson~\cite{Bohr1975}, the wobbling motion
as a small amplitude vibration has the character of a harmonic oscillation with
frequency $\Omega_{\rm wob}$. As well-known, the oscillation frequency $\Omega$ for a
harmonic oscillator system is related to the mass parameter $B$ of the oscillator and
the stiffness parameter $C$ of the harmonic oscillator potential by
\begin{align}
  \Omega=\sqrt{\frac{C}{B}}.
\end{align}
Therefore, once the stiffness parameter $C$ and oscillation frequency $\Omega$
are determined, the mass parameter $B$ can be obtained.

To extract the stiffness parameter $C$ of the collective potential
$V(\varphi)$, one can expand the collective potential
$V(\varphi)$ by Taylor series at $\varphi=0^\circ$ up to $\sim \varphi^2$
terms, i.e., the harmonic approximation (HA) is adopted.
For the total Routhian (\ref{eq10}) of simple wobbler, one can find
that its minimum along $\theta$ direction is always at $\theta=90^\circ$
for any value of $\varphi$. Therefore, the collective potential becomes
\begin{align}
  \label{eq1}
  V(\varphi)&=-\frac{1}{2}\omega^2(\mathcal{J}_1\cos^2\varphi+\mathcal{J}_2\sin^2\varphi)\\
  \label{eq2}
            &\approx -\frac{1}{2}\mathcal{J}_1\omega^2+\frac{1}{2}\omega^2(\mathcal{J}_1
            -\mathcal{J}_2)\varphi^2, \quad\quad \textrm{for $\varphi\to 0^\circ$}.
\end{align}
The Eq.~(\ref{eq2}) suggests that the collective potential can be regarded
as the sum of a rotational energy term along 1-axis with frequency
$\omega$ and a harmonic oscillation potential term along $\varphi$ direction
with  stiffness parameter $C=\omega^2(\mathcal{J}_1 - \mathcal{J}_2)$.
Thus the wobbling frequency $\Omega_{\rm wob}$ and the mass parameter $B$ are
related each other by
\begin{align}\label{eq3}
\hbar\Omega_{\rm wob}=\hbar\sqrt{\frac{C}{B}}
                     =\hbar\omega\sqrt{\frac{\mathcal{J}_1-\mathcal{J}_2}{B}}.
\end{align}

To determine the mass parameter $B$ in Eq.~(\ref{eq3}), we further recall the
wobbling frequency (\ref{eq6}) given by Bohr and Mottelson~\cite{Bohr1975}
\begin{align}\label{eq4}
  \hbar\Omega_{\rm wob}
  &=2I\sqrt{\Big(\frac{\hbar^2}{2\mathcal{J}_2}-\frac{\hbar^2}{2\mathcal{J}_1}\Big)
    \Big(\frac{\hbar^2}{2\mathcal{J}_3}-\frac{\hbar^2}{2\mathcal{J}_1}\Big)}\notag\\
  &=\frac{\hbar^2 I}{\mathcal{J}_1}\sqrt{\frac{(\mathcal{J}_1-\mathcal{J}_2)
    (\mathcal{J}_1-\mathcal{J}_3)}{\mathcal{J}_3\mathcal{J}_2}}\notag\\
  &=\hbar\omega\sqrt{\frac{(\mathcal{J}_1-\mathcal{J}_2)(\mathcal{J}_1-\mathcal{J}_3)}
    {\mathcal{J}_3\mathcal{J}_2}}.
\end{align}
Combining Eqs.~(\ref{eq3}) and (\ref{eq4}), the mass parameter is obtained
for simple wobbler
\begin{align}\label{eq11}
  B=\frac{\mathcal{J}_2\mathcal{J}_3}{\mathcal{J}_1-\mathcal{J}_3}.
\end{align}
It is determined only by the moments of inertia of three principal axes and
independent of rotational frequency.

%\subsubsection{Longitudinal and transverse wobblers}

For longitudinal and transverse wobblers, we introduce the harmonic frozen
alignment (HFA) approximation as Ref.~\cite{Frauendorf2014PRC}, i.e., the
angular momentum of the odd particle is assumed to be firmly aligned with
the short-axis (1-axis) and can be considered as a number. Then for a given
rotational frequency $\omega$, the moment of inertia of 1-axis is treated
as a $\omega$-dependent effective moment of inertia
\begin{align}
  \mathcal{J}_1^*(\omega)=\frac{\mathcal{J}_1\omega+j}{\omega}=\mathcal{J}_1
  +\frac{j}{\omega}.
\end{align}
The odd-particle contributes a $\omega$-dependent
term to the effective moment of inertia, which will decrease with the increasing
rotational frequency.

Similar to simple wobbler, it can be also found that for the longitudinal and
transverse wobblers the collective potential obtained from the total Routhian
$E^\prime(\theta,\varphi)$ (\ref{eq9}) is minimized at $\theta=90^\circ$ for
any given $\varphi$. Therefore, the collective potential
is written as
\begin{align}
  \label{eq7}
  V(\varphi)
   &=\langle \hat{h}_{\rm def} \rangle-\omega j\cos\varphi -\frac{1}{2}\omega^2(\mathcal{J}_1\cos^2\varphi
    +\mathcal{J}_2\sin^2\varphi)\\
   &\approx \langle \hat{h}_{\rm def} \rangle -\omega j(1-\frac{\varphi^2}{2})
    -\frac{1}{2}\mathcal{J}_1\omega^2+\frac{1}{2}\omega^2(\mathcal{J}_1-\mathcal{J}_2)\varphi^2,
     \quad\quad \textrm{for $\varphi\to 0$}\notag\\
   &=\langle \hat{h}_{\rm def} \rangle -\frac{1}{2}\omega j
    -\frac{1}{2}\Big(\mathcal{J}_1+\frac{j}{\omega}\Big)\omega^2
    +\frac{1}{2}\omega^2\Big[\Big(\mathcal{J}_1+\frac{j}{\omega}\Big)
    -\mathcal{J}_2\Big]\varphi^2\notag\\
   \label{eq8}
   &=\langle \hat{h}_{\rm def} \rangle-\frac{1}{2}\omega j-\frac{1}{2}\mathcal{J}_1^*\omega^2
   +\frac{1}{2}\omega^2\Big[\mathcal{J}_1^*(\omega)-\mathcal{J}_2\Big]\varphi^2.
\end{align}
This formula is similar to Eq.~(\ref{eq2}) except that the moment of inertia of
1-axis $\mathcal{J}_1$ has been replaced by the effect moment of inertia
$\mathcal{J}_1^*(\omega)$, thereby one expects the mass parameter for longitudinal
and transverse wobblers has the similar form as simple wobbler
\begin{align}\label{eq12}
  B(\omega)=\frac{\mathcal{J}_2\mathcal{J}_3}{\mathcal{J}_1^*(\omega)-\mathcal{J}_3}
  =\frac{\mathcal{J}_2\mathcal{J}_3}{(\mathcal{J}_1-\mathcal{J}_3)
  +\displaystyle\frac{j}{\omega}}.
\end{align}
Differing from the mass parameter (\ref{eq11}) for simple wobbler, it is determined
not only by the moments of inertia of three principal axes, but also by the angular
momentum of the odd particle and the rotational frequency. As the rotational
frequency increases, the mass parameter for longitudinal and transverse wobblers
will increase as well.

The wobbling frequency for the longitudinal and transverse wobbling motions
can be then obtained from Eq.~(\ref{eq3})
\begin{align}\label{eq5}
  \hbar\Omega_{\rm wob}
  &=\sqrt{\frac{\mathcal{J}_1^*(\omega)-\mathcal{J}_2}{B(\omega)}}\hbar\omega\notag\\
  &=\hbar\sqrt{\frac{\big[(\mathcal{J}_1-\mathcal{J}_3)\omega+j\big]
  \big[(\mathcal{J}_1-\mathcal{J}_2)\omega+j\big]}
  {\mathcal{J}_2\mathcal{J}_3}}.
\end{align}
This formula is nothing but the HFA formula in Ref.~\cite{Frauendorf2014PRC} by replacing
the spin with $\mathcal{J}_1\omega+j$. For longitudinal wobbling motion, since
$\mathcal{J}_1>\mathcal{J}_2,\mathcal{J}_3$, the wobbling frequency increases with
the rotational frequency. While for transverse wobbling motion, since
$\mathcal{J}_2>\mathcal{J}_1$, the wobbling frequency decreases
with the rotational frequency, and will reach to zero at a critical
rotational frequency $\hbar\omega_{\rm c}=j/(\mathcal{J}_2-\mathcal{J}_1)$.

\section{Numerical details}\label{sec3}

In the following calculations, a triaxial rotor with the deformation
parameters $\beta=0.25$ and $\gamma=-30^\circ$ is considered to investigate
the simple wobbling motion. Following the notation as in Ref.~\cite{Ring1980book},
for such deformation, three principal axes $1$, $2$, and $3$-axis respectively
correspond to short ($s$), intermediate ($i$), and long ($l$) axis. For the
investigation of the longitudinal and transverse wobbling motions, the triaxial
rotor is assumed to be further coupled with a $h_{11/2}$ proton particle.
Thus the proton aligns its angular momentum along short axis (namely,
1-axis). The longitudinal (transverse) wobblers is achieved by choosing
1-axis to be (perpendicular to) the axis with largest moments of inertia.

With regard to the moments of inertia, both the rigid body type
\begin{align}\label{eq14}
  \mathcal{J}_k^{\rm rig}
  &=\frac{2}{5}mAR_0^2\displaystyle\Big[1-\sqrt{\frac{5}{4\pi}}
  \beta\cos(\gamma-\frac{2\pi}{3}k)\Big]\notag\\
  &=\mathcal{J}_0^{\rm rig}\displaystyle \Big[1-\sqrt{\frac{5}{4\pi}}
  \beta\cos(\gamma-\frac{2\pi}{3}k)\Big], \quad k=1, 2, 3
\end{align}
and the irrotational flow type
\begin{align}\label{eq15}
  \mathcal{J}_k^{\rm irr}
  &=\frac{3}{2\pi}mAR_0^2\beta^2\sin^2(\gamma-\frac{2\pi}{3}k)\notag\\
  &=\mathcal{J}_0^{\rm irr}\sin^2(\gamma-\frac{2\pi}{3}k), \quad k=1, 2, 3
\end{align}
are often assumed~\cite{Ring1980book}. $\mathcal{J}_k^{\rm rig}$ shows less
dependence on the deformation $\beta$ than $\mathcal{J}_k^{\rm irr}$ ($\sim \beta^2$).
In the $\gamma$-dependence, $\mathcal{J}^{\rm irr}$ vanishes about the symmetry axes
while $\mathcal{J}^{\rm rig}$ not and the largest moment of inertia axes of them
are different. For the present deformation parameters $\beta=0.25$ and $\gamma=-30^\circ$,
the largest moment of inertia axis is 1-axis ($s$-axis) for rigid body type while
2-axis ($i$-axis) for irrotational flow type.

In present investigation, the wobbling angle $\varphi$ in collective Hamiltonian
is restricted to $-\pi/2 \leq \varphi \leq \pi/2$, or in other words the wobbling motion
happens around 1-axis. For simple wobbler, the rigid body type of moments of inertia
(\ref{eq14}) is adopted for 1-axis being the axis with the largest moment of inertia.
Similarly, the rigid body type of moments of inertia is also applied to longitudinal wobbler
so that the orientation of proton angular momentum (1-axis) being parallel to the axis
with largest moments of inertia. For transverse wobbler, while the orientation of proton
angular momentum (1-axis) is required to be perpendicular to the axis with largest moments
of inertia, the irrotational flow type of moments of inertia (\ref{eq15}) is adopted.
In the calculations, the constants $\mathcal{J}_0^{\rm rig}$ and $\mathcal{J}_0^{\rm irr}$
in Eqs. (\ref{eq14}) and (\ref{eq15}) are respectively taken as $\mathcal{J}_0^{\rm rig}
=256\pi/15~\hbar^2/\rm MeV$ and $\mathcal{J}_0^{\rm irr}=40~\hbar^2/\rm MeV$.

\section{Results and discussion}\label{sec4}

\subsection{Simple wobbler}

We first present the results of our calculations for the simple wobbling motion by means
of the TAC model and collective Hamiltonian. As described in Sec.~\ref{sec2}, the collective
potential and the mass parameter included in the collective Hamiltonian are respectively
calculated by TAC model and Eq.~(\ref{eq11}). The obtained wobbling energies will be
compared with the HA formula and the exact TRM.

\subsubsection{Collective potential}

\begin{figure}[!th]
  \begin{center}
    \includegraphics[width=6 cm]{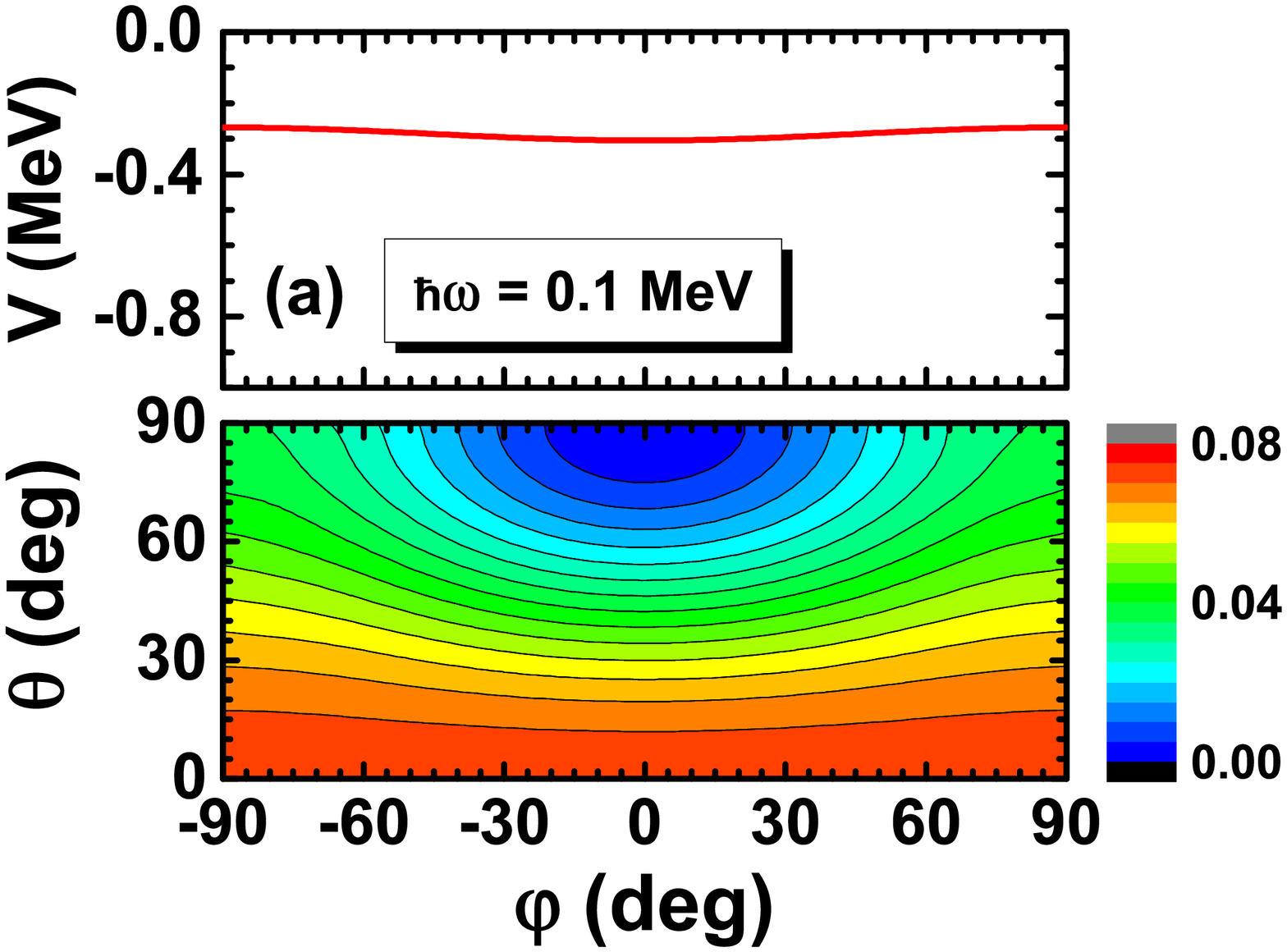}\quad
    \includegraphics[width=6 cm]{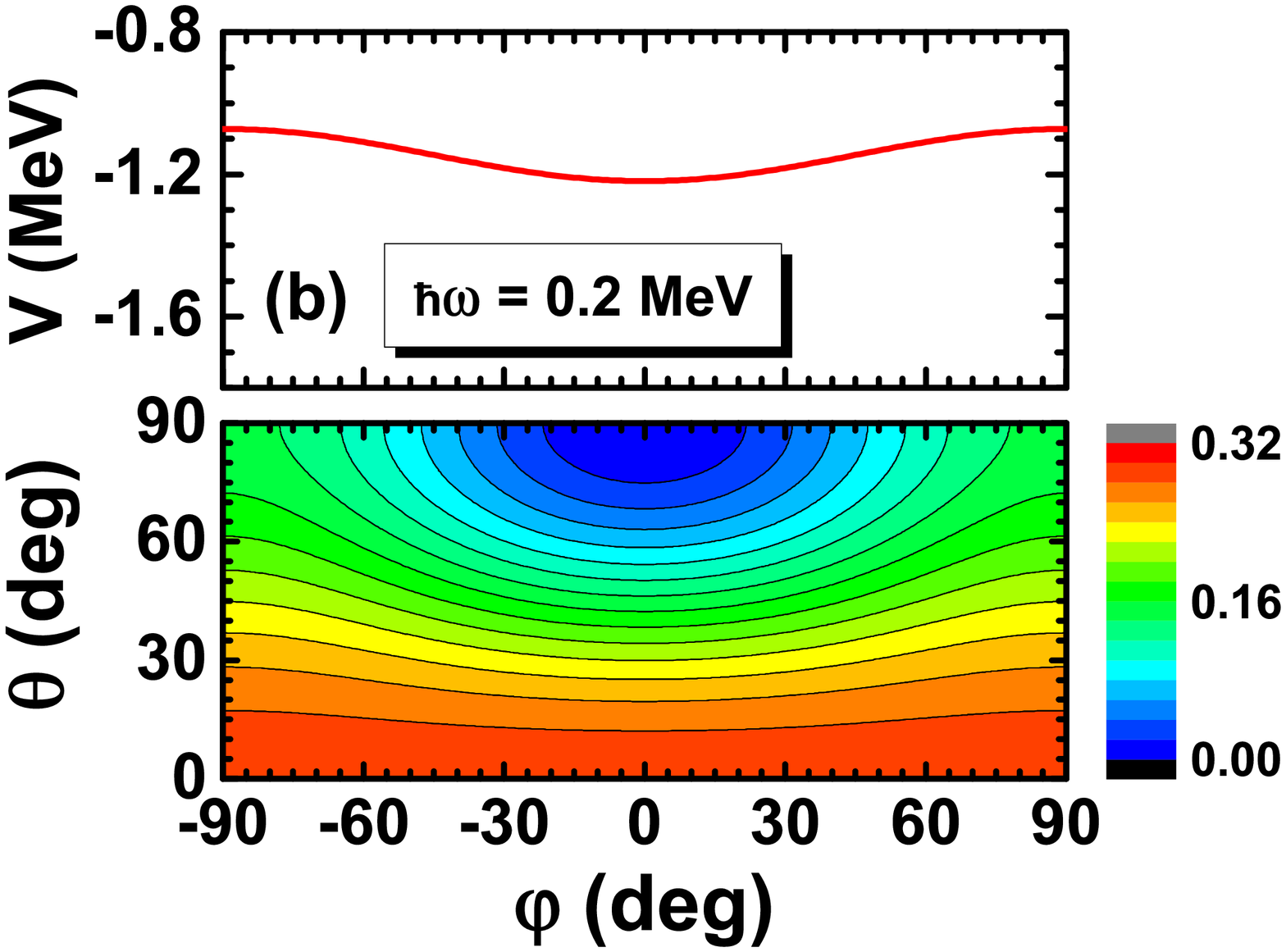}\\
    \includegraphics[width=6 cm]{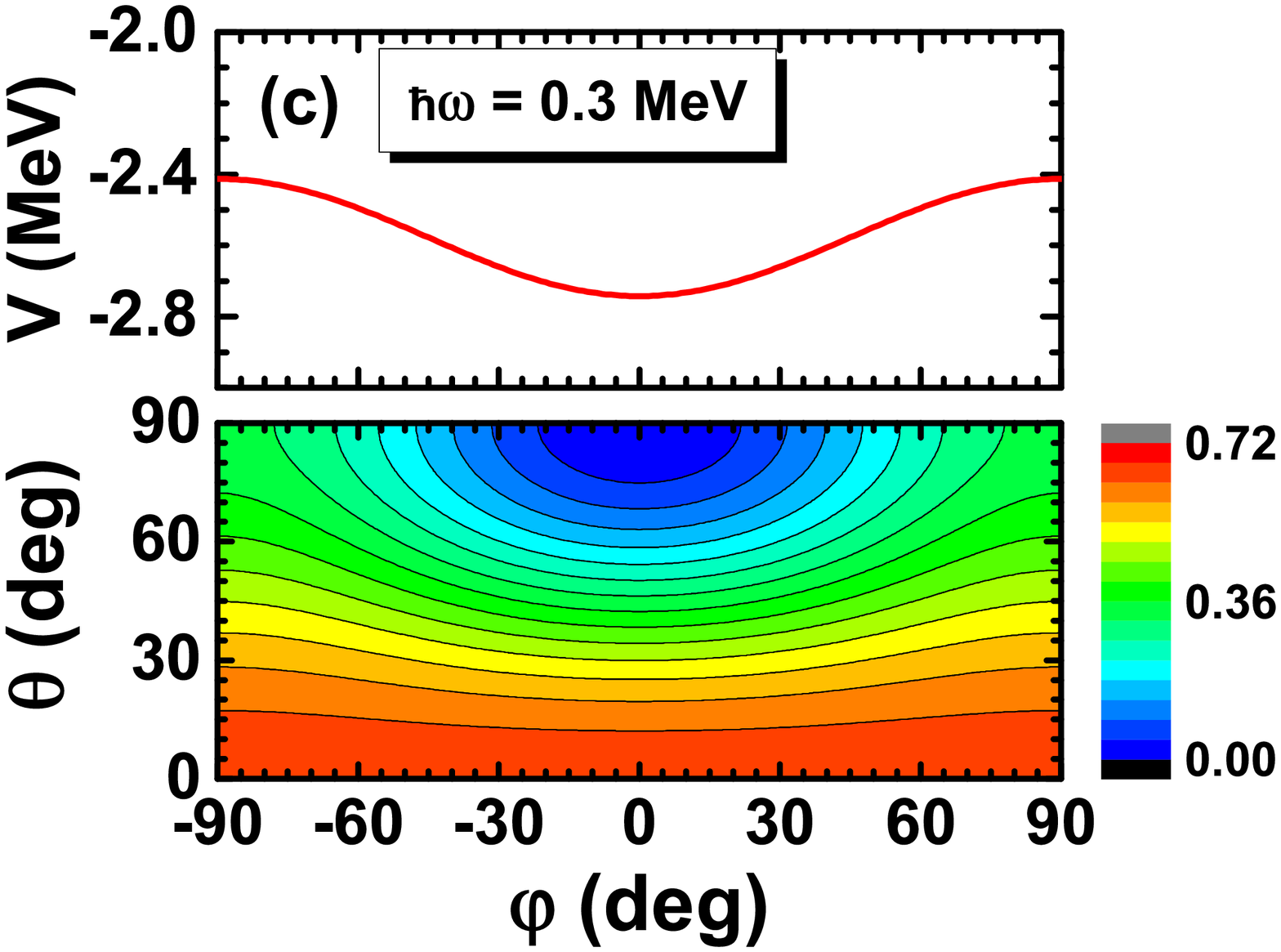}\quad
    \includegraphics[width=6 cm]{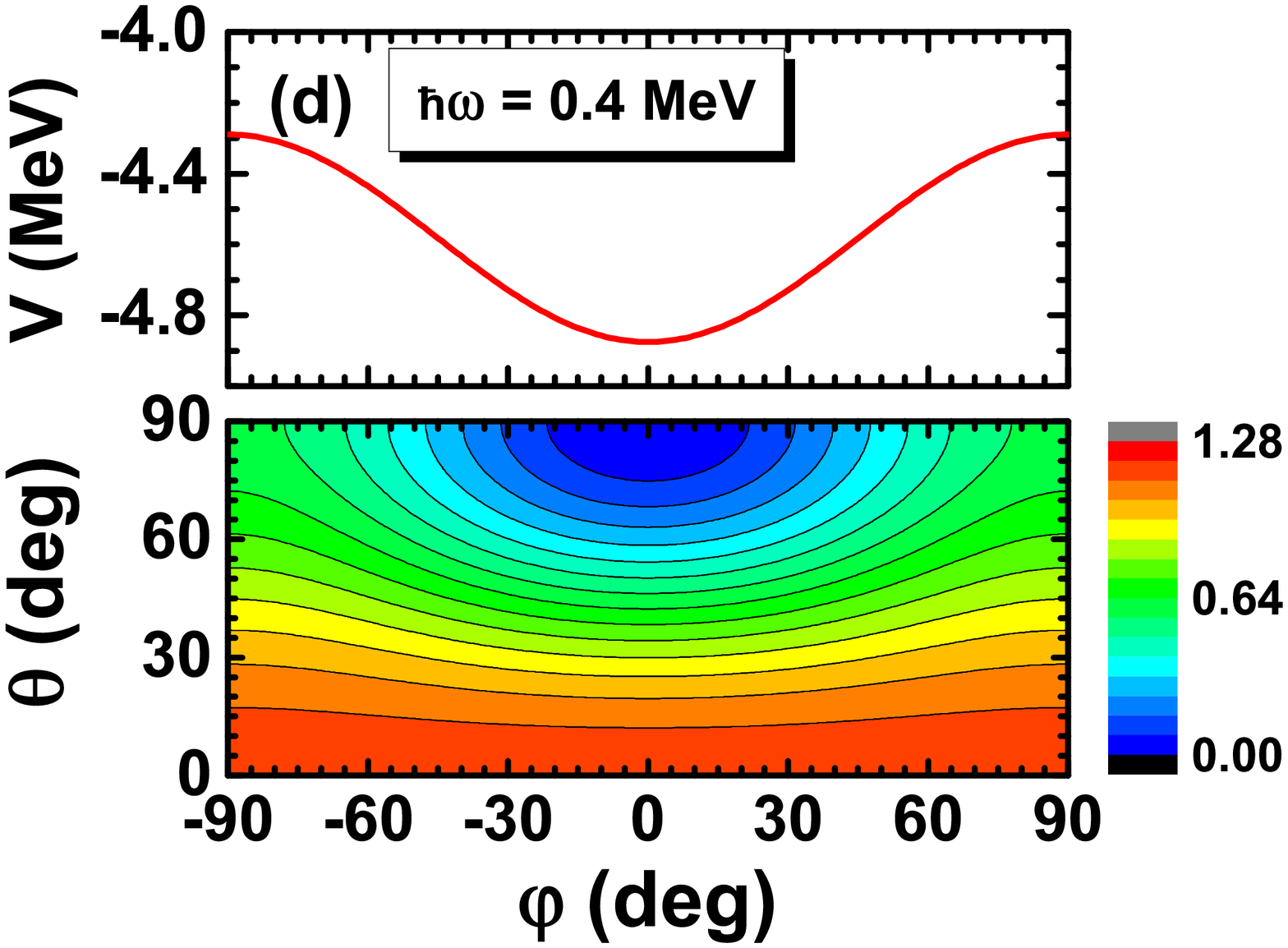}
    \caption{(Color online) Lower panels: Contour plots of total Routhian surface
             $E^\prime(\theta,\varphi)$ for a triaxial rigid body rotor with $\gamma=-30^\circ$
             at the frequencies $\hbar\omega=0.1$, $0.2$, $0.3$ and $0.4~\rm MeV$. All energies
             at each rotational frequency are normalized with respect to the
             absolute minimum. Upper panels: The collective potential $V(\varphi)$ as a function of
             $\varphi$ extracted from the corresponding total Routhian surface calculation.}\label{fig2}
  \end{center}
\end{figure}

In contour plots of Fig.~\ref{fig2}(a)-(d), the total Routhian $E^\prime(\theta,\varphi)$
(\ref{eq10}) in the ($\theta$, $\varphi$) plane at the rotational frequencies
$\hbar\omega=0.1$, $0.2$, $0.3$, and $0.4~\rm MeV$ are shown. All the potential energy
surfaces are symmetrical with respect to $\varphi=0^\circ$ line. With the increasing rotational
frequency, the minima in the potential energy surfaces always locate at $(\theta=90^\circ,
\varphi=0^\circ)$, which corresponds to uniform rotation about the axis with the largest
moment of inertia.

Minimizing the Routhian $E^\prime(\theta,\varphi)$ with $\theta$ for given $\varphi$,
we find that the minimum along $\theta$ direction is always at $\theta=90^\circ$ for
any value of $\varphi$ at each rotational frequency. The corresponding extracted
collective potentials $V(\varphi)$ are shown in the upper panels of Fig.~\ref{fig2}(a)-(d)
respectively for $\hbar\omega=0.1$, $0.2$, $0.3$, and $0.4~\rm MeV$. Again, the potential
energy is symmetrical about $\varphi=0^\circ$ in correspondence with the results
displayed in the lower panels of Fig.~\ref{fig2}(a)-(d). For all cases, the potential
$V(\varphi)$ is a harmonic oscillator type that has only one minimum at $\varphi=0^\circ$,
corresponding to the rotation about 1-axis. The stiffness of the collective potential
becomes larger as the rotational frequency increases. This is directly reflected by the
increase of energy difference between $\varphi=\pm 90^\circ$ and $\varphi=0^\circ$. For example, the
value is only $\sim 30~\rm keV$ at $\hbar\omega=0.1~\rm MeV$ while reaches to $\sim 600~\rm keV$
at $\hbar\omega=0.4~\rm MeV$.

\subsubsection{Collective levels and wave functions}

\begin{figure}[!th]
  \begin{center}
   \includegraphics[width=4.55 cm]{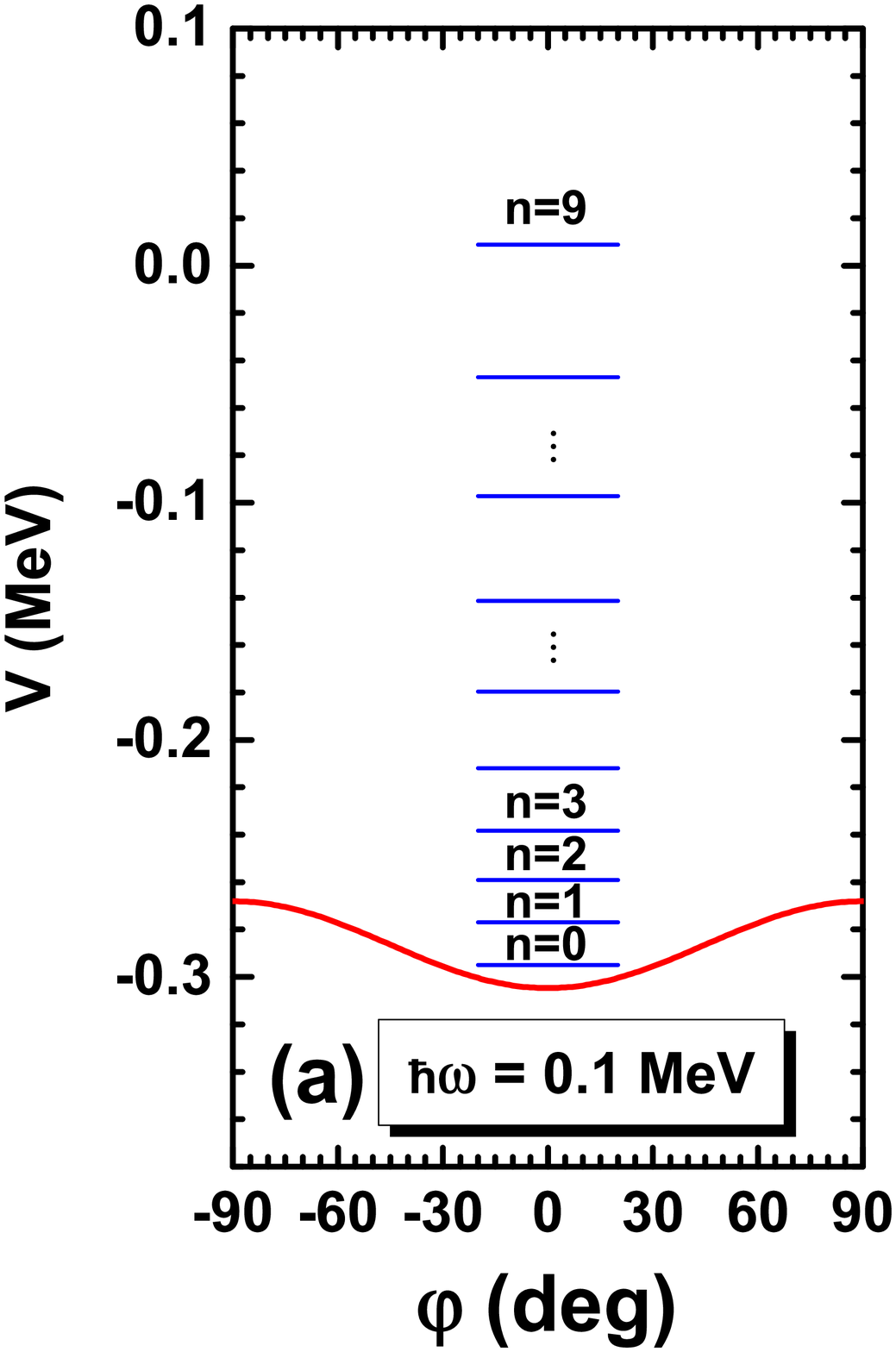}\quad\quad
   \includegraphics[width=4.55 cm]{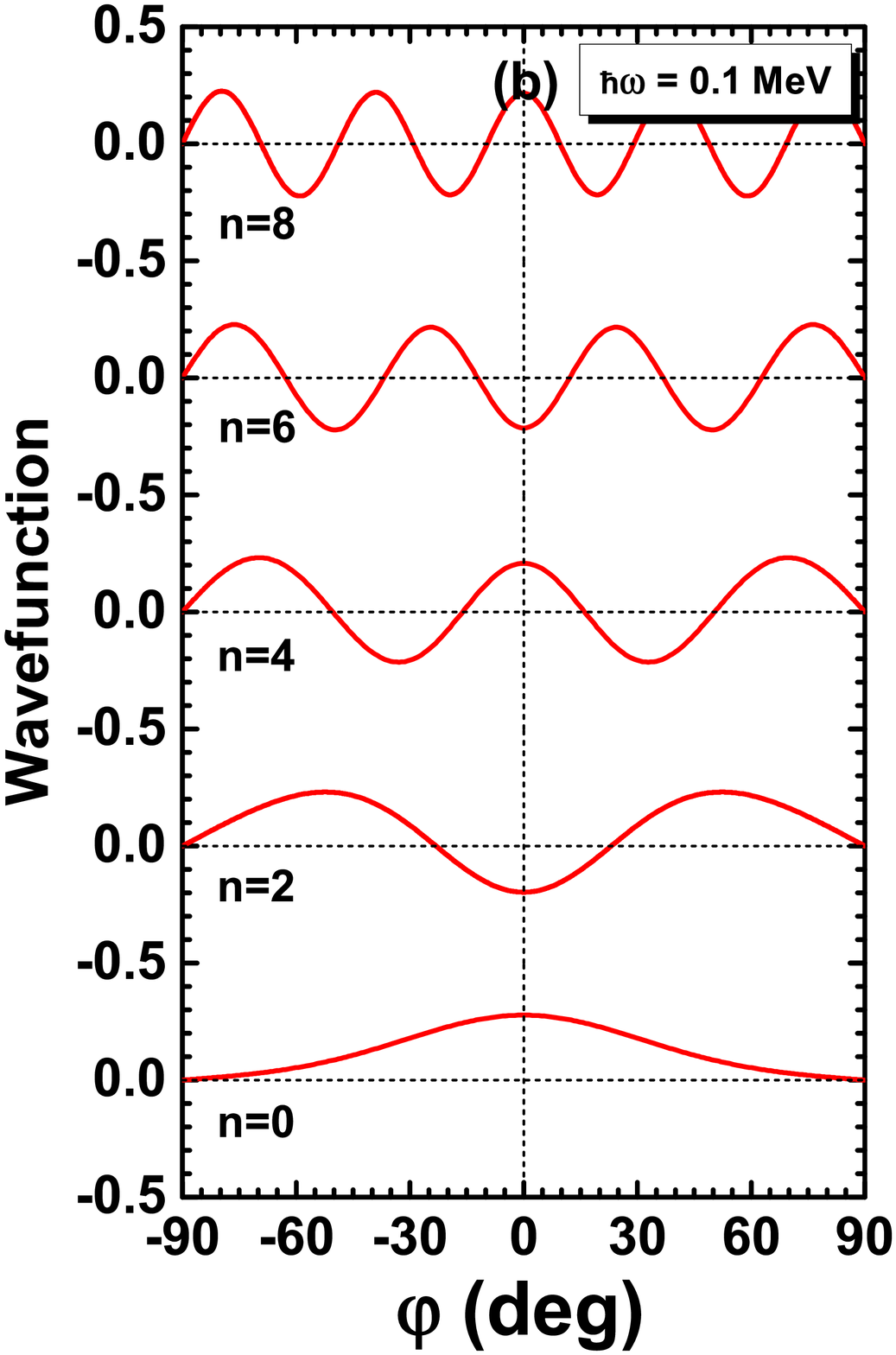}
   \includegraphics[width=4.55 cm]{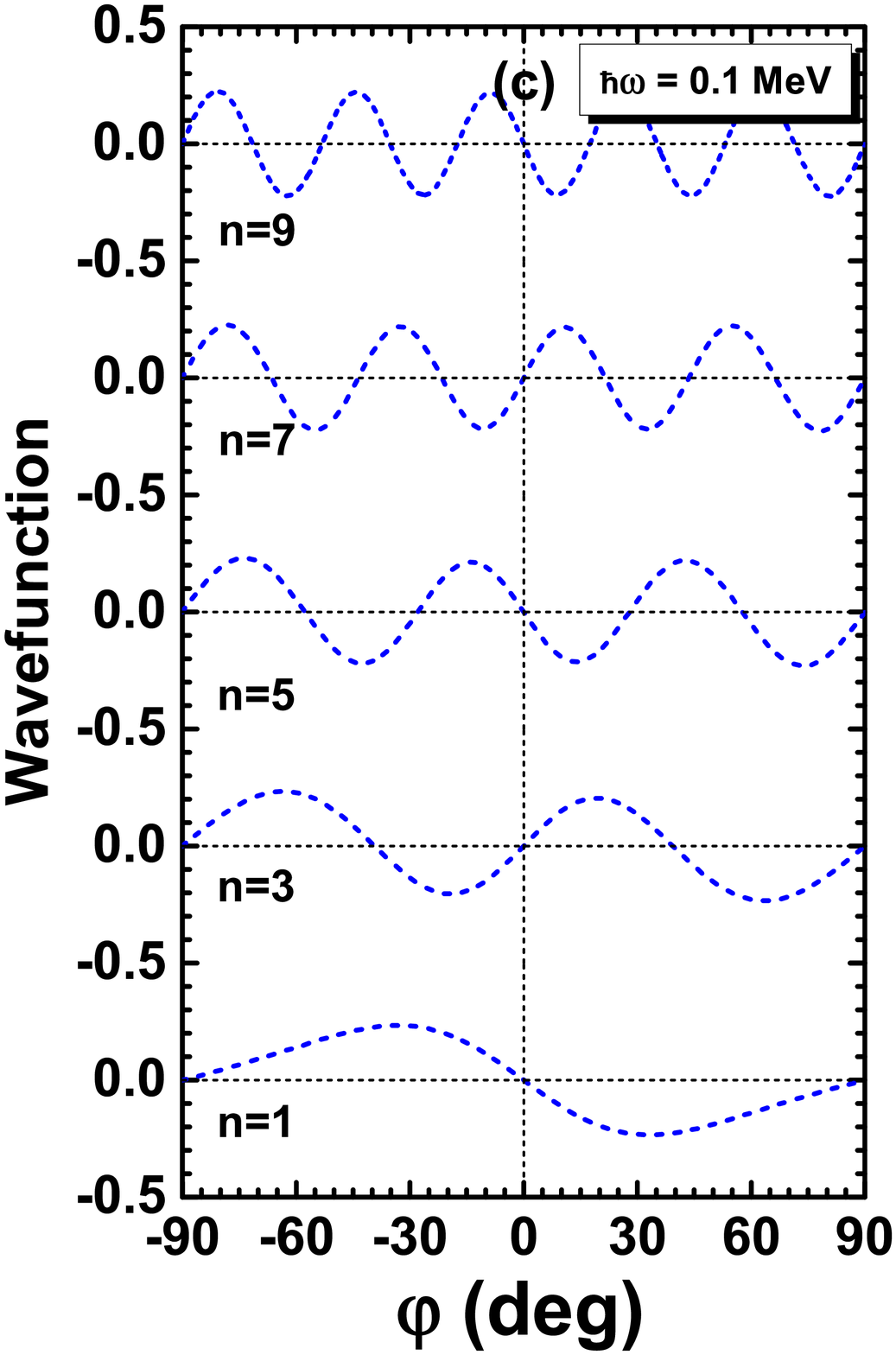}\\
   ~~\\
   \includegraphics[width=4.55 cm]{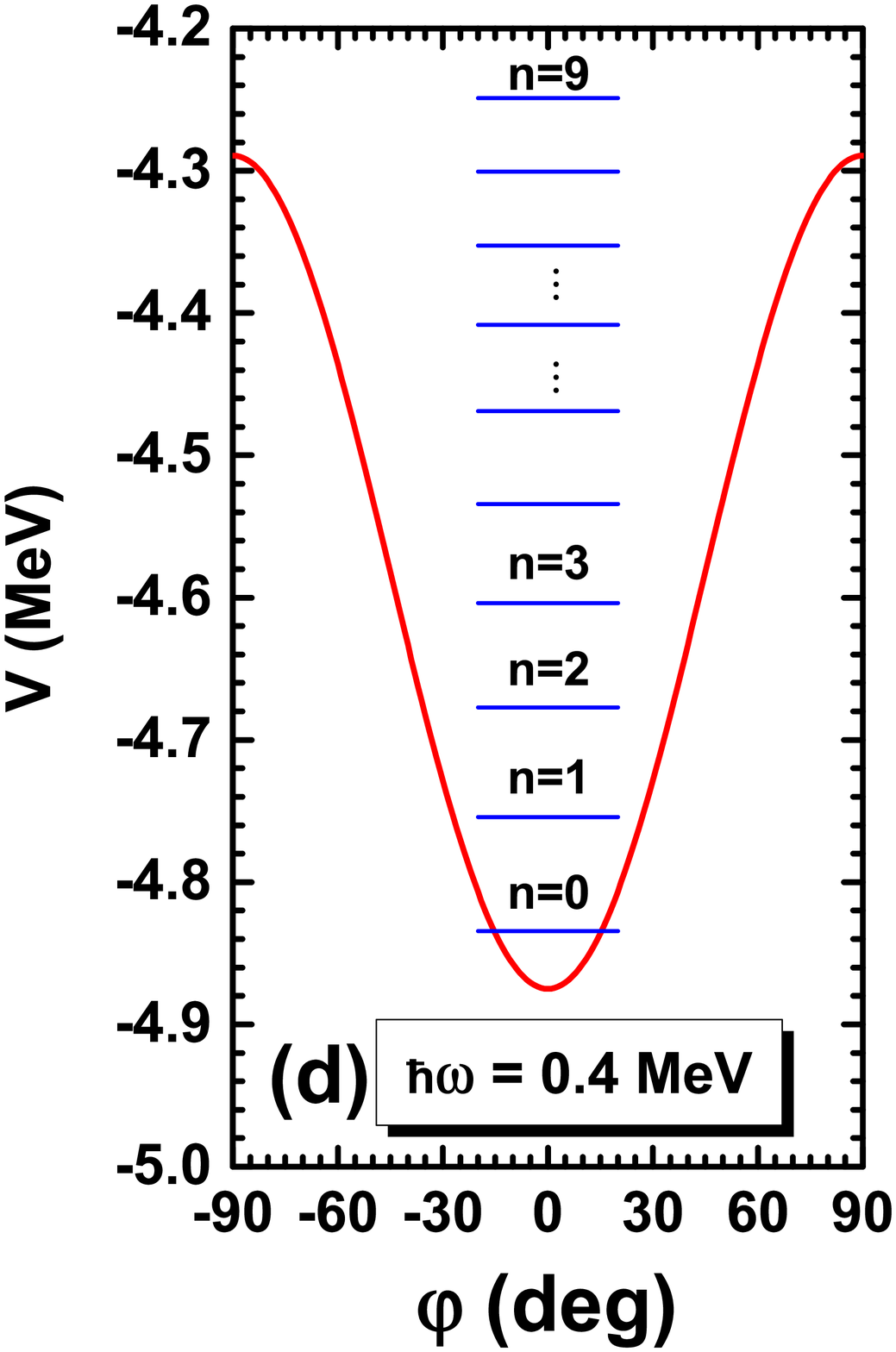}\quad\quad
   \includegraphics[width=4.55 cm]{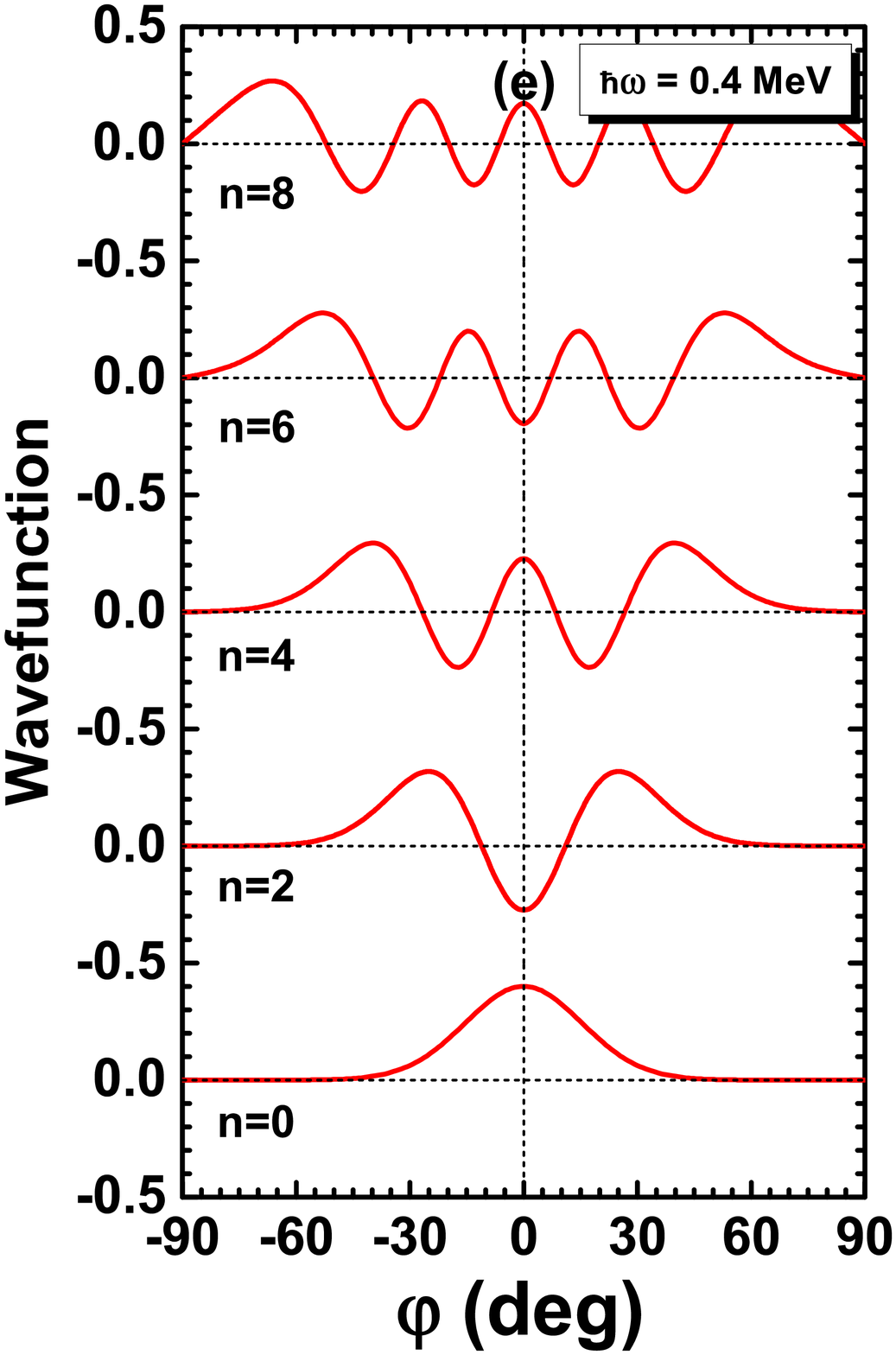}
   \includegraphics[width=4.55 cm]{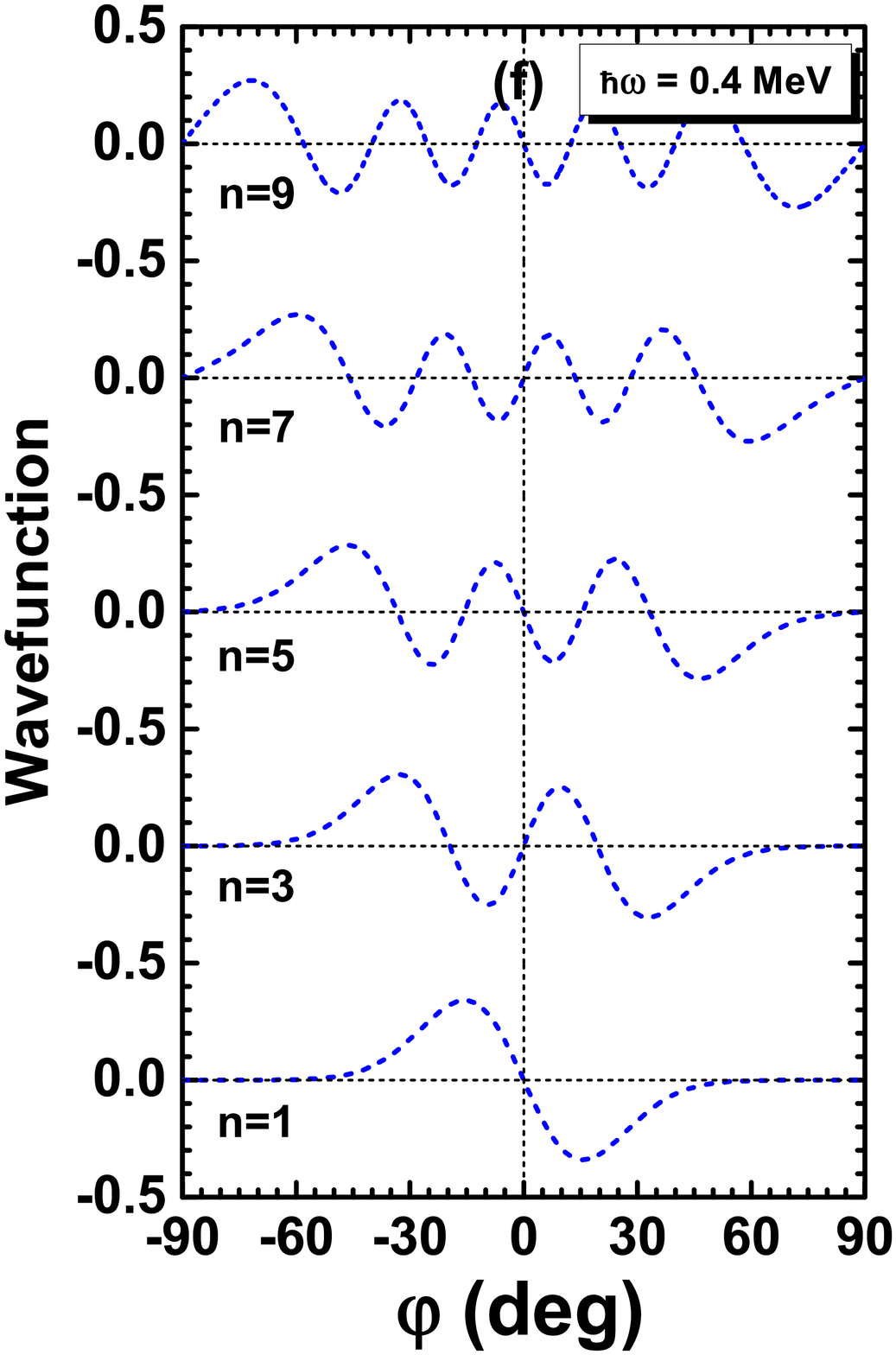}
  \caption{(Color online) The collective levels and wave functions obtained from
           the collective Hamiltonian. Upper panel: The ten lowest energy levels labeled
           as $n=0$-$9$ (left) and the corresponding wave functions for even-$n$ (middle)
           and odd-$n$ (right) states at the frequency $\hbar\omega=0.1~\rm MeV$. Lower panel:
           The ten lowest energy levels labeled as $n=0$-$9$ (left) and the corresponding
           wave functions for even-$n$ (middle) and odd-$n$ (right) states at the frequency
           $\hbar\omega=0.4~\rm MeV$.}\label{fig3}
  \end{center}
\end{figure}

The collective potential obtained above and the mass parameter obtained using
Eq.~(\ref{eq11}) are combined to construct the collective Hamiltonian for
investigating the simple wobbling motion. Diagonalizing the collective Hamiltonian,
the collective energy levels and wave functions at each cranking frequency are yielded.
Taking $\hbar\omega=0.1~\rm MeV$ and $0.4~\rm MeV$ for example, the obtained ten lowest
wobbling energy levels and corresponding wave functions are presented Fig.~\ref{fig3}.
It is obviously seen that the wave functions are symmetric for even-$n$ levels and
antisymmetric for odd-$n$ levels with respect to $\varphi\to-\varphi$ transformation.
Thus the broken signature symmetry in the TAC model is restored in the collective
Hamiltonian by the quantization of wobbling angle $\varphi$ and the consideration
of quantum fluctuation along $\varphi$ motion. In addition, it is also shown that
the wave function of the most favored wobbling energy levels are symmetric.

\begin{figure}[!th]
  \begin{center}
    \includegraphics[width=7 cm]{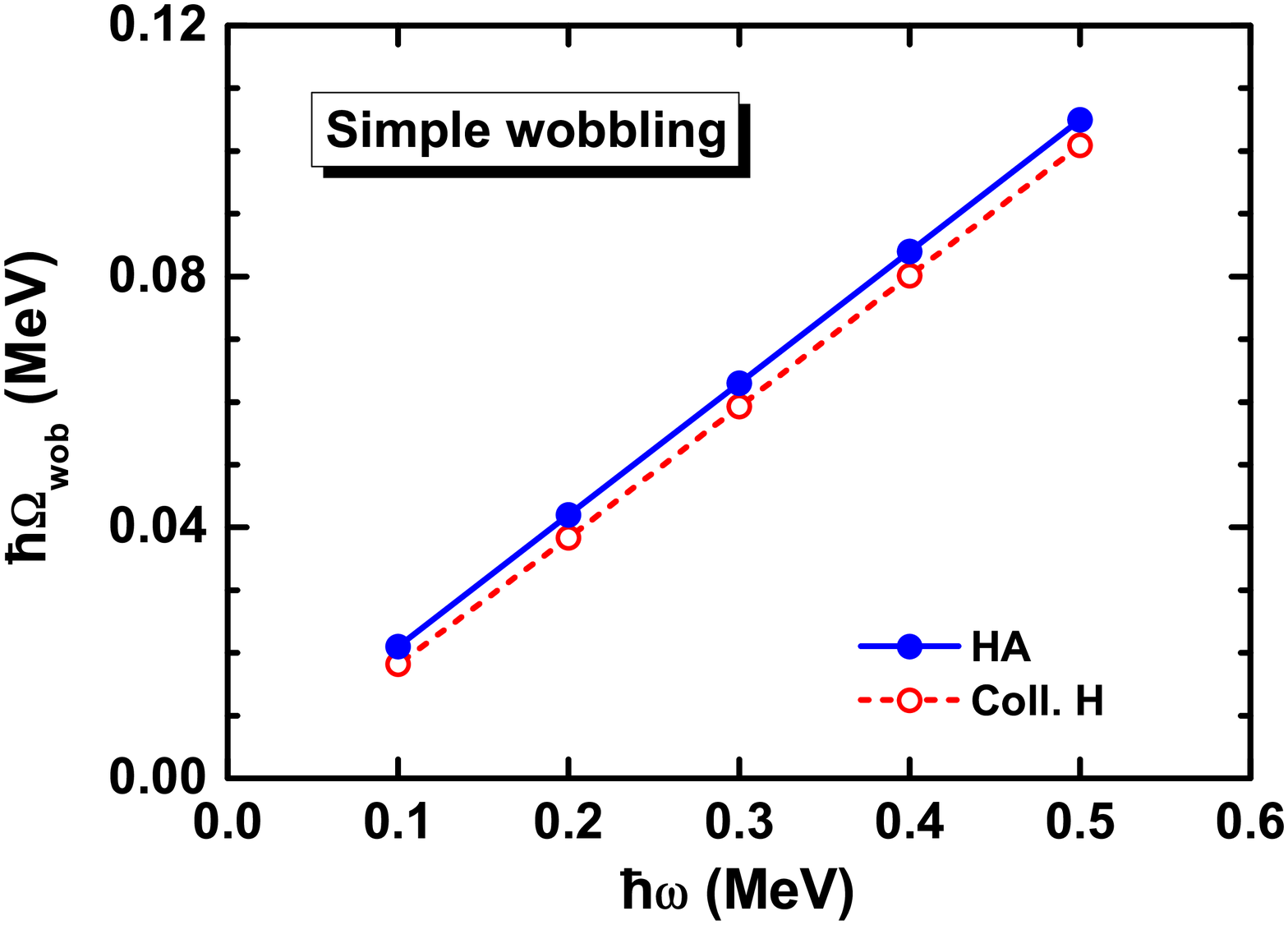}
    \caption{(Color online) The simple wobbling frequency $\hbar\Omega_{\rm wob}$ obtained
             by collective Hamiltonian in comparison with those of HA formula
             (\ref{eq4}).}\label{fig4}
  \end{center}
\end{figure}

\begin{figure}[!th]
  \begin{center}
    \includegraphics[width=7 cm]{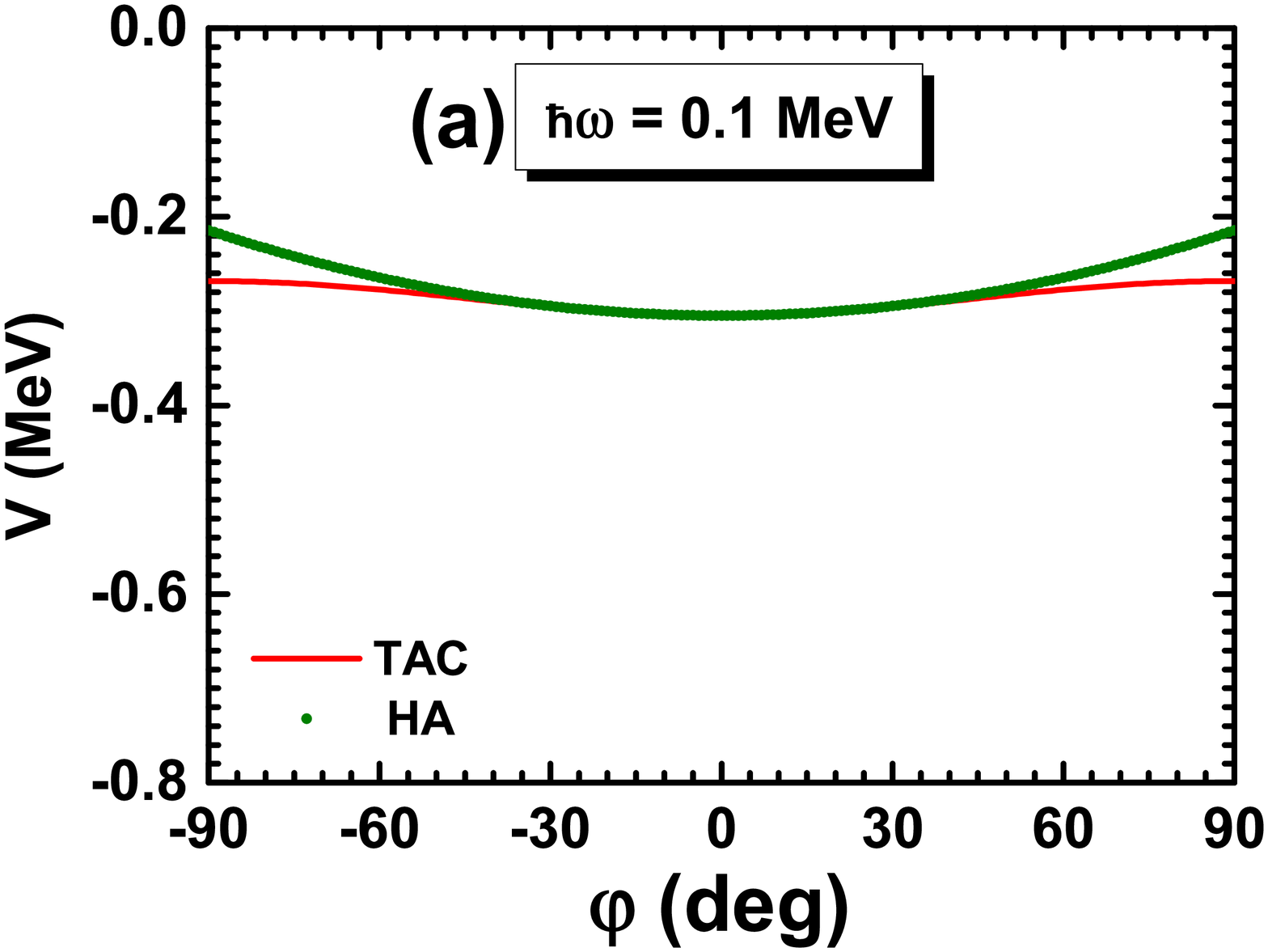}
    \includegraphics[width=7 cm]{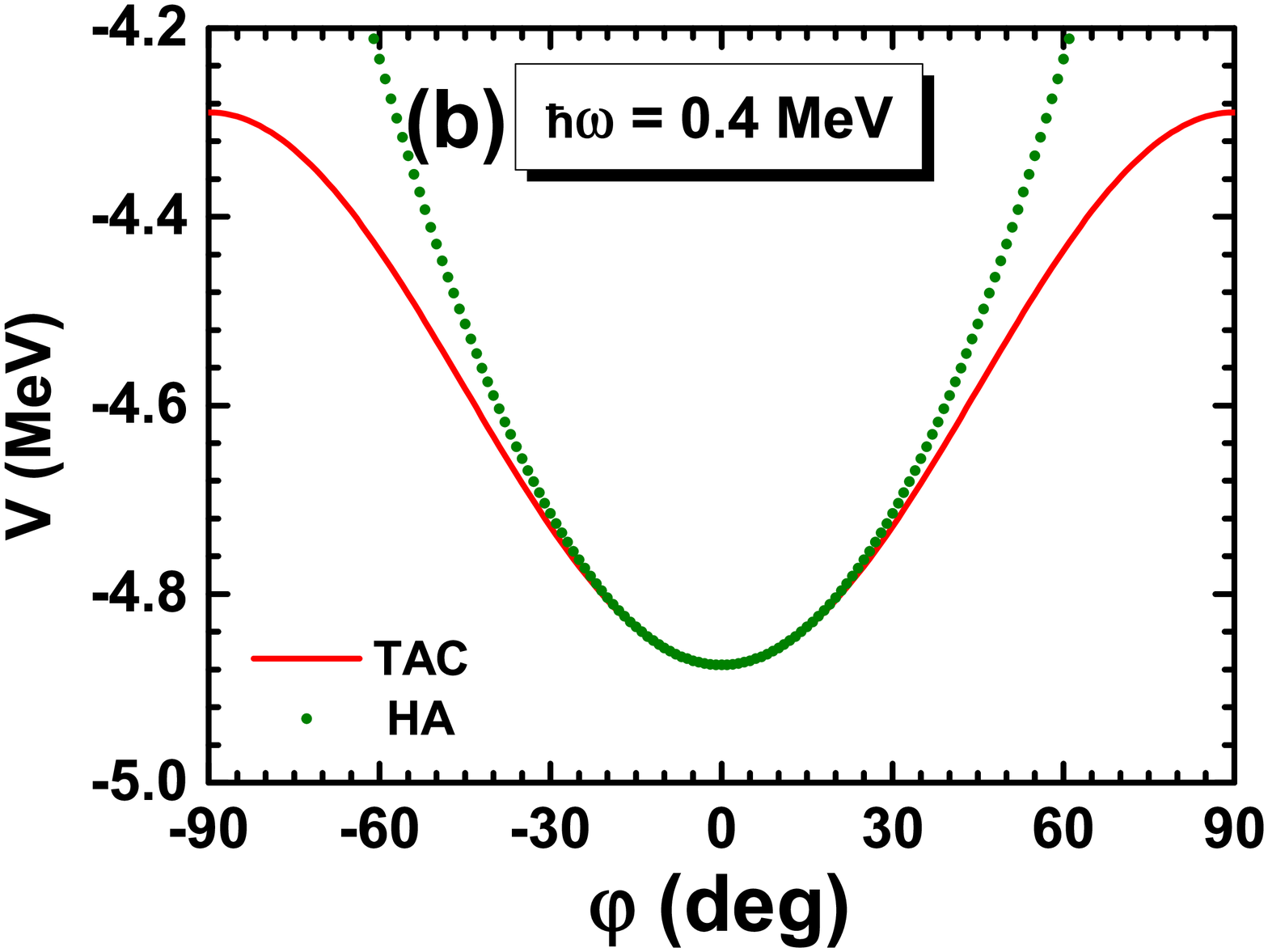}
    \caption{(Color online) The collective potentials obtained by HA (\ref{eq2}) in
             comparison with those by TAC (\ref{eq1}) at frequencies $\hbar\omega=0.1$
             and $0.4~\rm MeV$.}\label{fig5}
  \end{center}
\end{figure}

The wobbling frequency $\hbar\Omega_{\rm wob}$ defined as the energy difference
between the lowest two levels for a certain rotational frequency in the collective
Hamiltonian is shown in Fig.~\ref{fig4} in comparison with those from HA formula (\ref{eq4}).
It is seen that both collective Hamiltonian and HA give the linear increasing trend of
wobbling frequency with respect to rotational frequency. For the HA results, this is just
the expected since the coefficient $\sqrt{\frac{(\mathcal{J}_1-\mathcal{J}_2)
(\mathcal{J}_1-\mathcal{J}_3)}{\mathcal{J}_3\mathcal{J}_2}}$ in the HA formula
(\ref{eq4}) is a positive constant values. For the collective Hamiltonian results, this
can be also readily understood according to the stiffness of the collective potential, as
shown in the upper panels of Fig.~\ref{fig2}(a)-(d), which becomes larger with increasing
of rotational frequency. The wobbling frequency given by HA formula is a bit larger than
that by collective Hamiltonian results from the fact that the simple harmonic approximation
for the collective potential would overestimate the stiffness of the potential, as
shown in Fig.~\ref{fig5}.

\subsubsection{Comparison with TRM solutions}

\begin{figure}[!th]
  \begin{center}
    \includegraphics[width=7 cm]{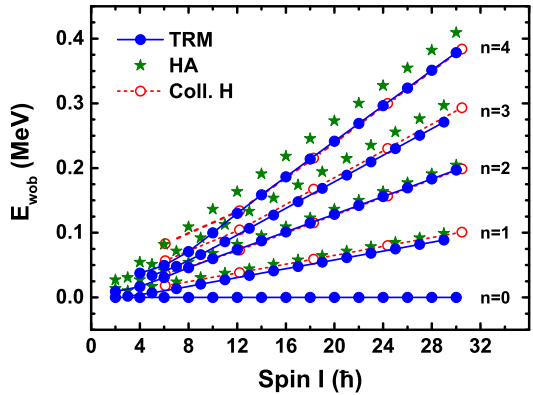}
    \caption{(Color online) Energy spectra of four simple wobbling bands
             $n=1,2,3,4$ relative to the $n=0$ yrast sequence obtained by collective
             Hamiltonian in comparison with TRM and HA. In the TRM, the wobbling energies
             for even-$n$ wobbling bands are calculated as $E_{\rm wob}^n=E_n(I)-E_0(I)$,
             while for odd-$n$ wobbling bands $E_{\rm wob}^n(I)=E_n(I)-[E_0(I+1)+E_0(I-1)]/2$.}
    \label{fig6}
  \end{center}
\end{figure}

The simple wobbler solutions discussed here can be exactly obtained
by TRM. To study the accuracy of collective Hamiltonian scheme, in Fig.~\ref{fig6},
the energies of the four lowest wobbling bands $n=1,2,3,4$ relative to the $n=0$
yrast sequence obtained by collective Hamiltonian are displayed in comparison with
those from TRM and HA. In the TRM, the states possesses $D_2$ symmetry so that the
spectrum is restricted to the states with $(-1)^n=(-1)^I$~\cite{Bohr1975}, i.e., only
even spins for even-$n$ wobbling bands while only odd appear for odd-$n$ wobbling bands.
The even-$n$ wobbling bands are more energetically favored than odd-($n+1$) wobbling
bands. Hence, the wobbling energies are calculated in different ways for even-$n$
and odd-$n$ wobbling bands. For even-$n$ wobbling bands, the wobbling energies
are directly calculated as the energy difference with respect to $n=0$
wobbling bands $E_{\rm wob}^n=E_n(I)-E_0(I)$, while for odd-$n$ wobbling bands
calculated as the energy difference with respect to the interpolated energies
by $n=0$ wobbling band $E_{\rm wob}^n(I)=E_n(I)-[E_0(I+1)+E_0(I-1)]/2$. The spin in
the TRM is treated as a good quantum number, while in the collective Hamiltonian it is
not but a expectation value of angular momentum operator on the rotational state
with given rotational frequency.

It is observed from Fig.~\ref{fig6} the increasing trend of wobbling energies
with spin for each wobbling bands. With the increasing of $n$, the HA results
gradually deviate from TRM, which indicates that the wobbling motion gradually
deviates from harmonic oscillation character. The collective Hamiltonian excellently
reproduce the TRM results even for the large-$n$ wobbling bands. The collective
Hamiltonian based on TAC approach, however, provides a new perspective to
interpret the variation trend of wobbling frequency with spin by
exploring the variation trend of stiffness of the collective potential.

\subsection{Longitudinal wobbler}

Now we discuss the longitudinal wobbler, where a $h_{11/2}$ proton particle
is assumed to couple to a triaxial rotor and its angular momentum is parallel
to the axis with the largest moment of inertia. The rigid body type of moments
of inertia~(\ref{eq14}) are used here too.

\subsubsection{Collective potential}
\begin{figure}[!th]
  \begin{center}
    \includegraphics[width=6 cm]{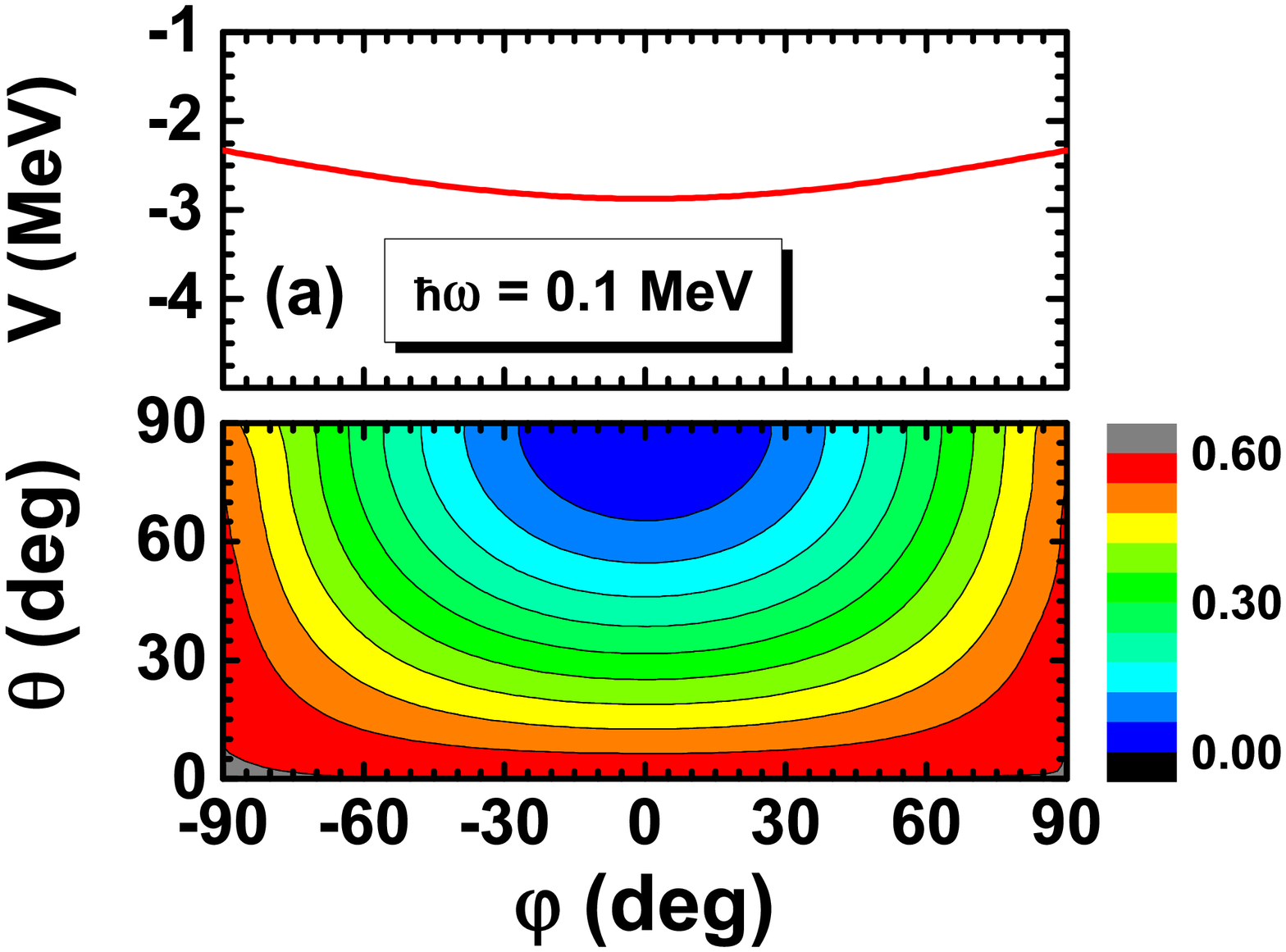}\quad
    \includegraphics[width=6 cm]{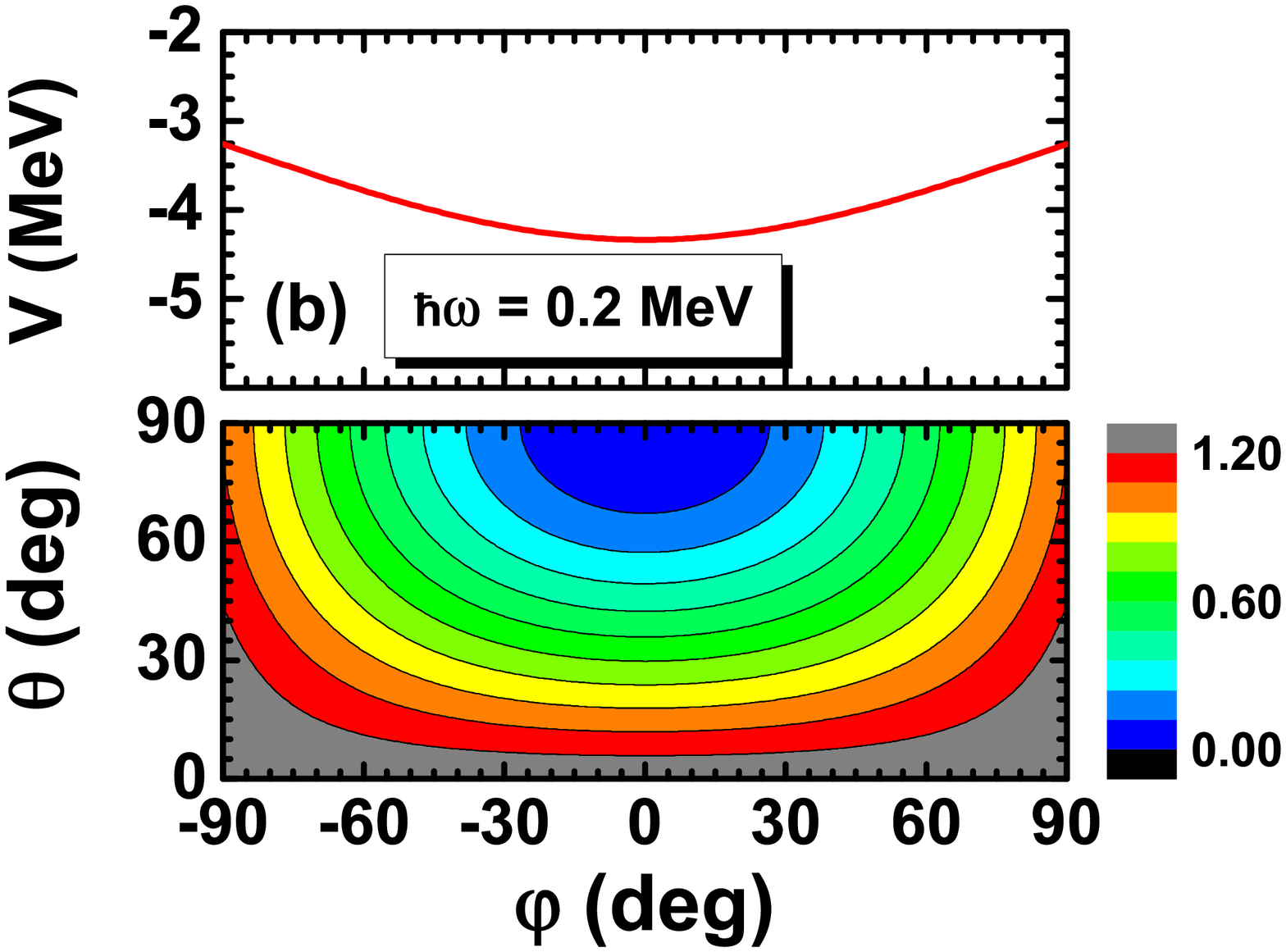}\\
    \includegraphics[width=6 cm]{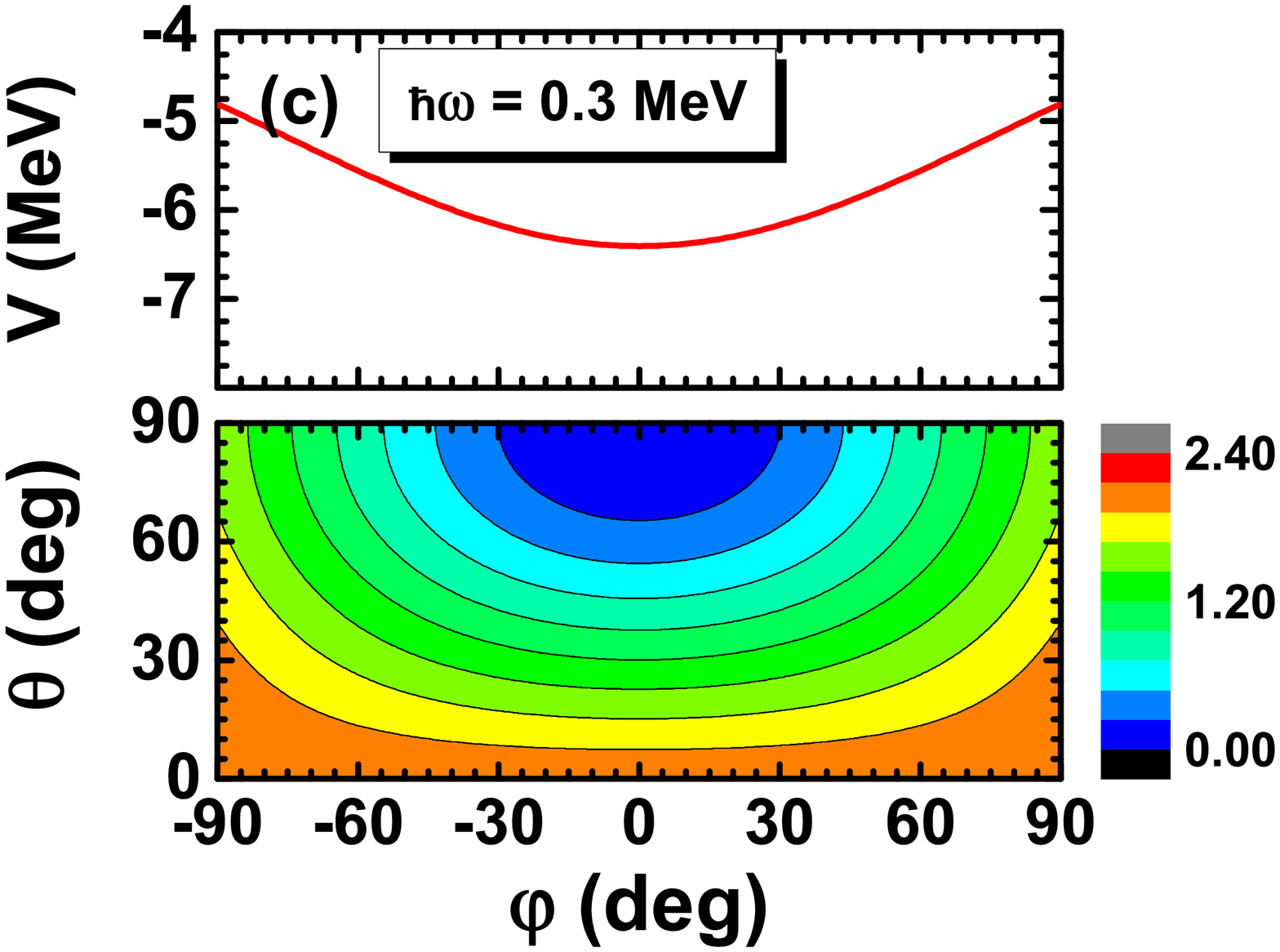}\quad
    \includegraphics[width=6 cm]{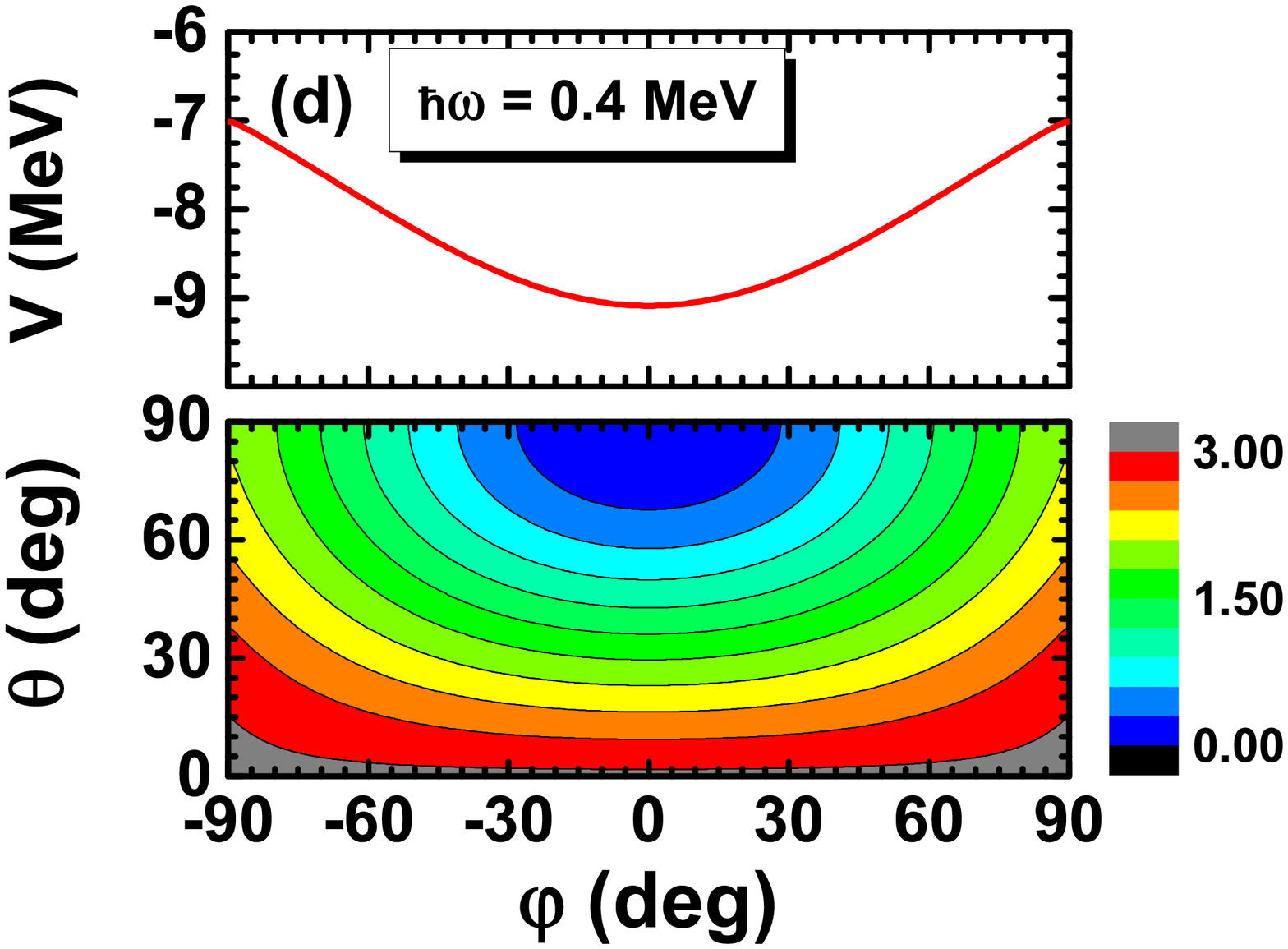}
    \caption{(Color online) Same as Fig.~\ref{fig2} but for longitudinal
             wobbling motion, where a proton $h_{11/2}$ particle coupled to a
             triaxial rigid body rotor with $\gamma=-30^\circ$.}\label{fig7}
  \end{center}
\end{figure}

In the contour plots of Fig.~\ref{fig7}(a)-(d), the total Routhians for longitudinal
wobbling motions obtained TAC are shown at the rotational frequencies $\hbar\omega=0.1$,
$0.2$, $0.3$, and $0.4~\rm MeV$. Similar to the case of simple wobbling, the
total Routhian is also symmetrical with respect to the $\varphi=0^\circ$ line
and the minima always locate at $(\theta=90^\circ, \varphi=0^\circ)$ regardless
of how fast the nucleus rotates. This is very clear since both the proton particle
and triaxial rotor angular momenta in the longitudinal wobbling system are oriented
along the short axis, the axis with the largest moment of inertia.

With the total Routhian, the extracted collective potentials $V(\varphi)$ are
presented at $\hbar\omega=0.1$, $0.2$, $0.3$, and $0.4~\rm MeV$ in
the upper panels of Fig.~\ref{fig7}(a)-(d). It is clearly seen that the collective
potentials presented here are very similar as those presented in the upper panels of
Fig.~\ref{fig2}(a)-(d) for simple wobbling motions, while the only difference
is that the stiffness here become larger. Therefore, similar discussions for simple
wobbling motion still hold true here. It is worth to stress that the deeper potentials
here are attributed to the proton particle and its contribution would become larger at
larger rotational frequency. For example, at $\hbar\omega=0.1~\rm MeV$, the energy
difference between $\varphi=\pm 90^\circ$ and $\varphi=0^\circ$ is $\sim 540~\rm keV$
and reaches to $\sim 2080~\rm keV$ at $\hbar\omega=0.4~\rm MeV$. Comparing with the
simple wobbling motions, one obtains the contribution from proton increases from
$\sim 510~\rm keV$ at $\hbar\omega=0.1~\rm MeV$ to $1480~\rm keV$ at $0.4~\rm MeV$.

\subsubsection{Mass parameter}

\begin{figure}[!th]
  \begin{center}
    \includegraphics[width=7 cm]{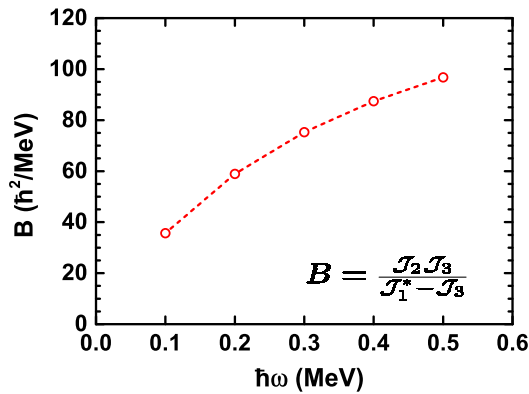}
    \caption{(Color online) The calculated mass parameter as a function of rotational
             frequency $\hbar\omega$ for longitudinal wobbling motion.}\label{fig8}
  \end{center}
\end{figure}

The mass parameter for longitudinal wobbling motion is calculated by Eq.~(\ref{eq12})
and shown in Fig.~\ref{fig8}. As discussed in Sec.~\ref{sec2}, since the effective
moments of inertia for 1-axis decreases with the rotational frequency, the mass
parameter increases with increasing rotational frequency. This increasing characteristic
is different from the simple wobbler, where the mass parameter is constant at any
rotational frequency.

\subsubsection{Collective levels and wave functions}

\begin{figure}[!th]
  \begin{center}
   \includegraphics[width=4.55 cm]{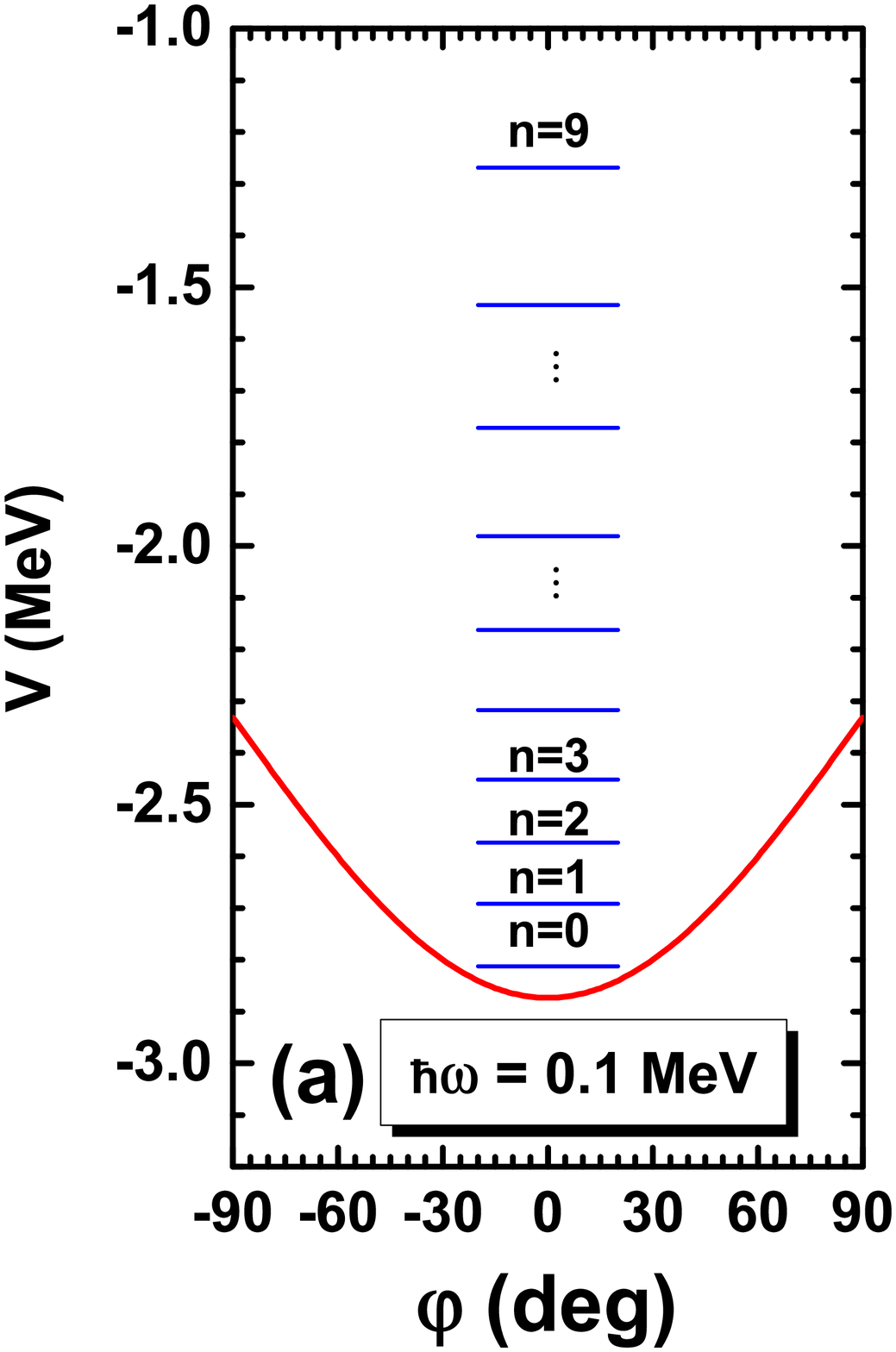}\quad\quad
   \includegraphics[width=4.55 cm]{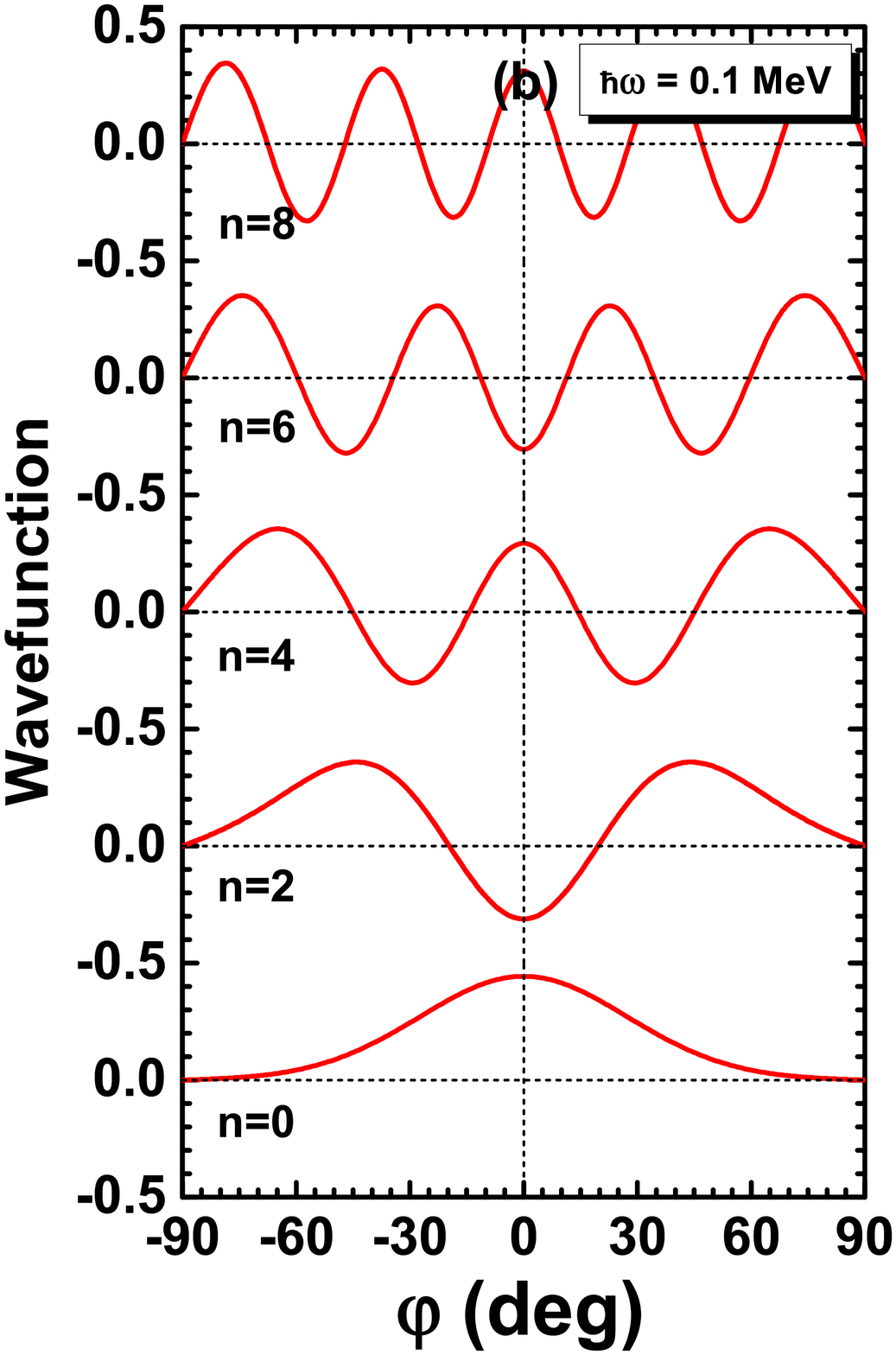}
   \includegraphics[width=4.55 cm]{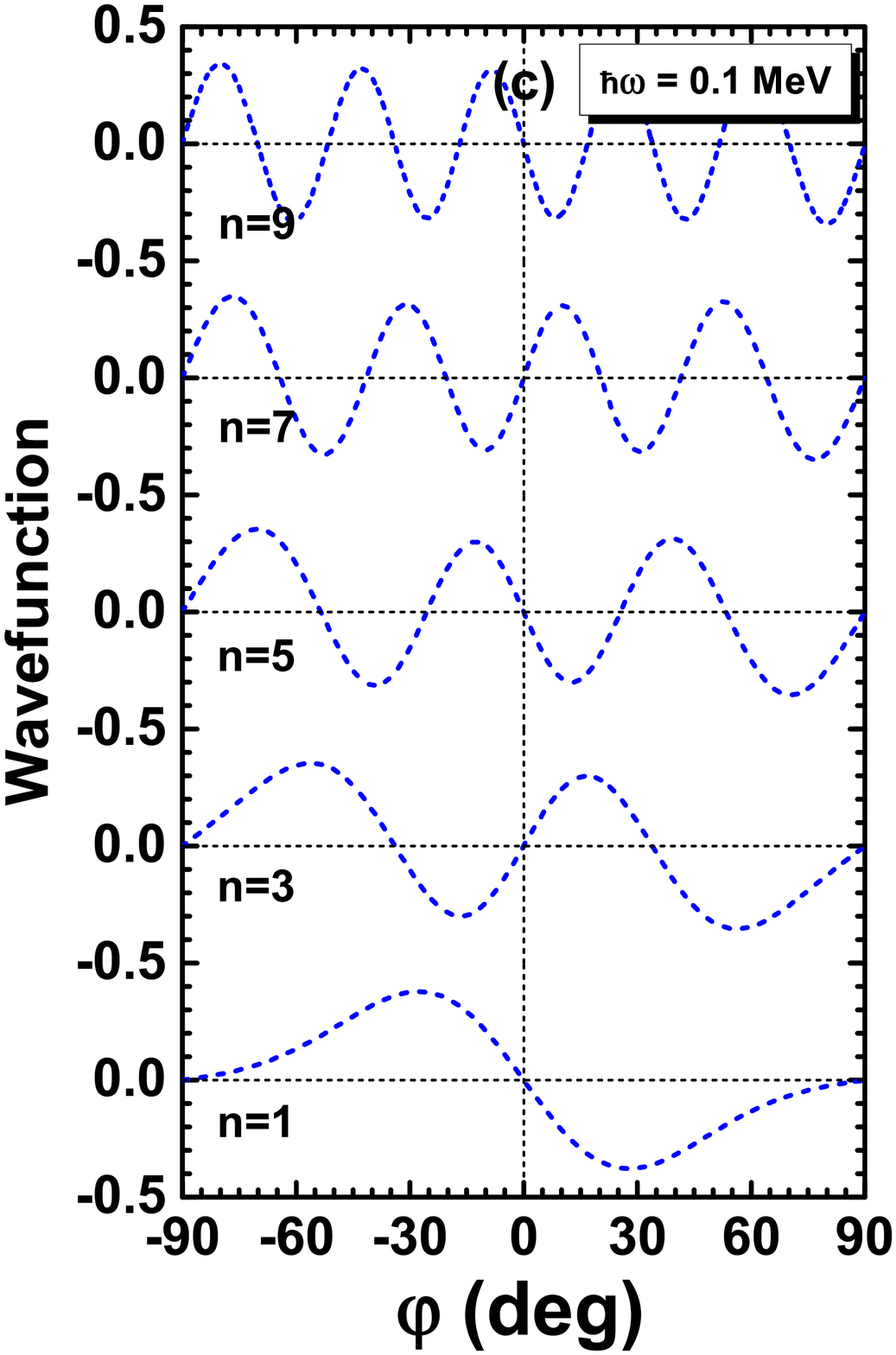}\\
   ~~\\
   \includegraphics[width=4.55 cm]{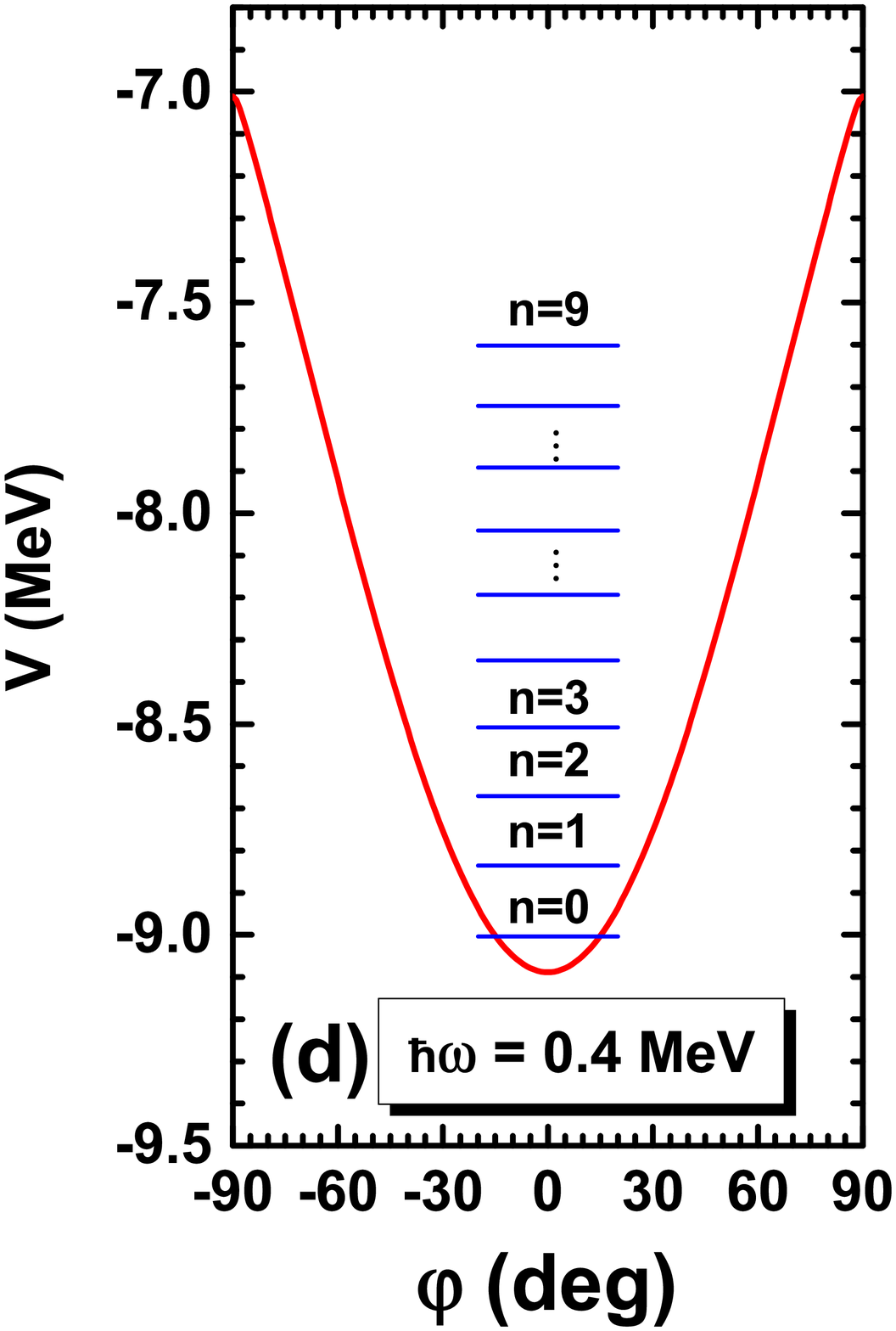}\quad\quad
   \includegraphics[width=4.55 cm]{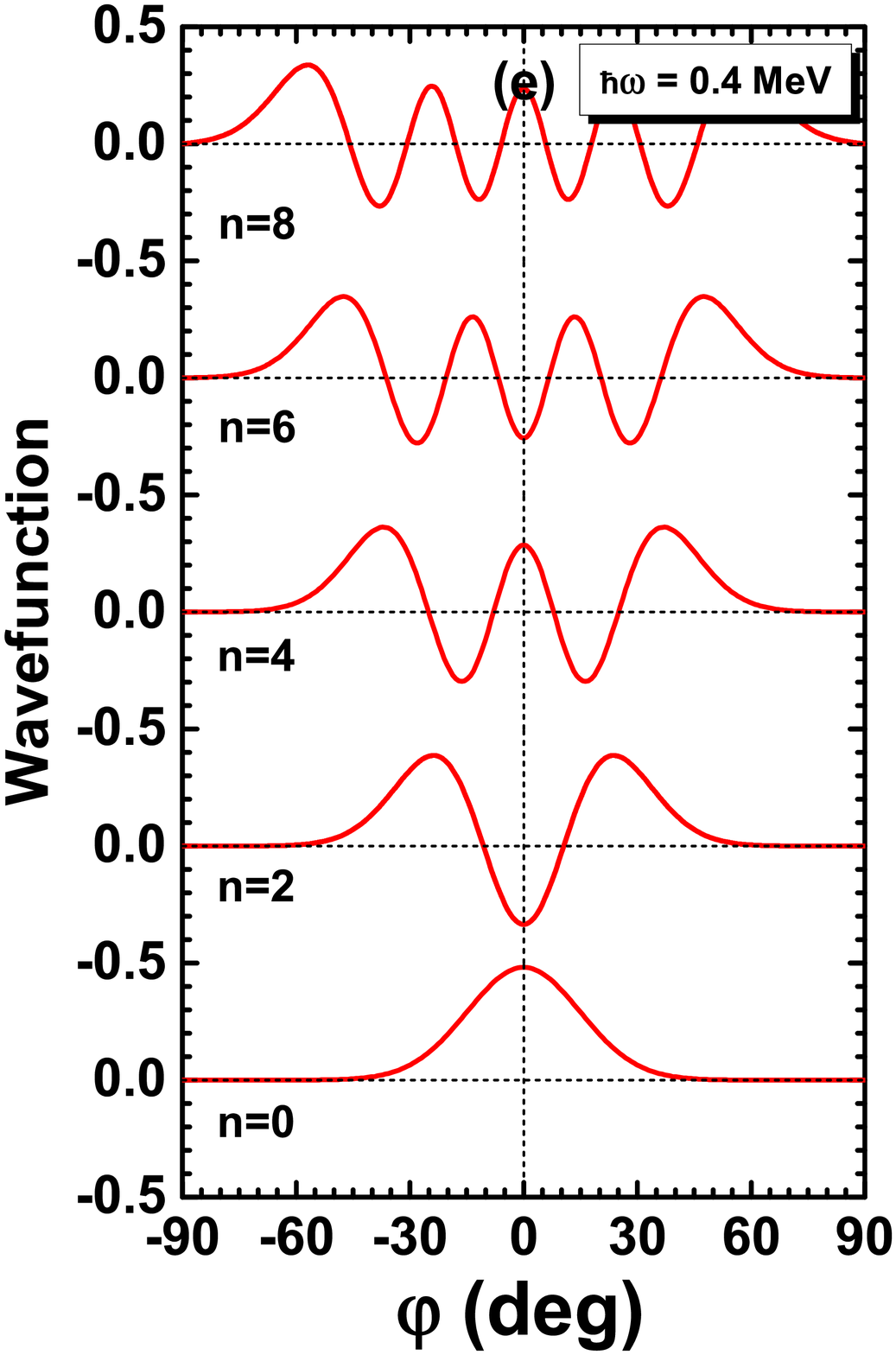}
   \includegraphics[width=4.55 cm]{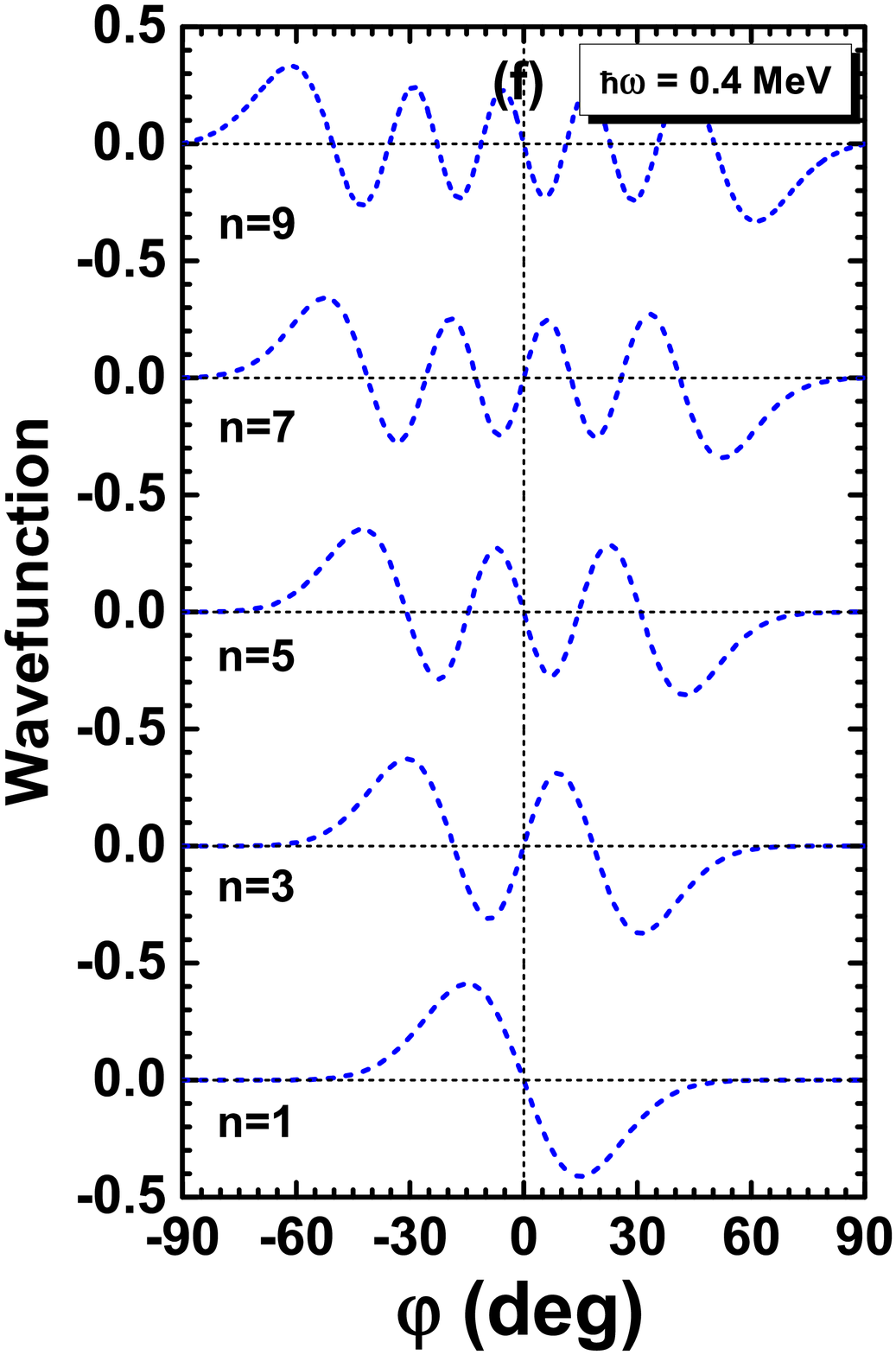}
  \caption{(Color online) Same as Fig.~\ref{fig3} but for longitudinal wobbling
           motion.}\label{fig9}
  \end{center}
\end{figure}

The obtained collective energy levels and corresponding wave functions are
illustrated in Fig.~\ref{fig9} for $\hbar\omega=0.1$ and $0.4~\rm MeV$. Again,
the wave functions presented here are similar as those presented in Fig.~\ref{fig3}
for simple wobbling motions.

\begin{figure}[!th]
  \begin{center}
    \includegraphics[width=7 cm]{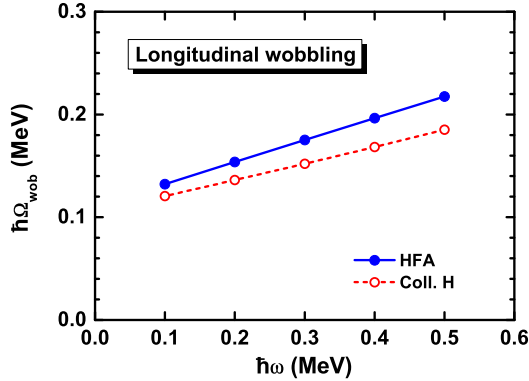}
    \caption{(Color online) The longitudinal wobbling frequency $\hbar\Omega_{\rm wob}$
             obtained by collective Hamiltonian in comparison with those of HFA
             approximation (\ref{eq5}).}\label{fig10}
  \end{center}
\end{figure}

\begin{figure}[!th]
  \begin{center}
    \includegraphics[width=7 cm]{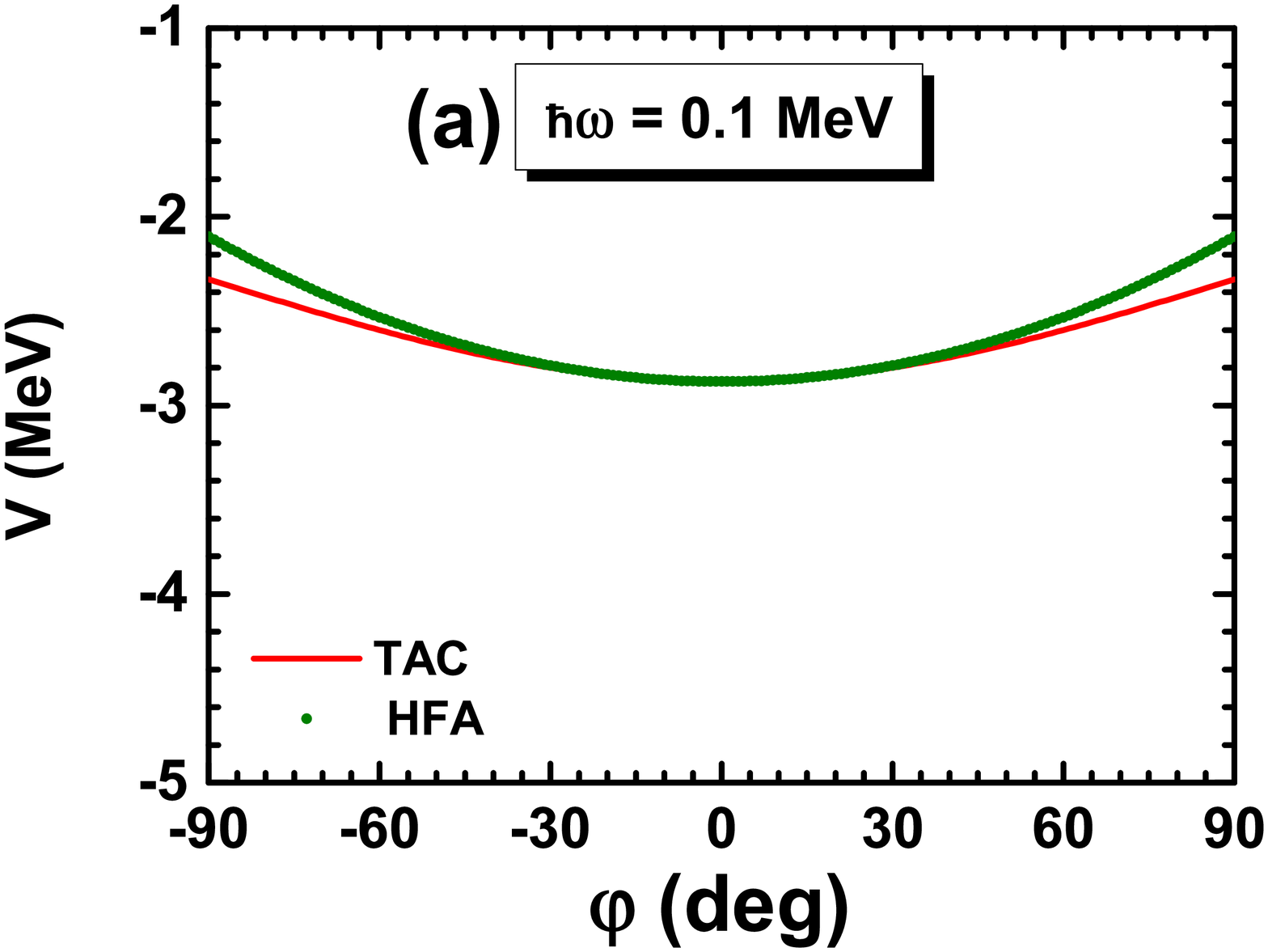}
    \includegraphics[width=7 cm]{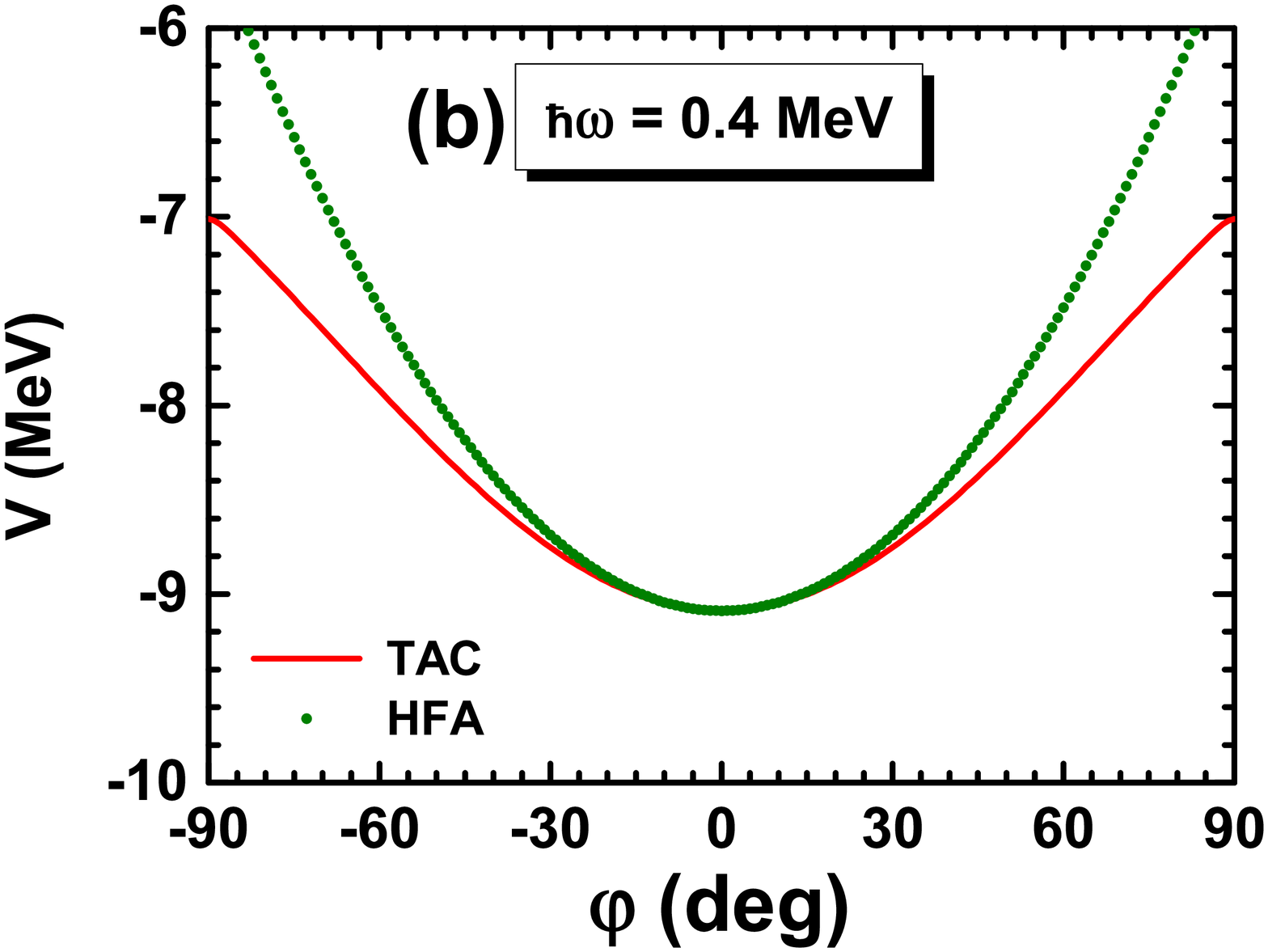}
    \caption{(Color online) The collective potentials obtained by HFA
             (\ref{eq8}) in comparison with those by TAC (\ref{eq7})
             at frequencies $\hbar\omega=0.1$ and $0.4~\rm MeV$.}\label{fig11}
  \end{center}
\end{figure}

In Fig.~\ref{fig10}, the obtained wobbling frequency calculated by the collective
Hamiltonian, is in comparison with the results obtained by HFA approximation
(\ref{eq5}). It is found that both collective Hamiltonian and HFA give the
increased wobbling frequency as function of rotational frequency. However, the HFA
results are larger than the collective Hamiltonian ones over the whole range of
rotational frequency. To understand the origin of the differences between HFA and
collective Hamiltonian, the collective potential obtained by HFA approximation
in comparison with the results obtained by TAC at $\hbar\omega=0.1~\rm MeV$ and
$\hbar\omega=0.4~\rm MeV$ are shown in Fig.~\ref{fig11}. It is seen that the stiffness
of collective potential calculated by HFA are larger than the collective Hamiltonian
at both $\hbar\omega=0.1$ and $0.4~\rm MeV$. Since the mass parameter in the
collective Hamiltonian is the same as the in the HFA, the wobbling frequency
of HFA is larger than collective Hamiltonian. Besides the harmonic approximation
as for simple wobbling motion, the HFA further introduces that the proton particle
rigidly aligns its angular momentum along short axis. Hence it deviates larger
from the collective Hamiltonian for the longitudinal wobbling motion ($\sim 20~\rm keV$,
see Fig.~\ref{fig10}) than HA for the simple wobbling motion ($\sim 5~\rm keV$,
see Fig.~\ref{fig4}).

\subsubsection{Comparison with PRM solutions}

\begin{figure}[!th]
  \begin{center}
    \includegraphics[width=7 cm]{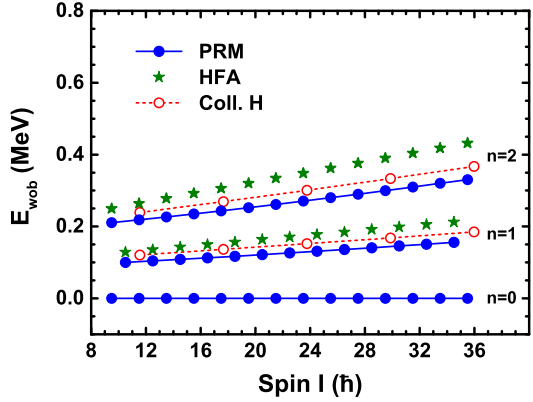}
    \caption{(Color online) Energy spectra of two longitudinal wobbling bands
             $n=1,2$ relative to the $n=0$ yrast sequence obtained by collective
             Hamiltonian in comparison with PRM and HFA. In the PRM, the wobbling
             energies for even-$n$, the wobbling energies are calculated as
             $E_{\rm wob}^n=E_n(I)-E_0(I)$, while for odd-$n$ $E_{\rm wob}^n(I)
             =E_n(I)-[E_0(I+1)+E_0(I-1)]/2$.}\label{fig12}
  \end{center}
\end{figure}

The exact solutions for longitudinal wobbling motion can be obtained by PRM.
In order to investigate the quality of the collective Hamiltonian, the energies
of the two lowest wobbling bands $n=1,2$ relative to the $n=0$ yrast sequence
obtained by collective Hamiltonian are shown in Fig.~\ref{fig12} in comparison
with those from PRM. In PRM, for odd-$n$, the wobbling energies are calculated
as $E_{\rm wob}^n=E_n(I)-E_0(I)$, while for even-$n$ wobbling bands are calculated
as $E_{\rm wob}^n(I)=E_n(I)-[E_0(I+1)+E_0(I-1)]/2$. It is found that the
collective Hamiltonian can reproduce the PRM very well. With the increasing
spin, the wobbling energy increases. The results calculated by HFA are also
shown in Fig.~\ref{fig12}. The wobbling energies given by HFA are
larger than those obtained by both PRM and collective Hamiltonian.

Both HFA and collective Hamiltonian are approximate solutions with respect to PRM.
In the HFA approximation, the harmonic oscillator potential and the frozen alignment
of proton particle are assumed. In the collective Hamiltonian, however, only the
mass parameter is calculated with the HFA approximation, while the collective potential
is calculated by TAC model without prior assuming the frozen alignment with respect to
any axis for proton particle. The PRM exactly diagonals the particle rotor coupling
Hamiltonian and thus gives the exact solutions. From this point of view, the collective
Hamiltonian has improved the descriptions for the collective potential and
provides a more accurate solution than HFA.

\subsection{Transverse wobbler}

For transverse wobbling motions, the proton particle angular momentum is
supposed to be perpendicular to the axis with the largest moment of inertia.
In the present investigation, the irrotational flow type of moments of inertia
(\ref{eq15}) is employed to satisfy this requirement.

\subsubsection{Collective potential}

\begin{figure}[!th]
  \begin{center}
    \includegraphics[width=6 cm]{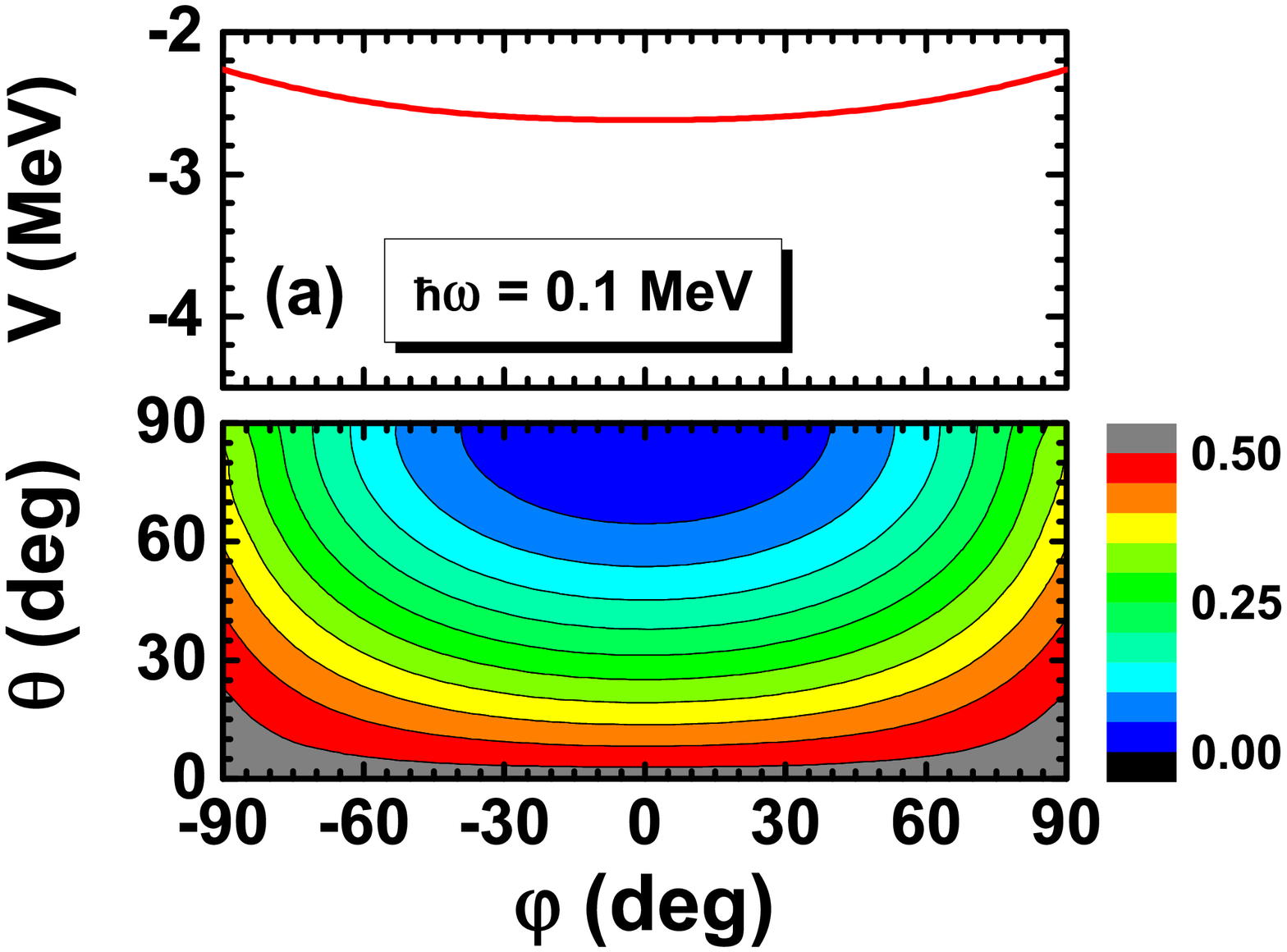}\quad
    \includegraphics[width=6 cm]{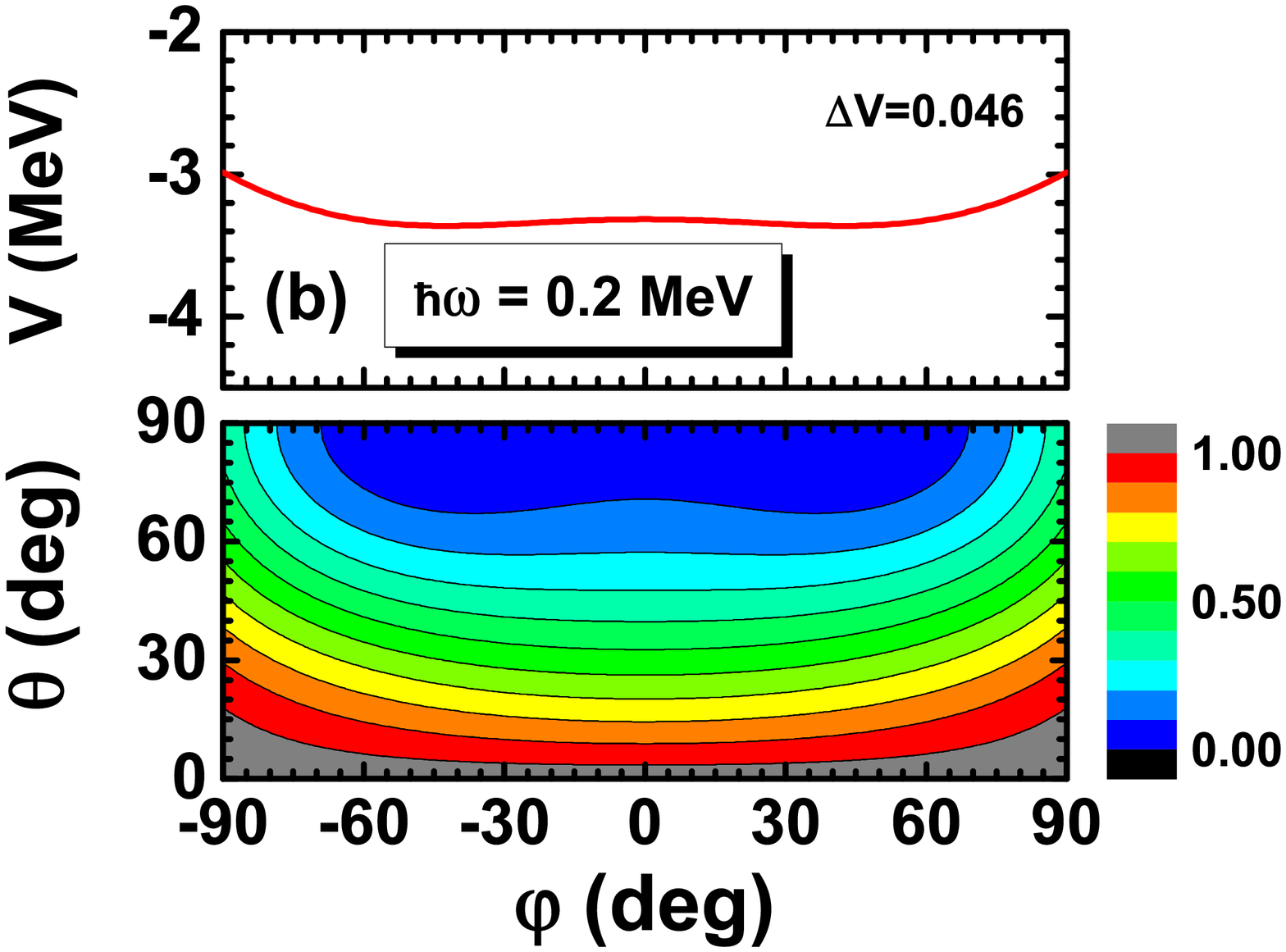}\\
    \includegraphics[width=6 cm]{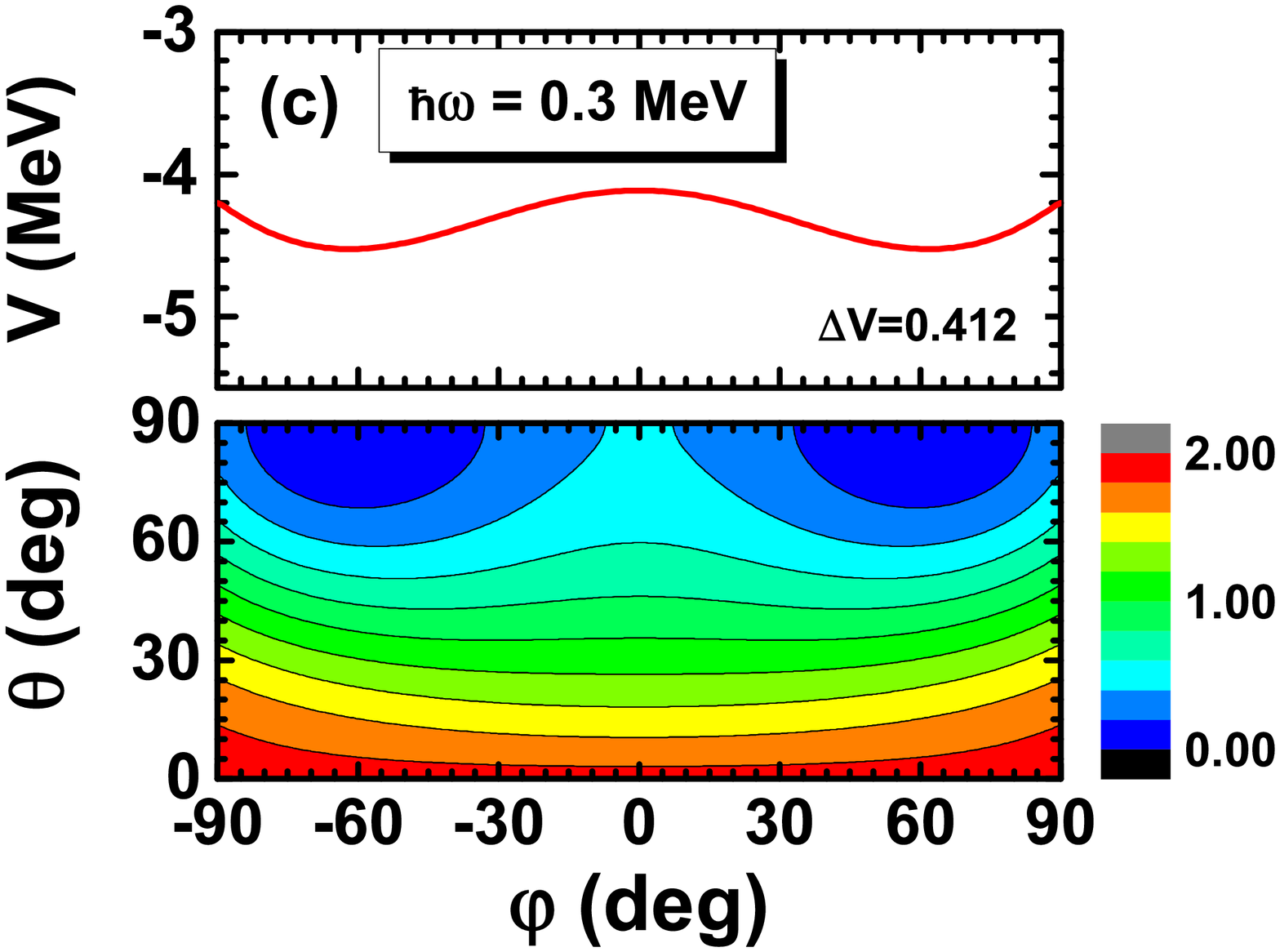}\quad
    \includegraphics[width=6 cm]{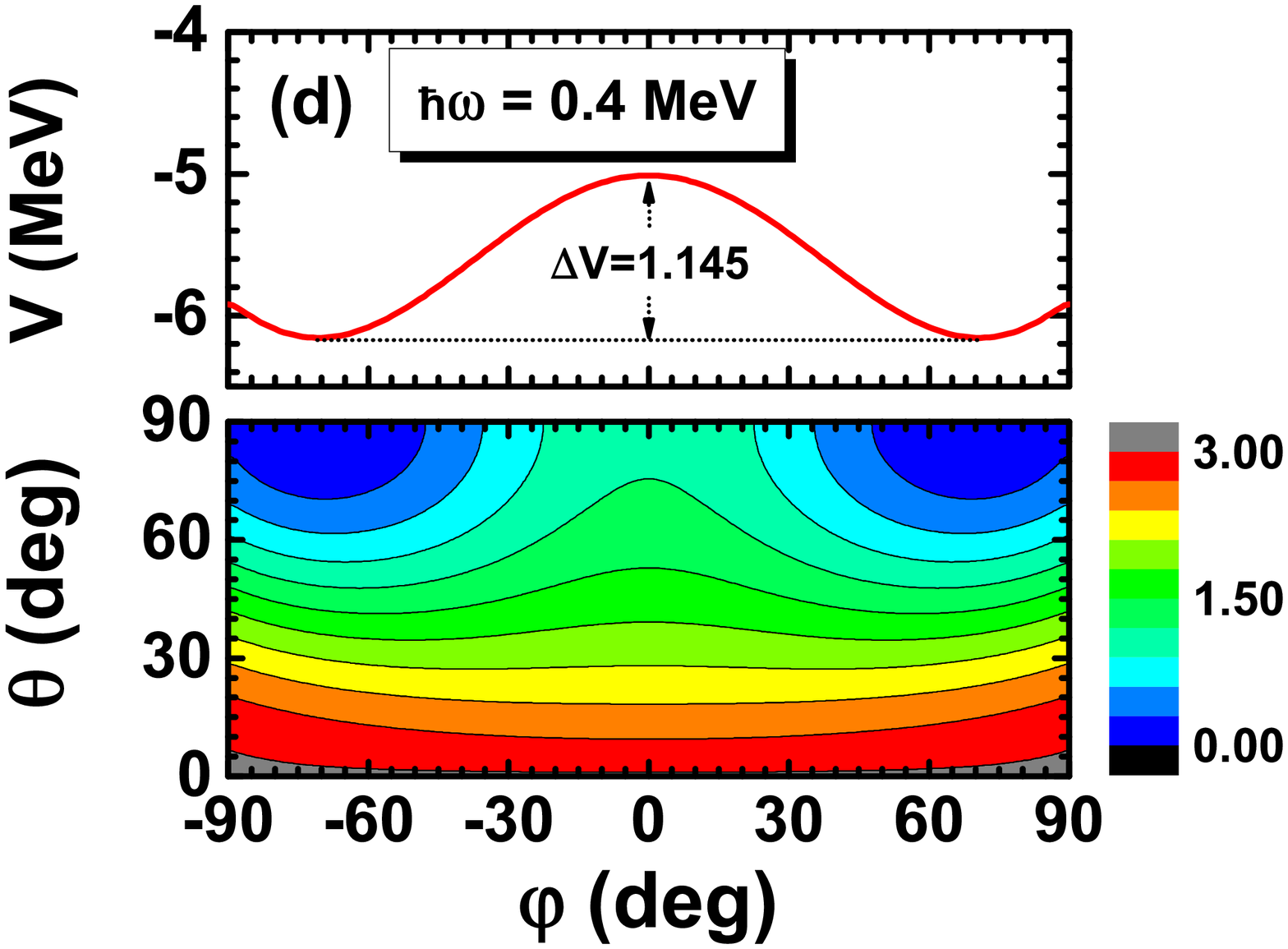}
    \caption{(Color online) Same as Fig.~\ref{fig2} but for transverse wobbling,
             where a proton $h_{11/2}$ particle coupled to a triaxial irrotational flow
             rotor with $\gamma=-30^\circ$.}\label{fig13}
  \end{center}
\end{figure}

The total Routhians calculated by TAC for a
$h_{11/2}$ proton particle coupled to a triaxial irrotational flow rotor with
$\gamma=-30^\circ$ in the $(\theta,\varphi)$ plane are displayed at the rotational
frequencies $\hbar\omega=0.1$, $0.2$, $0.3$, $0.4~\rm MeV$ in contour plots of
Fig.~\ref{fig13}(a)-(d). The potential energy surfaces are also symmetric with
the $\varphi=0^\circ$ line. In contrast to the simple and longitudinal wobbling
motions, the minima in the potential energy surfaces change from $\varphi=0^\circ$
to $\varphi\neq 0^\circ$ with the increasing frequency. As discussed in
Ref.~\cite{Frauendorf2014PRC}, this implies the axis of uniform rotation is tilted
from $s$ axis into the $s$-$i$ plane.

The extracted collective potentials $V(\varphi)$ for transverse wobbling motion
are shown in the upper panels of Fig.~\ref{fig13}(a)-(d). For $\hbar\omega=0.1~\rm MeV$,
the potential $V(\varphi)$ is a harmonic oscillator type which has only one minimum
at $\varphi=0^\circ$, which corresponds to the uniform rotation around 1-axis.
For $\hbar\omega\geq 0.20~\rm MeV$, the potential $V(\varphi)$ has two symmetrical
minima, which corresponds to the tilted rotation. Due to the appearance of the
potential barrier, the tilted solutions are achieved in the body-fixed frame.
The heights of barrier defined as $\Delta V=V(0)-V_{\rm min}$ (in MeV) with $V_{\rm min}$
being the value of the potential at the minimum presented also in the figure.
It is found that the potential barrier increases with the rotational frequency,
e.g., from $0.046~\rm MeV$ at $\hbar\omega=0.20~\rm MeV$ to $1.145~\rm MeV$ at
$\hbar\omega=0.40~\rm MeV$.

\subsubsection{Mass parameter}

\begin{figure}[!th]
  \begin{center}
    \includegraphics[width=7 cm]{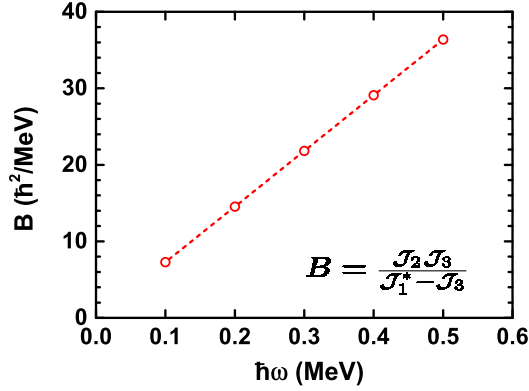}
    \caption{(Color online) Same as Fig.~\ref{fig8} but for transverse wobbling
             motion.}\label{fig14}
  \end{center}
\end{figure}

The obtained mass parameter calculated by Eq.~(\ref{eq12}) as a function of
rotational frequency is shown in Fig.~\ref{fig14}. Since here the irrotational
flow type of moments of inertia~(\ref{eq15}) with $\gamma=-30^\circ$
assumed, i.e., $\mathcal{J}_1=\mathcal{J}_3$, the deduced mass parameter is linear
dependence on rotational frequency. The mass parameter~(\ref{eq12}) is derived based
on the assumption of harmonic frozen alignment approximation, therefore, it is strictly
speaking valid only at the wobbling motion region and will becomes invalid in the
tilted rotation region. Nevertheless, as a rough approximation, the mass parameter
formula~(\ref{eq12}) is used for the calculations over the whole range of rotational
frequency.

\subsubsection{Collective levels and wave functions}

\begin{figure}[!th]
  \begin{center}
   \includegraphics[width=4.55 cm]{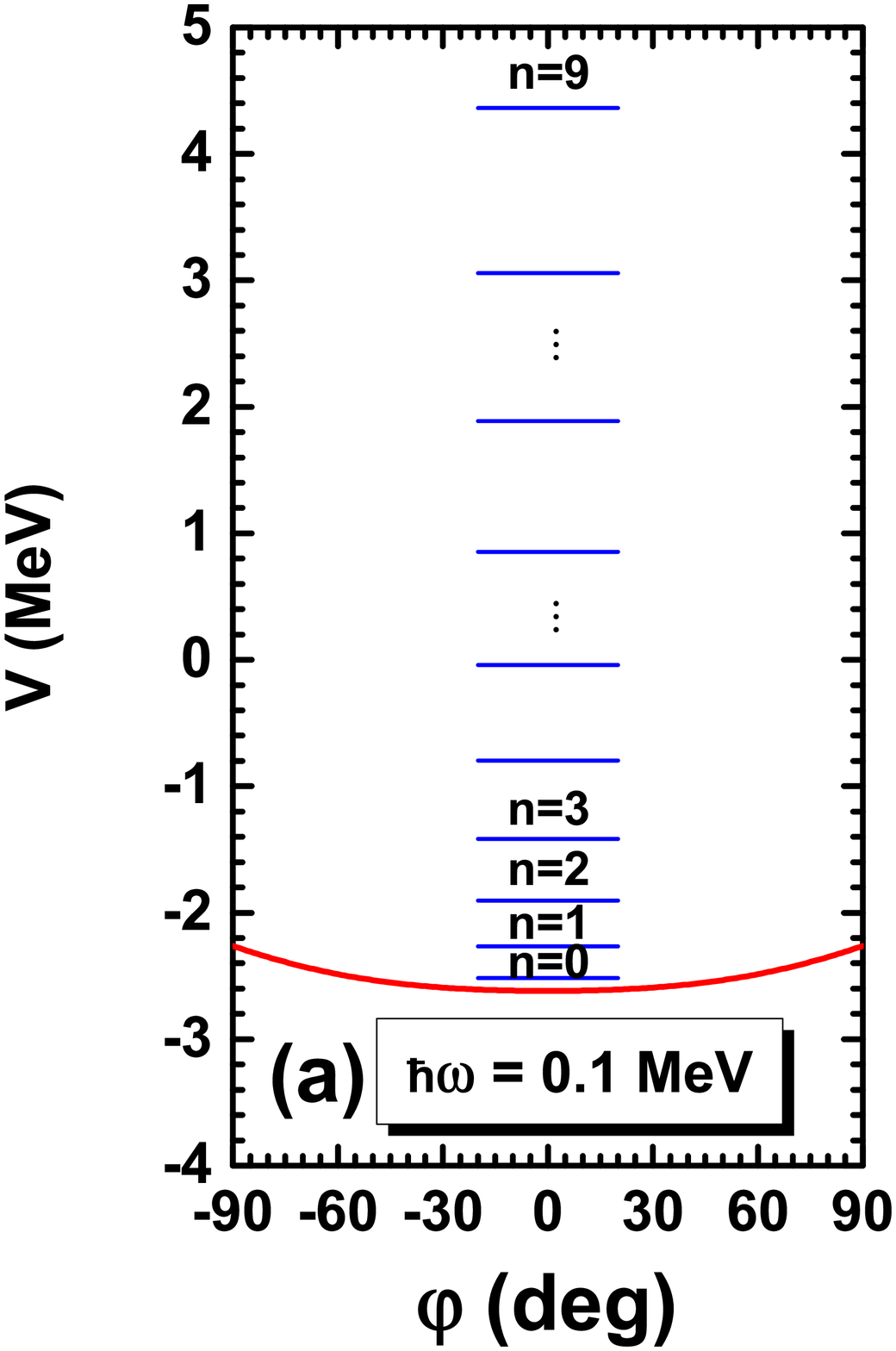}\quad\quad
   \includegraphics[width=4.55 cm]{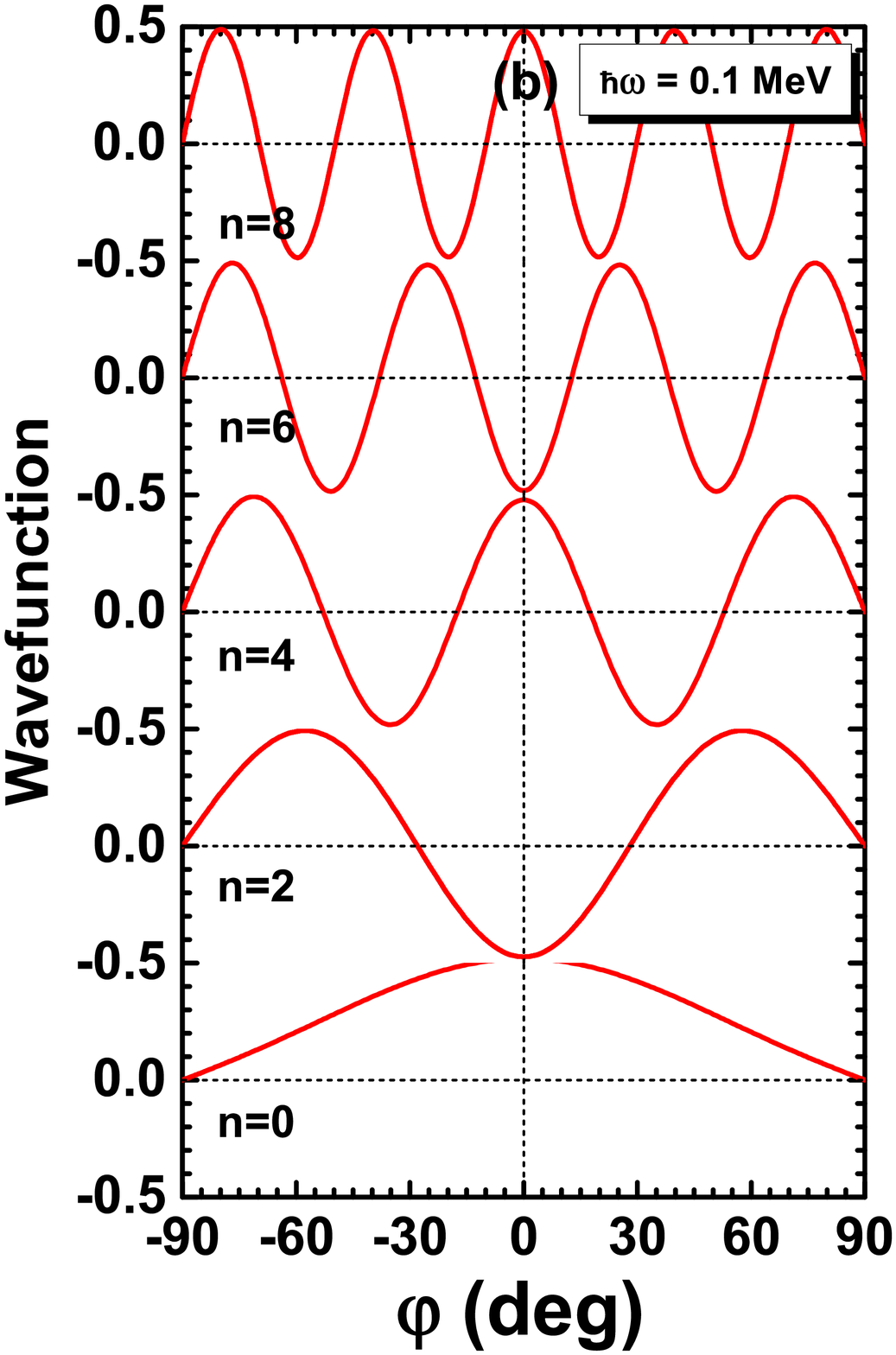}
   \includegraphics[width=4.55 cm]{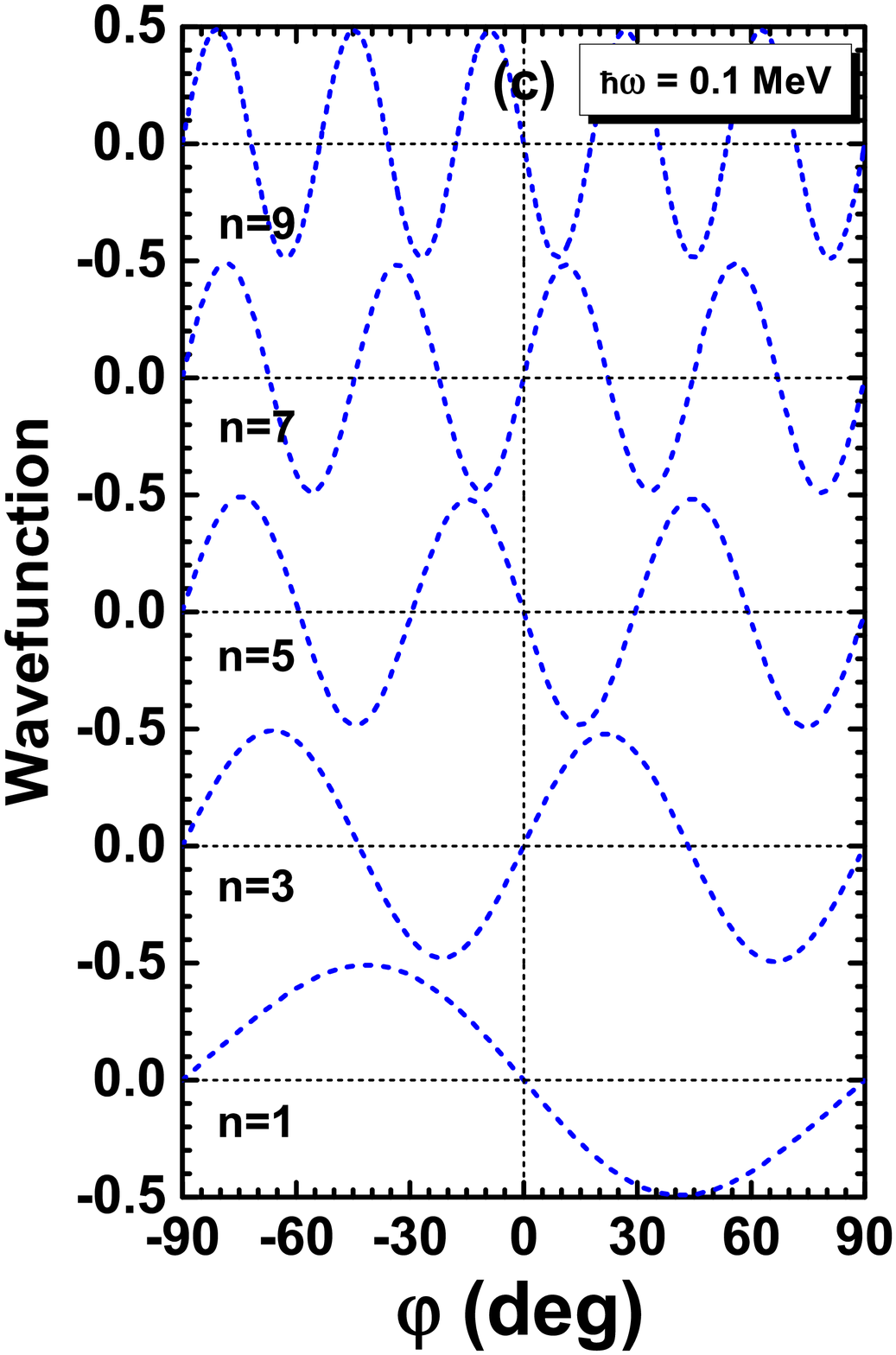}\\
   ~~\\
   \includegraphics[width=4.55 cm]{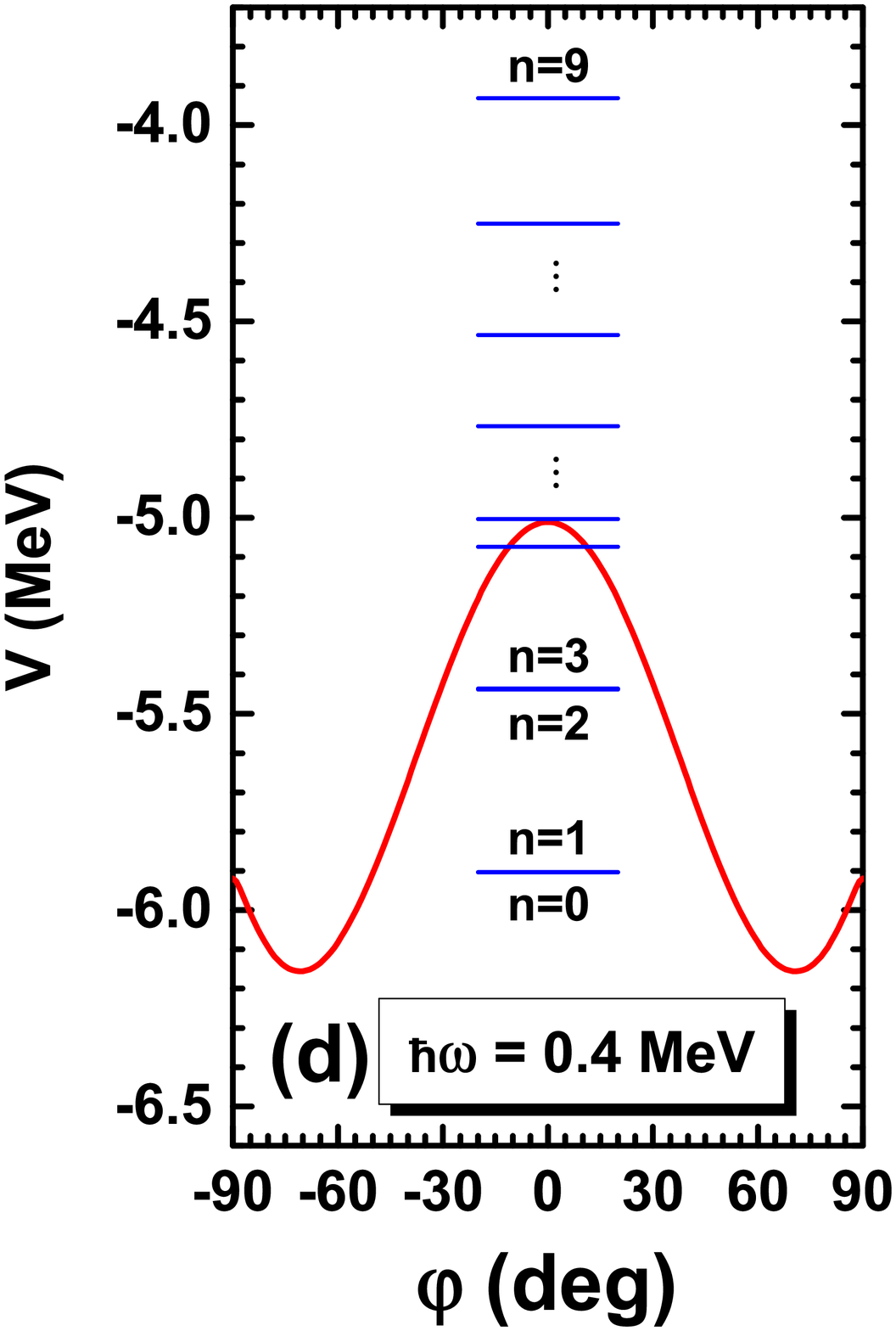}\quad\quad
   \includegraphics[width=4.55 cm]{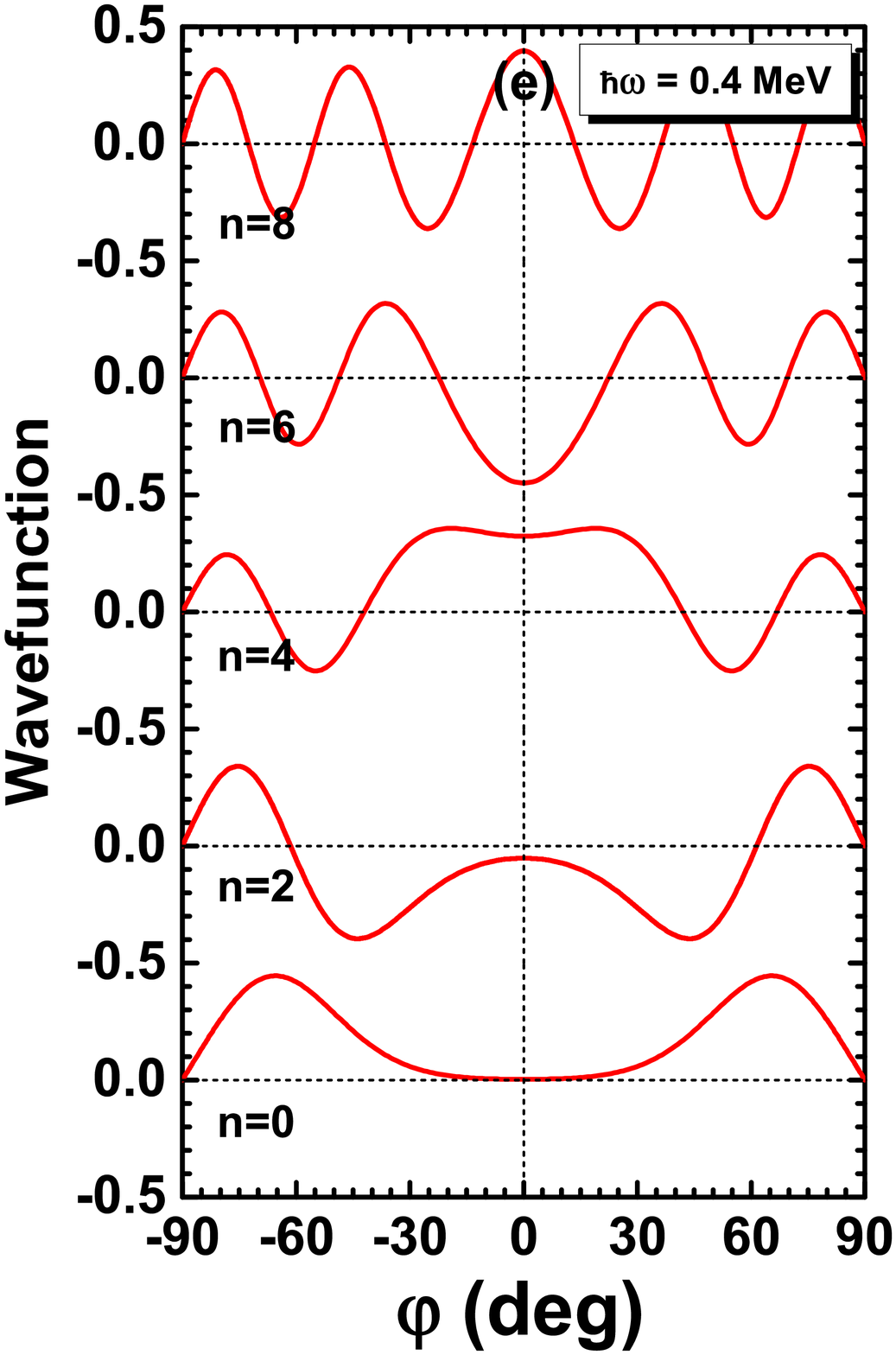}
   \includegraphics[width=4.55 cm]{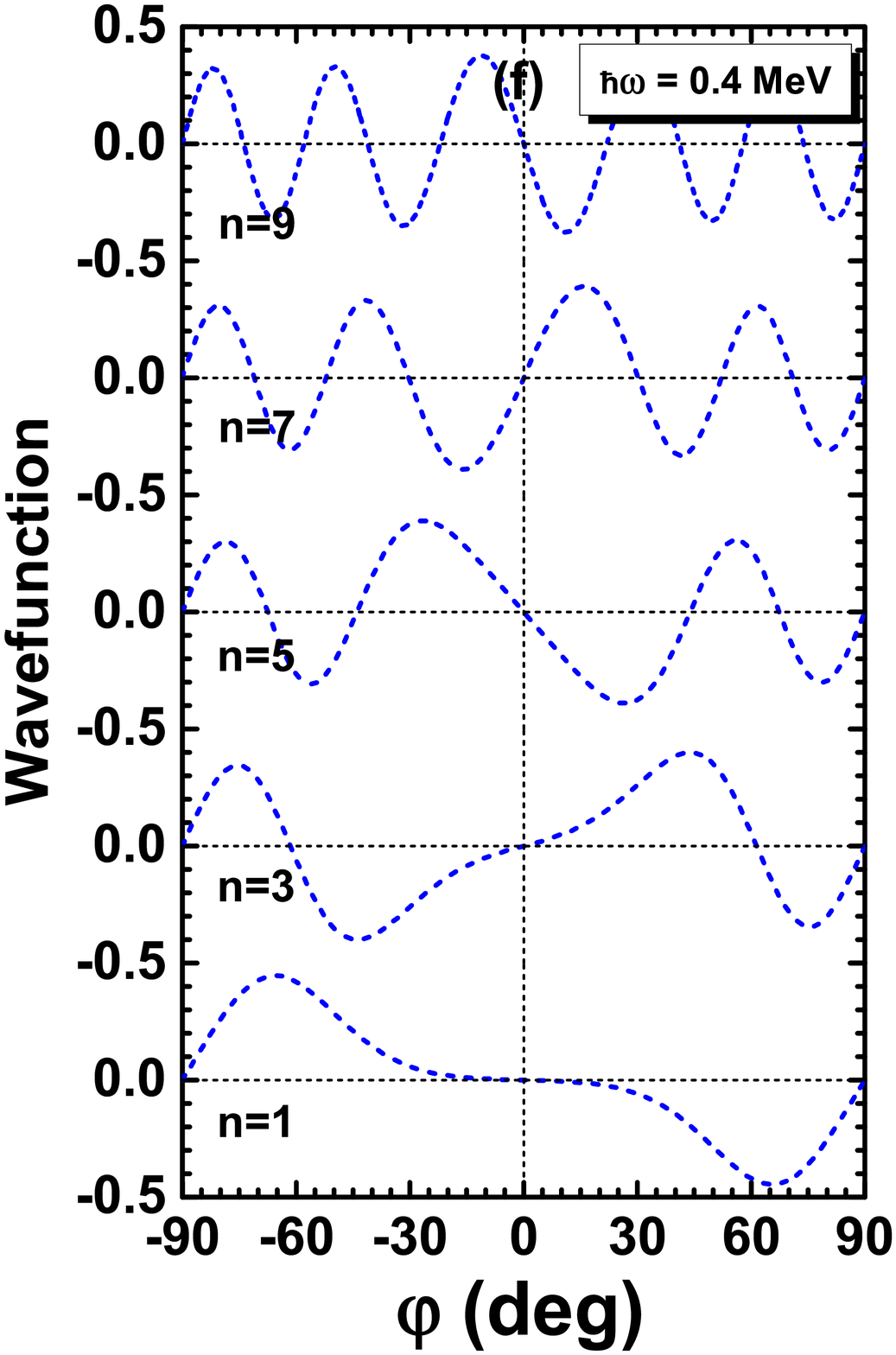}
   \caption{(Color online) Same as Fig.~\ref{fig3} but for transverse wobbling
            motion.}\label{fig15}
  \end{center}
\end{figure}

The obtained collective levels and wave functions are shown in Fig.~\ref{fig15}
for $\hbar\omega=0.1$, $0.4~\rm MeV$. Similar as the simple and longitudinal
wobbling motions, the wave functions are symmetric for even-$n$
levels and antisymmetric for odd-$n$ levels. For $n=0$, the peak of the wave
function locates around $\varphi=0^\circ$ at $\hbar\omega=0.1~\rm MeV$, while
moves towards to $\varphi=90^\circ$ at $\hbar\omega=0.4~\rm MeV$. In addition,
when the rotational frequency increases, the probability distributions
determined by the absolute square of wave functions tend to show similar pattern
for $n=0$ and $n=1$ levels. This is consistent with that their energy
differences, as shown in left panel of Fig.~\ref{fig15}, tend to zero.

\begin{figure}[!th]
  \begin{center}
    \includegraphics[width=7 cm]{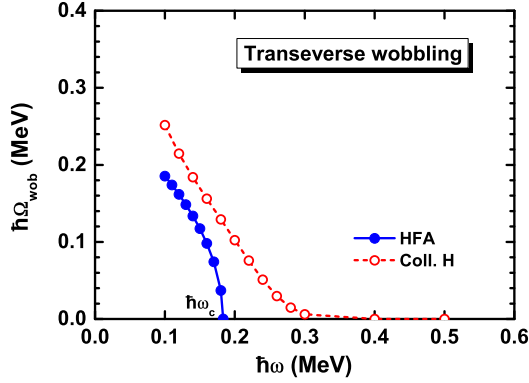}
    \caption{(Color online) Same as Fig.~\ref{fig10} but for transverse
             wobbling motion.}\label{fig16}
  \end{center}
\end{figure}

The calculated wobbling frequencies are shown in Fig.~\ref{fig16}. It can be seen
from Fig.~\ref{fig16}, the wobbling frequency decreases with the rotational frequency.
This decreasing is attributed to the increase of the potential barrier,
as shown in the upper panels of Fig.~\ref{fig13}, which will suppress the tunneling
probability between the two symmetrical TAC solutions. At $\hbar\omega\geq0.3~\rm MeV$,
the wobbling frequency tends to zero, which implies the transverse wobbling motion is
terminated. For comparison, the wobbling frequencies calculated by HFA are also shown
in Fig.~\ref{fig16}. The decreasing trend is clearly observed. As discussed above, at
the critical rotational frequency $\hbar\omega_{\rm c}=j_\pi/(\mathcal{J}_2-\mathcal{J}_1)
\approx 0.183~\rm MeV$ the wobbling frequency becomes zero and above it the HFA formula
becomes invalid. Comparing with the collective Hamiltonian, HFA gives about $50~\rm keV$
smaller values of wobbling frequency. It is also worthy to mention that the HFA gives a
more rapid decreasing trend than the collective Hamiltonian since the quantum fluctuations
are not taken into account in the HFA beyond the region of transverse wobbling motion.

\subsubsection{Comparison with PRM solutions}

\begin{figure}[!th]
  \begin{center}
    \includegraphics[width=7 cm]{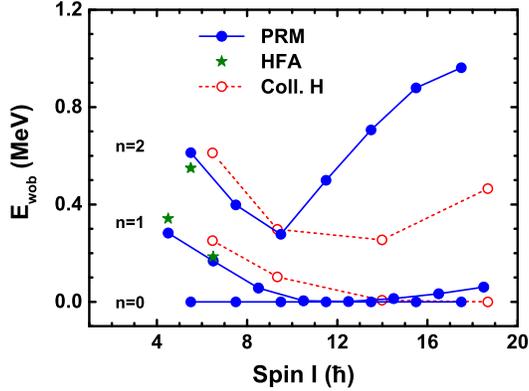}
    \caption{(Color online) Same as Fig.~\ref{fig12} but for transverse
             wobbling motion.}\label{fig17}
  \end{center}
\end{figure}

In Fig.~\ref{fig17}, the energies of the two lowest wobbling bands $n=1,2$
relative to the $n=0$ yrast sequence obtained by collective Hamiltonian
are shown in comparison with the PRM solutions and HFA results. It is found
that the collective Hamiltonian can reproduce the PRM results well at the
region of wobbling motions. For $I\geq 16.5\hbar$, the wobbling energies
of $n=1$ increase in the PRM, which indicates the onset of transitions
from the transverse to longitudinal wobbling motions as discussed in
Ref.~\cite{Frauendorf2014PRC}. This transition, however, is not reproduced
by the present collective Hamiltonian since the boundary conditions of wave
functions at $\varphi=\pm 90^\circ$ are assumed to be zero. Further investigation
on this topic will be done in the future.

\section{Summary}\label{sec5}

In summary, three types of wobble modes for the nucleus have been studied in the
framework of collective Hamiltonian. The simple wobbler is a pure triaxial rotor
assumed with rigid body type of moments of inertia. With an odd $h_{11/2}$ proton
of particle character coupling to the triaxial rotor, the longitudinal wobbler is
achieved by arranging the moments of inertia as rigid body type, while the transverse
wobbler achieved as irrotational body type. The collective potential in the collective
Hamiltonian are calculated based on TAC approach. The mass parameter are obtained by
HA for simple wobbling motion, while by HFA approximation for longitudinal and
transverse wobbling motions.

Diagonalizing the collective Hamiltonian, the energies and the wave functions of
the wobbling states are yielded. The obtained wobbling energies of simple wobbler
are compared with the results calculated by HA and TRM, while those of longitudinal
and transverse wobblers energies are compared with HFA and PRM. It is found that the
results of collective Hamiltonian are in good agreement with those exact solutions
by TRM or PRM.

In accord with those obtained by HA or HFA formula~\cite{Frauendorf2014PRC}, it is
observed that the wobbling frequency increases with the rotational frequency for
the simple and longitudinal wobbling motions, while decreases for the transverse
wobbling motion. It is presented here that these variation trends of the wobbling
frequency are in association with the stiffness of the collective potentials.
It should be mentioned that the present work has provided a new way to understand
the wobbling phenomena, which in particular may further contribute to the
investigation of nuclear wobbling based on a realistic TAC theory such as
tilted axis cranking density functional theory~\cite{J.Meng2013FP}.

%%%%%%%%%%%%%%%%%%%%%%%%%%%%%%%%%%%%%%%%%%%%%%%%%%%%%%%%%%%%%%%%
%                    Acknowlegments
%%%%%%%%%%%%%%%%%%%%%%%%%%%%%%%%%%%%%%%%%%%%%%%%%%%%%%%%%%%%%%%%

\section*{ACKNOWLEDGMENTS}

This work was supported in part by the Major State 973 Program of China (Grant No.
2013CB834400), the National Natural Science Foundation of China (Grants No. 11175002,
No. 11335002, No. 11375015, No. 11345004, No. 11105005), Research Fund for the
Doctoral Program of Higher Education (Grant No. 20110001110087).

%%%%%%%%%%%%%%%%%%%%%%%%%%%%%%%%%%%%%%%%%%%%%%%%%%%%%%%%%%%%%%%%%%%%%%%%
%                  begin refereee
%%%%%%%%%%%%%%%%%%%%%%%%%%%%%%%%%%%%%%%%%%%%%%%%%%%%%%%%%%%%%%%%%%%%%%%%
%\bibliography{../../../../M_Mybib/MYBIB}% Produces the bibliography via BibTeX.

%\bibliography{MYBIB}

\begin{thebibliography}{36}
\expandafter\ifx\csname natexlab\endcsname\relax\def\natexlab#1{#1}\fi
\expandafter\ifx\csname bibnamefont\endcsname\relax
  \def\bibnamefont#1{#1}\fi
\expandafter\ifx\csname bibfnamefont\endcsname\relax
  \def\bibfnamefont#1{#1}\fi
\expandafter\ifx\csname citenamefont\endcsname\relax
  \def\citenamefont#1{#1}\fi
\expandafter\ifx\csname url\endcsname\relax
  \def\url#1{\texttt{#1}}\fi
\expandafter\ifx\csname urlprefix\endcsname\relax\def\urlprefix{URL }\fi
\providecommand{\bibinfo}[2]{#2}
\providecommand{\eprint}[2][]{\url{#2}}

\bibitem[{\citenamefont{Bohr and Mottelson}(1975)}]{Bohr1975}
\bibinfo{author}{\bibfnamefont{A.}~\bibnamefont{Bohr}} \bibnamefont{and}
  \bibinfo{author}{\bibfnamefont{B.~R.} \bibnamefont{Mottelson}},
  \emph{\bibinfo{title}{Nuclear structure}}, vol.~\bibinfo{volume}{II}
  (\bibinfo{publisher}{Benjamin, New York}, \bibinfo{year}{1975}).

\bibitem[{\citenamefont{Bengtsson et~al.}(1984)\citenamefont{Bengtsson, Frisk,
  May, and Pinston}}]{Bengtsson1984NPA}
\bibinfo{author}{\bibfnamefont{R.}~\bibnamefont{Bengtsson}},
  \bibinfo{author}{\bibfnamefont{H.}~\bibnamefont{Frisk}},
  \bibinfo{author}{\bibfnamefont{F.}~\bibnamefont{May}}, \bibnamefont{and}
  \bibinfo{author}{\bibfnamefont{J.}~\bibnamefont{Pinston}},
  \bibinfo{journal}{Nucl. Phys. A} \textbf{\bibinfo{volume}{415}},
  \bibinfo{pages}{189 } (\bibinfo{year}{1984}).

\bibitem[{\citenamefont{Hamamoto and Sagawa}(1988)}]{Hamamoto1988PLB}
\bibinfo{author}{\bibfnamefont{I.}~\bibnamefont{Hamamoto}} \bibnamefont{and}
  \bibinfo{author}{\bibfnamefont{H.}~\bibnamefont{Sagawa}},
  \bibinfo{journal}{Phys. Lett. B} \textbf{\bibinfo{volume}{201}},
  \bibinfo{pages}{415 } (\bibinfo{year}{1988}).

\bibitem[{\citenamefont{Frauendorf and Meng}(1997)}]{Frauendorf1997NPA}
\bibinfo{author}{\bibfnamefont{S.}~\bibnamefont{Frauendorf}} \bibnamefont{and}
  \bibinfo{author}{\bibfnamefont{J.}~\bibnamefont{Meng}},
  \bibinfo{journal}{Nucl. Phys. A} \textbf{\bibinfo{volume}{617}},
  \bibinfo{pages}{131} (\bibinfo{year}{1997}).

\bibitem[{\citenamefont{Frauendorf}(2001)}]{Frauendorf2001RMP}
\bibinfo{author}{\bibfnamefont{S.}~\bibnamefont{Frauendorf}},
  \bibinfo{journal}{Rev. Mod. Phys.} \textbf{\bibinfo{volume}{73}},
  \bibinfo{pages}{463} (\bibinfo{year}{2001}).

\bibitem[{\citenamefont{Meng and Zhang}(2010)}]{Meng2010JPG}
\bibinfo{author}{\bibfnamefont{J.}~\bibnamefont{Meng}} \bibnamefont{and}
  \bibinfo{author}{\bibfnamefont{S.~Q.} \bibnamefont{Zhang}},
  \bibinfo{journal}{J. Phys. G: Nucl. Part. Phys.}
  \textbf{\bibinfo{volume}{37}}, \bibinfo{pages}{064025}
  (\bibinfo{year}{2010}).

\bibitem[{\citenamefont{Frauendorf and D\"onau}(2014)}]{Frauendorf2014PRC}
\bibinfo{author}{\bibfnamefont{S.}~\bibnamefont{Frauendorf}} \bibnamefont{and}
  \bibinfo{author}{\bibfnamefont{F.}~\bibnamefont{D\"onau}},
  \bibinfo{journal}{Phys. Rev. C} \textbf{\bibinfo{volume}{89}},
  \bibinfo{pages}{014322} (\bibinfo{year}{2014}).

\bibitem[{\citenamefont{\O{}deg\aa{}rd
  et~al.}(2001)\citenamefont{\O{}deg\aa{}rd, Hagemann, Jensen, Bergstr\"om,
  Herskind, Sletten, T\"orm\"anen, Wilson, Tj\o{}m, Hamamoto
  et~al.}}]{Odegard2001PRL}
\bibinfo{author}{\bibfnamefont{S.~W.} \bibnamefont{\O{}deg\aa{}rd}},
  \bibinfo{author}{\bibfnamefont{G.~B.} \bibnamefont{Hagemann}},
  \bibinfo{author}{\bibfnamefont{D.~R.} \bibnamefont{Jensen}},
  \bibinfo{author}{\bibfnamefont{M.}~\bibnamefont{Bergstr\"om}},
  \bibinfo{author}{\bibfnamefont{B.}~\bibnamefont{Herskind}},
  \bibinfo{author}{\bibfnamefont{G.}~\bibnamefont{Sletten}},
  \bibinfo{author}{\bibfnamefont{S.}~\bibnamefont{T\"orm\"anen}},
  \bibinfo{author}{\bibfnamefont{J.~N.} \bibnamefont{Wilson}},
  \bibinfo{author}{\bibfnamefont{P.~O.} \bibnamefont{Tj\o{}m}},
  \bibinfo{author}{\bibfnamefont{I.}~\bibnamefont{Hamamoto}},
  \bibnamefont{et~al.}, \bibinfo{journal}{Phys. Rev. Lett.}
  \textbf{\bibinfo{volume}{86}}, \bibinfo{pages}{5866} (\bibinfo{year}{2001}).

\bibitem[{\citenamefont{Jensen et~al.}(2002{\natexlab{a}})\citenamefont{Jensen,
  Hagemann, Hamamoto, \O{}deg\aa{}rd, Herskind, Sletten, Wilson, Spohr,
  H\"ubel, Bringel et~al.}}]{Jensen2002PRL}
\bibinfo{author}{\bibfnamefont{D.~R.} \bibnamefont{Jensen}},
  \bibinfo{author}{\bibfnamefont{G.~B.} \bibnamefont{Hagemann}},
  \bibinfo{author}{\bibfnamefont{I.}~\bibnamefont{Hamamoto}},
  \bibinfo{author}{\bibfnamefont{S.~W.} \bibnamefont{\O{}deg\aa{}rd}},
  \bibinfo{author}{\bibfnamefont{B.}~\bibnamefont{Herskind}},
  \bibinfo{author}{\bibfnamefont{G.}~\bibnamefont{Sletten}},
  \bibinfo{author}{\bibfnamefont{J.~N.} \bibnamefont{Wilson}},
  \bibinfo{author}{\bibfnamefont{K.}~\bibnamefont{Spohr}},
  \bibinfo{author}{\bibfnamefont{H.}~\bibnamefont{H\"ubel}},
  \bibinfo{author}{\bibfnamefont{P.}~\bibnamefont{Bringel}},
  \bibnamefont{et~al.}, \bibinfo{journal}{Phys. Rev. Lett.}
  \textbf{\bibinfo{volume}{89}}, \bibinfo{pages}{142503}
  (\bibinfo{year}{2002}{\natexlab{a}}).

\bibitem[{\citenamefont{Jensen et~al.}(2002{\natexlab{b}})\citenamefont{Jensen,
  Hagemann, Hamamoto, \O{}deg\aa{}rd, Bergstrom, Herskind, Sletten, Tormanen,
  Wilson, Tjom et~al.}}]{Jensen2002NPA}
\bibinfo{author}{\bibfnamefont{D.~R.} \bibnamefont{Jensen}},
  \bibinfo{author}{\bibfnamefont{G.~B.} \bibnamefont{Hagemann}},
  \bibinfo{author}{\bibfnamefont{I.}~\bibnamefont{Hamamoto}},
  \bibinfo{author}{\bibfnamefont{S.~W.} \bibnamefont{\O{}deg\aa{}rd}},
  \bibinfo{author}{\bibfnamefont{M.}~\bibnamefont{Bergstrom}},
  \bibinfo{author}{\bibfnamefont{B.}~\bibnamefont{Herskind}},
  \bibinfo{author}{\bibfnamefont{G.}~\bibnamefont{Sletten}},
  \bibinfo{author}{\bibfnamefont{S.}~\bibnamefont{Tormanen}},
  \bibinfo{author}{\bibfnamefont{J.~N.} \bibnamefont{Wilson}},
  \bibinfo{author}{\bibfnamefont{P.~O.} \bibnamefont{Tjom}},
  \bibnamefont{et~al.}, \bibinfo{journal}{Nucl. Phys. A}
  \textbf{\bibinfo{volume}{703}}, \bibinfo{pages}{3}
  (\bibinfo{year}{2002}{\natexlab{b}}).

\bibitem[{\citenamefont{Sch\"{o}nwa{\ss}er
  et~al.}(2003)\citenamefont{Sch\"{o}nwa{\ss}er, H\"{u}bel, Hagemann,
  Bednarczyk, Benzoni, Bracco, Bringel, Chapman, Curien, Domscheit
  et~al.}}]{Schonwasser2003PLB}
\bibinfo{author}{\bibfnamefont{G.}~\bibnamefont{Sch\"{o}nwa{\ss}er}},
  \bibinfo{author}{\bibfnamefont{H.}~\bibnamefont{H\"{u}bel}},
  \bibinfo{author}{\bibfnamefont{G.~B.} \bibnamefont{Hagemann}},
  \bibinfo{author}{\bibfnamefont{P.}~\bibnamefont{Bednarczyk}},
  \bibinfo{author}{\bibfnamefont{G.}~\bibnamefont{Benzoni}},
  \bibinfo{author}{\bibfnamefont{A.}~\bibnamefont{Bracco}},
  \bibinfo{author}{\bibfnamefont{P.}~\bibnamefont{Bringel}},
  \bibinfo{author}{\bibfnamefont{R.}~\bibnamefont{Chapman}},
  \bibinfo{author}{\bibfnamefont{D.}~\bibnamefont{Curien}},
  \bibinfo{author}{\bibfnamefont{J.}~\bibnamefont{Domscheit}},
  \bibnamefont{et~al.}, \bibinfo{journal}{Phys. Lett. B}
  \textbf{\bibinfo{volume}{552}}, \bibinfo{pages}{9} (\bibinfo{year}{2003}).

\bibitem[{\citenamefont{Bringel et~al.}(2005)\citenamefont{Bringel, Hagemann,
  H\"{u}bel, Al-khatib, Bednarczyk, B\"{u}rger, Curien, Gangopadhyay, Herskind,
  Jensen et~al.}}]{Bringel2005EPJA}
\bibinfo{author}{\bibfnamefont{P.}~\bibnamefont{Bringel}},
  \bibinfo{author}{\bibfnamefont{G.}~\bibnamefont{Hagemann}},
  \bibinfo{author}{\bibfnamefont{H.}~\bibnamefont{H\"{u}bel}},
  \bibinfo{author}{\bibfnamefont{A.}~\bibnamefont{Al-khatib}},
  \bibinfo{author}{\bibfnamefont{P.}~\bibnamefont{Bednarczyk}},
  \bibinfo{author}{\bibfnamefont{A.}~\bibnamefont{B\"{u}rger}},
  \bibinfo{author}{\bibfnamefont{D.}~\bibnamefont{Curien}},
  \bibinfo{author}{\bibfnamefont{G.}~\bibnamefont{Gangopadhyay}},
  \bibinfo{author}{\bibfnamefont{B.}~\bibnamefont{Herskind}},
  \bibinfo{author}{\bibfnamefont{D.}~\bibnamefont{Jensen}},
  \bibnamefont{et~al.}, \bibinfo{journal}{Eur. Phys. J. A}
  \textbf{\bibinfo{volume}{24}}, \bibinfo{pages}{167} (\bibinfo{year}{2005}).

\bibitem[{\citenamefont{Amro et~al.}(2003)\citenamefont{Amro, Ma, Hagemann,
  Diamond, Domscheit, Fallon, Gorgen, Herskind, Hubel, Jensen
  et~al.}}]{Amro2003PLB}
\bibinfo{author}{\bibfnamefont{H.}~\bibnamefont{Amro}},
  \bibinfo{author}{\bibfnamefont{W.~C.} \bibnamefont{Ma}},
  \bibinfo{author}{\bibfnamefont{G.~B.} \bibnamefont{Hagemann}},
  \bibinfo{author}{\bibfnamefont{R.~M.} \bibnamefont{Diamond}},
  \bibinfo{author}{\bibfnamefont{J.}~\bibnamefont{Domscheit}},
  \bibinfo{author}{\bibfnamefont{P.}~\bibnamefont{Fallon}},
  \bibinfo{author}{\bibfnamefont{A.}~\bibnamefont{Gorgen}},
  \bibinfo{author}{\bibfnamefont{B.}~\bibnamefont{Herskind}},
  \bibinfo{author}{\bibfnamefont{H.}~\bibnamefont{Hubel}},
  \bibinfo{author}{\bibfnamefont{D.~R.} \bibnamefont{Jensen}},
  \bibnamefont{et~al.}, \bibinfo{journal}{Phys. Lett. B}
  \textbf{\bibinfo{volume}{553}}, \bibinfo{pages}{197} (\bibinfo{year}{2003}).

\bibitem[{\citenamefont{Hagemann}(2004)}]{Hagemann2004EPJA}
\bibinfo{author}{\bibfnamefont{G.~B.} \bibnamefont{Hagemann}},
  \bibinfo{journal}{Eur. Phys. J. A} \textbf{\bibinfo{volume}{20}},
  \bibinfo{pages}{183} (\bibinfo{year}{2004}).

\bibitem[{\citenamefont{Hartley et~al.}(2009)\citenamefont{Hartley, Janssens,
  Riedinger, Riley, Aguilar, Carpenter, Chiara, Chowdhury, Darby, Garg
  et~al.}}]{Hartley2009PRC}
\bibinfo{author}{\bibfnamefont{D.~J.} \bibnamefont{Hartley}},
  \bibinfo{author}{\bibfnamefont{R.~V.~F.} \bibnamefont{Janssens}},
  \bibinfo{author}{\bibfnamefont{L.~L.} \bibnamefont{Riedinger}},
  \bibinfo{author}{\bibfnamefont{M.~A.} \bibnamefont{Riley}},
  \bibinfo{author}{\bibfnamefont{A.}~\bibnamefont{Aguilar}},
  \bibinfo{author}{\bibfnamefont{M.~P.} \bibnamefont{Carpenter}},
  \bibinfo{author}{\bibfnamefont{C.~J.} \bibnamefont{Chiara}},
  \bibinfo{author}{\bibfnamefont{P.}~\bibnamefont{Chowdhury}},
  \bibinfo{author}{\bibfnamefont{I.~G.} \bibnamefont{Darby}},
  \bibinfo{author}{\bibfnamefont{U.}~\bibnamefont{Garg}}, \bibnamefont{et~al.},
  \bibinfo{journal}{Phys. Rev. C} \textbf{\bibinfo{volume}{80}},
  \bibinfo{pages}{041304} (\bibinfo{year}{2009}).

\bibitem[{\citenamefont{Zhu et~al.}(2009)\citenamefont{Zhu, Luo, Hamilton,
  Ramayya, Che, Jiang, Hwang, Wood, Stoyer, Donangelo
  et~al.}}]{S.J.Zhu2009IJMPE}
\bibinfo{author}{\bibfnamefont{S.~J.} \bibnamefont{Zhu}},
  \bibinfo{author}{\bibfnamefont{Y.~X.} \bibnamefont{Luo}},
  \bibinfo{author}{\bibfnamefont{J.~H.} \bibnamefont{Hamilton}},
  \bibinfo{author}{\bibfnamefont{A.~V.} \bibnamefont{Ramayya}},
  \bibinfo{author}{\bibfnamefont{X.~L.} \bibnamefont{Che}},
  \bibinfo{author}{\bibfnamefont{Z.}~\bibnamefont{Jiang}},
  \bibinfo{author}{\bibfnamefont{J.~K.} \bibnamefont{Hwang}},
  \bibinfo{author}{\bibfnamefont{J.~L.} \bibnamefont{Wood}},
  \bibinfo{author}{\bibfnamefont{M.~A.} \bibnamefont{Stoyer}},
  \bibinfo{author}{\bibfnamefont{R.}~\bibnamefont{Donangelo}},
  \bibnamefont{et~al.}, \bibinfo{journal}{Int. J. Mod. Phys. E}
  \textbf{\bibinfo{volume}{18}}, \bibinfo{pages}{1717} (\bibinfo{year}{2009}).

\bibitem[{\citenamefont{Hamamoto}(2002)}]{Hamamoto2002PRC}
\bibinfo{author}{\bibfnamefont{I.}~\bibnamefont{Hamamoto}},
  \bibinfo{journal}{Phys. Rev. C} \textbf{\bibinfo{volume}{65}},
  \bibinfo{pages}{044305} (\bibinfo{year}{2002}).

\bibitem[{\citenamefont{Hamamoto and Mottelson}(2003)}]{Hamamoto2003PRC}
\bibinfo{author}{\bibfnamefont{I.}~\bibnamefont{Hamamoto}} \bibnamefont{and}
  \bibinfo{author}{\bibfnamefont{B.~R.} \bibnamefont{Mottelson}},
  \bibinfo{journal}{Phys. Rev. C} \textbf{\bibinfo{volume}{68}},
  \bibinfo{pages}{034312} (\bibinfo{year}{2003}).

\bibitem[{\citenamefont{Tanabe and Sugawara-Tanabe}(2006)}]{Tanabe2006PRC}
\bibinfo{author}{\bibfnamefont{K.}~\bibnamefont{Tanabe}} \bibnamefont{and}
  \bibinfo{author}{\bibfnamefont{K.}~\bibnamefont{Sugawara-Tanabe}},
  \bibinfo{journal}{Phys. Rev. C} \textbf{\bibinfo{volume}{73}},
  \bibinfo{pages}{034305} (\bibinfo{year}{2006}).

\bibitem[{\citenamefont{Tanabe and Sugawara-Tanabe}(2008)}]{Tanabe2008PRC}
\bibinfo{author}{\bibfnamefont{K.}~\bibnamefont{Tanabe}} \bibnamefont{and}
  \bibinfo{author}{\bibfnamefont{K.}~\bibnamefont{Sugawara-Tanabe}},
  \bibinfo{journal}{Phys. Rev. C} \textbf{\bibinfo{volume}{77}},
  \bibinfo{pages}{064318} (\bibinfo{year}{2008}).

\bibitem[{\citenamefont{Shimizu and Matsuzaki}(1995)}]{Shimizu1995NPA}
\bibinfo{author}{\bibfnamefont{Y.~R.} \bibnamefont{Shimizu}} \bibnamefont{and}
  \bibinfo{author}{\bibfnamefont{M.}~\bibnamefont{Matsuzaki}},
  \bibinfo{journal}{Nucl. Phys. A} \textbf{\bibinfo{volume}{588}},
  \bibinfo{pages}{559} (\bibinfo{year}{1995}).

\bibitem[{\citenamefont{Matsuzaki et~al.}(2002)\citenamefont{Matsuzaki,
  Shimizu, and Matsuyanagi}}]{Matsuzaki2002PRC}
\bibinfo{author}{\bibfnamefont{M.}~\bibnamefont{Matsuzaki}},
  \bibinfo{author}{\bibfnamefont{Y.~R.} \bibnamefont{Shimizu}},
  \bibnamefont{and}
  \bibinfo{author}{\bibfnamefont{K.}~\bibnamefont{Matsuyanagi}},
  \bibinfo{journal}{Phys. Rev. C} \textbf{\bibinfo{volume}{65}},
  \bibinfo{pages}{041303} (\bibinfo{year}{2002}).

\bibitem[{\citenamefont{Matsuzaki et~al.}(2003)\citenamefont{Matsuzaki,
  Shimizu, and Matsuyanagi}}]{Matsuzaki2003EPJA}
\bibinfo{author}{\bibfnamefont{M.}~\bibnamefont{Matsuzaki}},
  \bibinfo{author}{\bibfnamefont{Y.~R.} \bibnamefont{Shimizu}},
  \bibnamefont{and}
  \bibinfo{author}{\bibfnamefont{K.}~\bibnamefont{Matsuyanagi}},
  \bibinfo{journal}{Eur. Phys. J. A} \textbf{\bibinfo{volume}{20}},
  \bibinfo{pages}{189} (\bibinfo{year}{2003}).

\bibitem[{\citenamefont{Matsuzaki and Ohtsubo}(2004)}]{Matsuzaki2004PRCa}
\bibinfo{author}{\bibfnamefont{M.}~\bibnamefont{Matsuzaki}} \bibnamefont{and}
  \bibinfo{author}{\bibfnamefont{S.}~\bibnamefont{Ohtsubo}},
  \bibinfo{journal}{Phys. Rev. C} \textbf{\bibinfo{volume}{69}},
  \bibinfo{pages}{064317} (\bibinfo{year}{2004}).

\bibitem[{\citenamefont{Matsuzaki et~al.}(2004)\citenamefont{Matsuzaki,
  Shimizu, and Matsuyanagi}}]{Matsuzaki2004PRC}
\bibinfo{author}{\bibfnamefont{M.}~\bibnamefont{Matsuzaki}},
  \bibinfo{author}{\bibfnamefont{Y.~R.} \bibnamefont{Shimizu}},
  \bibnamefont{and}
  \bibinfo{author}{\bibfnamefont{K.}~\bibnamefont{Matsuyanagi}},
  \bibinfo{journal}{Phys. Rev. C} \textbf{\bibinfo{volume}{69}},
  \bibinfo{pages}{034325} (\bibinfo{year}{2004}).

\bibitem[{\citenamefont{Shimizu et~al.}(2005)\citenamefont{Shimizu, Matsuzaki,
  and Matsuyanagi}}]{Shimizu2005PRC}
\bibinfo{author}{\bibfnamefont{Y.~R.} \bibnamefont{Shimizu}},
  \bibinfo{author}{\bibfnamefont{M.}~\bibnamefont{Matsuzaki}},
  \bibnamefont{and}
  \bibinfo{author}{\bibfnamefont{K.}~\bibnamefont{Matsuyanagi}},
  \bibinfo{journal}{Phys. Rev. C} \textbf{\bibinfo{volume}{72}},
  \bibinfo{pages}{014306} (\bibinfo{year}{2005}).

\bibitem[{\citenamefont{Shimizu et~al.}(2008)\citenamefont{Shimizu, Shoji, and
  Matsuzaki}}]{Shimizu2008PRC}
\bibinfo{author}{\bibfnamefont{Y.~R.} \bibnamefont{Shimizu}},
  \bibinfo{author}{\bibfnamefont{T.}~\bibnamefont{Shoji}}, \bibnamefont{and}
  \bibinfo{author}{\bibfnamefont{M.}~\bibnamefont{Matsuzaki}},
  \bibinfo{journal}{Phys. Rev. C} \textbf{\bibinfo{volume}{77}},
  \bibinfo{pages}{024319} (\bibinfo{year}{2008}).

\bibitem[{\citenamefont{Shoji and Shimizu}(2009)}]{Shoji2009PTP}
\bibinfo{author}{\bibfnamefont{T.}~\bibnamefont{Shoji}} \bibnamefont{and}
  \bibinfo{author}{\bibfnamefont{Y.~R.} \bibnamefont{Shimizu}},
  \bibinfo{journal}{Progr. Theor. Phys.} \textbf{\bibinfo{volume}{121}},
  \bibinfo{pages}{319} (\bibinfo{year}{2009}).

\bibitem[{\citenamefont{Oi et~al.}(2000)\citenamefont{Oi, Ansari, Horibata, and
  Onishi}}]{Oi2000PLB}
\bibinfo{author}{\bibfnamefont{M.}~\bibnamefont{Oi}},
  \bibinfo{author}{\bibfnamefont{A.}~\bibnamefont{Ansari}},
  \bibinfo{author}{\bibfnamefont{T.}~\bibnamefont{Horibata}}, \bibnamefont{and}
  \bibinfo{author}{\bibfnamefont{N.}~\bibnamefont{Onishi}},
  \bibinfo{journal}{Phys. Lett. B} \textbf{\bibinfo{volume}{480}},
  \bibinfo{pages}{53} (\bibinfo{year}{2000}).

\bibitem[{\citenamefont{Chen et~al.}(2013)\citenamefont{Chen, Zhang, Zhao,
  Jolos, and Meng}}]{Q.B.Chen2013PRC}
\bibinfo{author}{\bibfnamefont{Q.~B.} \bibnamefont{Chen}},
  \bibinfo{author}{\bibfnamefont{S.~Q.} \bibnamefont{Zhang}},
  \bibinfo{author}{\bibfnamefont{P.~W.} \bibnamefont{Zhao}},
  \bibinfo{author}{\bibfnamefont{R.~V.} \bibnamefont{Jolos}}, \bibnamefont{and}
  \bibinfo{author}{\bibfnamefont{J.}~\bibnamefont{Meng}},
  \bibinfo{journal}{Phys. Rev. C} \textbf{\bibinfo{volume}{87}},
  \bibinfo{pages}{024314} (\bibinfo{year}{2013}).

\bibitem[{\citenamefont{Mukhopadhyay et~al.}(2007)\citenamefont{Mukhopadhyay,
  Almehed, Garg, Frauendorf, Li, Rao, Wang, Ghugre, Carpenter, Gros
  et~al.}}]{Mukhopadhyay2007PRL}
\bibinfo{author}{\bibfnamefont{S.}~\bibnamefont{Mukhopadhyay}},
  \bibinfo{author}{\bibfnamefont{D.}~\bibnamefont{Almehed}},
  \bibinfo{author}{\bibfnamefont{U.}~\bibnamefont{Garg}},
  \bibinfo{author}{\bibfnamefont{S.}~\bibnamefont{Frauendorf}},
  \bibinfo{author}{\bibfnamefont{T.}~\bibnamefont{Li}},
  \bibinfo{author}{\bibfnamefont{P.~V.~M.} \bibnamefont{Rao}},
  \bibinfo{author}{\bibfnamefont{X.}~\bibnamefont{Wang}},
  \bibinfo{author}{\bibfnamefont{S.~S.} \bibnamefont{Ghugre}},
  \bibinfo{author}{\bibfnamefont{M.~P.} \bibnamefont{Carpenter}},
  \bibinfo{author}{\bibfnamefont{S.}~\bibnamefont{Gros}}, \bibnamefont{et~al.},
  \bibinfo{journal}{Phys. Rev. Lett.} \textbf{\bibinfo{volume}{99}},
  \bibinfo{pages}{172501} (\bibinfo{year}{2007}).

\bibitem[{\citenamefont{Qi et~al.}(2009)\citenamefont{Qi, Zhang, Meng, Wang,
  and Frauendorf}}]{B.Qi2009PLB}
\bibinfo{author}{\bibfnamefont{B.}~\bibnamefont{Qi}},
  \bibinfo{author}{\bibfnamefont{S.~Q.} \bibnamefont{Zhang}},
  \bibinfo{author}{\bibfnamefont{J.}~\bibnamefont{Meng}},
  \bibinfo{author}{\bibfnamefont{S.~Y.} \bibnamefont{Wang}}, \bibnamefont{and}
  \bibinfo{author}{\bibfnamefont{S.}~\bibnamefont{Frauendorf}},
  \bibinfo{journal}{Phys. Lett. B} \textbf{\bibinfo{volume}{675}},
  \bibinfo{pages}{175} (\bibinfo{year}{2009}).

\bibitem[{\citenamefont{Oi}(2006)}]{Oi2006PLB}
\bibinfo{author}{\bibfnamefont{M.}~\bibnamefont{Oi}}, \bibinfo{journal}{Phys.
  Lett. B} \textbf{\bibinfo{volume}{634}}, \bibinfo{pages}{30}
  (\bibinfo{year}{2006}).

\bibitem[{\citenamefont{Pauli}(1933)}]{Pauli1933}
\bibinfo{author}{\bibfnamefont{W.}~\bibnamefont{Pauli}},
  \emph{\bibinfo{title}{in Handbuch der Physik}}, vol. \bibinfo{volume}{XXIV,
  p. 120} (\bibinfo{publisher}{Springer Verlag, Berlin}, \bibinfo{year}{1933}).

\bibitem[{\citenamefont{Ring and Schuck}(1980)}]{Ring1980book}
\bibinfo{author}{\bibfnamefont{P.}~\bibnamefont{Ring}} \bibnamefont{and}
  \bibinfo{author}{\bibfnamefont{P.}~\bibnamefont{Schuck}},
  \emph{\bibinfo{title}{The nuclear many body problem}}
  (\bibinfo{publisher}{Springer Verlag, Berlin}, \bibinfo{year}{1980}).

\bibitem[{\citenamefont{Meng et~al.}(2013)\citenamefont{Meng, Peng, Zhang, and
  Zhao}}]{J.Meng2013FP}
\bibinfo{author}{\bibfnamefont{J.}~\bibnamefont{Meng}},
  \bibinfo{author}{\bibfnamefont{J.}~\bibnamefont{Peng}},
  \bibinfo{author}{\bibfnamefont{S.~Q.} \bibnamefont{Zhang}}, \bibnamefont{and}
  \bibinfo{author}{\bibfnamefont{P.~W.} \bibnamefont{Zhao}},
  \bibinfo{journal}{Front. Phys.} \textbf{\bibinfo{volume}{8}},
  \bibinfo{pages}{55} (\bibinfo{year}{2013}).

\end{thebibliography}

%%%%%%%%%%%%%%%%%%%%%%%%%%%%%%%%%%%%%%%%%%%%%%%%%%%%%%%%%%%%%%%%%%%%%%%%
\end{CJK*}
\end{document}